\documentclass[twocolumn,showpacs,notitlepage,nofootinbib,floatfix,superscriptaddress,longbibliography]{revtex4-2} 
\usepackage{graphicx}
\usepackage{amssymb,amsmath,amsthm,amsfonts,upgreek}
\usepackage{hyperref} 
\usepackage{enumitem,float,varioref,mathtools}
\renewcommand{\hat}{\widehat}

\DeclareFontEncoding{LGR}{}{}
\DeclareFontSubstitution{LGR}{cmr}{m}{n}
\DeclareSymbolFont{upgreek}{LGR}{cmr}{m}{n}
\DeclareMathSymbol{\varchi}{\mathord}{upgreek}{`q}

\begin{document}

 \title{Protecting information via probabilistic cellular automata}
	\author{Annie Ray}
	\affiliation{Institute for Quantum Computing, University of Waterloo, Waterloo, ON N2L 3G1, Canada}
	\affiliation{Department of Physics and Astronomy, University of Waterloo, Waterloo, ON N2L 3G1, Canada}
	\affiliation{Perimeter Institute for Theoretical Physics, Waterloo, ON N2L 2Y5, Canada}
	\author{Raymond Laflamme}
	\affiliation{Institute for Quantum Computing, University of Waterloo, Waterloo, ON N2L 3G1, Canada}
	\affiliation{Department of Physics and Astronomy, University of Waterloo, Waterloo, ON N2L 3G1, Canada}
	\affiliation{Perimeter Institute for Theoretical Physics, Waterloo, ON N2L 2Y5, Canada}
	\author{Aleksander Kubica}
	\affiliation{AWS Center for Quantum Computing, Pasadena, CA 91125, USA}
	\affiliation{California Institute of Technology, Pasadena, CA 91125, USA}
	
\begin{abstract}
Probabilistic cellular automata describe the dynamics of classical spin models, which, for sufficiently small temperature $T$, can serve as classical memory capable of storing information even in the presence of nonzero external magnetic field $h$. 
In this article, we study a recently-introduced probabilistic cellular automaton, the sweep rule, and map out a region of two coexisting stable phases in the $(T,h)$ plane.
We also find that the sweep rule belongs to the weak two-dimensional Ising universality class. 
Our work is a step towards understanding how simple geometrically-local error-correction strategies can protect information encoded into complex noisy systems, such as topological quantum error-correcting codes.
\end{abstract}

\maketitle
	
Memory is a crucial component of computers, allowing to reliably store information for long durations of time and preserve it throughout computation.
The two-dimensional (2D) Ising ferromagnet~\cite{Lenz:460663, mccoyTwodimensionalIsingModel1973a} is a canonical toy model of classical memory.
In the absence of an external magnetic field, $h=0$, and for sufficiently small but nonzero temperature, $T>0$, the Ising model exhibits two coexisting phases with opposite magnetizations~\cite{peierlsIsingModelFerromagnetism1936, onsagerCrystalStatisticsTwoDimensional1944}, which can be used to encode one bit of information.
Although individual spins can flip due to thermal fluctuations, the sign of the magnetization of the system is unlikely to change and, subsequently, the encoded bit is protected for time that grows exponentially with the system size~\cite{dayIsingFerromagnetSelfcorrecting2012}.

Unfortunately, in the presence of any nonzero magnetic field, $h\neq 0$, the Ising model has only one stable phase (whose magnetization has the same sign as $h$), and thus cannot work as memory.
However, for sufficiently small temperature $T$, one can realize memory based on a classical spin model on the square lattice with periodic boundary conditions whose nonergodic dynamics is governed by a probabilistic cellular automaton (PCA)~\cite{wolframStatisticalMechanicsCellular1983}, such as Toom's rule~\cite{toom1980stable}.
This is possible because for such a spin model there is a region of nonzero measure in the $(T,h)$ plane, where two stable phases coexist~\cite{bennettRoleIrreversibilityStabilizing1985, slowinskiPhaseDiagramsMajority2015}.

\begin{figure}[ht!]
\centering
\includegraphics[width=0.9\columnwidth]{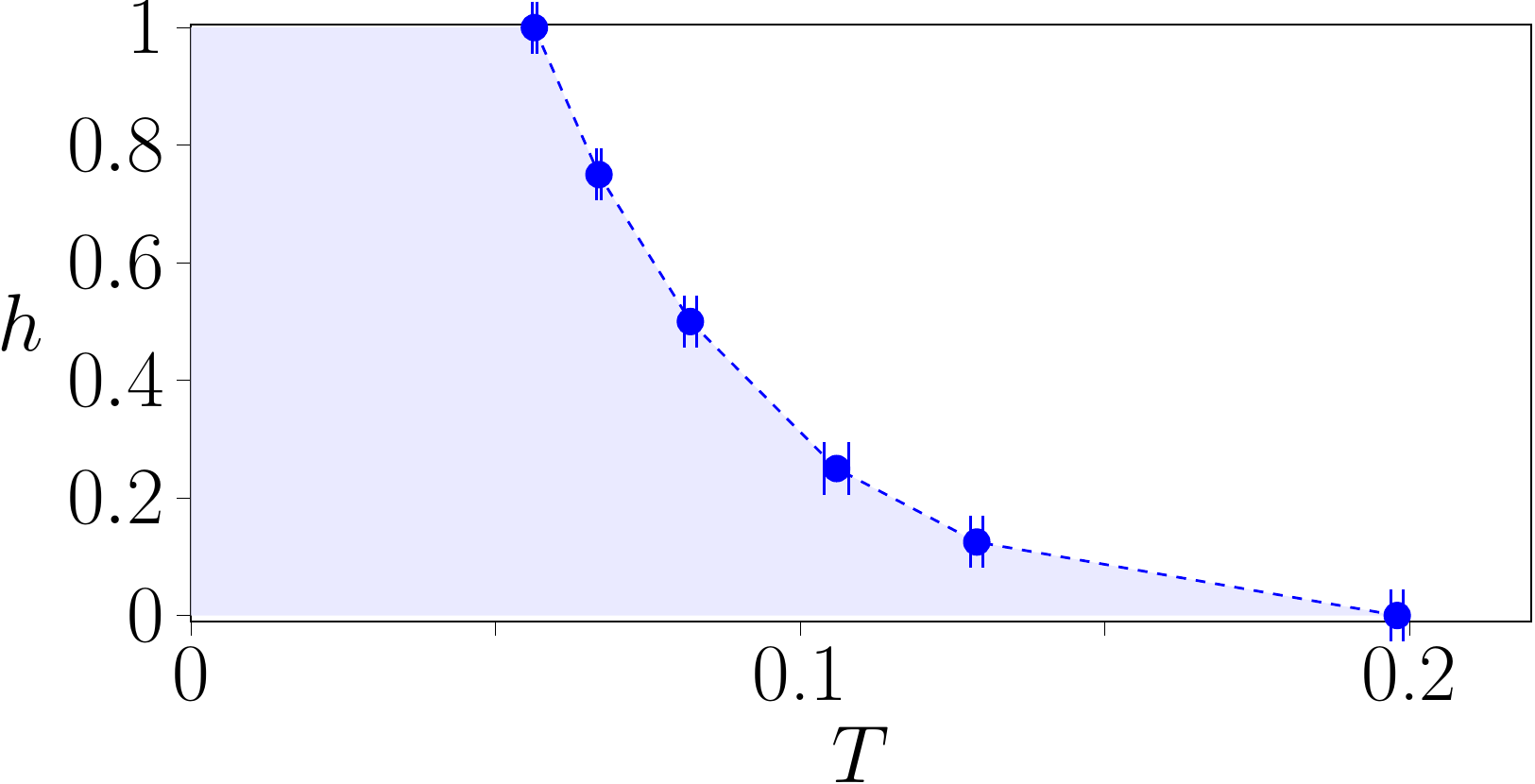}
\caption{Phase diagram of the sweep rule with noise parameters $p,q\in[0,1]$ in the $(T,h)$ plane, where $T=p+q$ and $h=(p-q)/(p+q)$ can be interpreted as temperature and bias.
The shaded region corresponds to a region, where two stable phases coexist and the sweep rule can work as memory.
The diagram is symmetric with respect to the $h=0$ axis.}
\label{PhaseDiag}
\end{figure}

Here, we study a classical memory based on the sweep rule~\cite{kubicaCellularAutomatonDecodersProvable2019}, which is a recently introduced generalization of Toom's rule applicable to spin models on any lattices in $d\geq 2$ dimensions.
In particular, we focus on the spin model on the triangular lattice with periodic boundary conditions and map out a region of two coexisting stable phases in the $(T,h)$ plane (see Fig.~\ref{PhaseDiag}).
We achieve this by studying the scaling of the memory lifetime.
We also find that, similarly to Toom's rule ~\cite{makowiecUniversalityClassProbabilistic2002}, the sweep rule belongs to the weak 2D Ising universality class, i.e., the ratios $\gamma/\nu$ and $\beta/\nu$ of its critical exponents are the same as for the Ising model.

The rest of the article is organized as follows.
In Sec.~\ref{ClassMem}, we describe the classical memory model based on the sweep rule.
In Sec.~\ref{Sim}, we numerically analyze the memory lifetime of the model.
To corroborate our results for $h=0$, we identify the phase transition of the sweep rule using statistical-mechanical techniques in Sec.~\ref{StatMech} and discuss a simple analytical model that captures the memory lifetime at high temperatures in Sec.~\ref{Analytical}.
We conclude by discussing the impact of our results in Sec.~\ref{Conclude}.

\section{Classical Memory Model}
\label{ClassMem}
Our model of classical memory consists of a 2D system of $N=2L^2$ classical two-state spins arranged on the faces of a triangular lattice of linear size $L$ with periodic boundary conditions.
The state of the $i^{\text{th}}$ spin is denoted by $s^{(t)}_{i} \in \{+1, -1\}$ for $i \in \{1, 2,...,N\}$ at discrete half-integer time steps $t = 0, 0.5, 1,\ldots$. At $t=0$, we \emph{encode} a logical bit in this system in one of two completely aligned configurations, either $\forall i: s^{(0)}_{i}=+1$ or $\forall i: s^{(0)}_{i}=-1$, corresponding to logical state 1 or 0, respectively.

\begin{figure}[ht!]
		\centering
        \includegraphics[width=0.9\columnwidth]{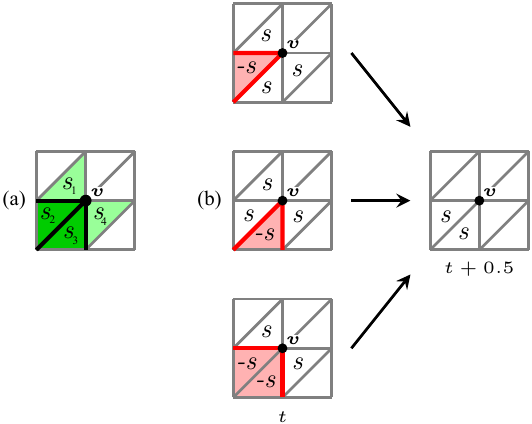}
		\caption{(a) Neighborhood of the vertex $v$ comprises the highlighted spins $s_{1}, s_{2}, s_{3}, s_{4}$. The deterministic rule updates $s_{2}, s_{3}$ (dark green) based on the restriction of the domain wall on the thick black edges. (b) Different configurations of the restricted domain wall (red edges) at time $t$ and the resulting updated spins at $t+0.5$.}
		\label{SweepRule}
\end{figure}

The system evolves under the sweep rule applied at each vertex of the lattice simultaneously (\emph{synchronous update}). The sweep rule is a PCA comprising a deterministic and a probabilistic update rule. 
A single time step $t \rightarrow t+1$ of this PCA can be divided into two half-steps. In the first half-step $t \rightarrow t+0.5$, the system is updated by the deterministic rule  and in the second half-step $t+0.5 \rightarrow t+1$, the system is updated by the probabilistic rule.
In a single time step, the sweep rule can only update the spins $s_2, s_3$ in the neighborhood of the vertex $v$ (see Fig.~\ref{SweepRule}). 
The deterministic rule updates $s_2, s_3$ based on the local configuration of the domain wall, which is defined as the set of edges of the lattice that separate oppositely aligned spins.
In particular, if the spins $s_1$ and $s_4$ are equal, then the deterministic rule updates spins $s_2, s_3$ to match them, i.e., if $s^{(t)}_{1}=s^{(t)}_{4}$, then set $s^{(t+0.5)}_{2}=s^{(t+0.5)}_{3}=s^{(t)}_{1}=s^{(t)}_{4}$.
The probabilistic rule then updates the spins $s_2$ and $s_3$ as follows
\begin{align}
    \text{Pr}\left(s^{(t+1)}_{i} = +1  \middle| s^{(t+0.5)}_{i} = -1\right) = p, \label{noise1} \\
    \text{Pr}\left(s^{(t+1)}_{i} = -1  \middle| s^{(t+0.5)}_{i} = +1\right) = q, \label{noise2}
\end{align}
where $p,q \in [0,1]$ are the two parameters of the model.
A single time step in the evolution of the system comprises synchronous update of all $N$ spins with the deterministic rule followed by the probabilistic rule.
In Appendix~\ref{Variants}, we also report properties of a variant of this PCA, the \emph{asynchronous} sweep rule, where in a single time step the sweep rule is applied sequentially at $N$ vertices selected via simple random sampling with replacement.
Since the sweep rule updates the state of the system based only on its current state, its evolution is a Markov process.

\begin{figure}[ht!]
	\centering
	\includegraphics[width=0.95\columnwidth]
		{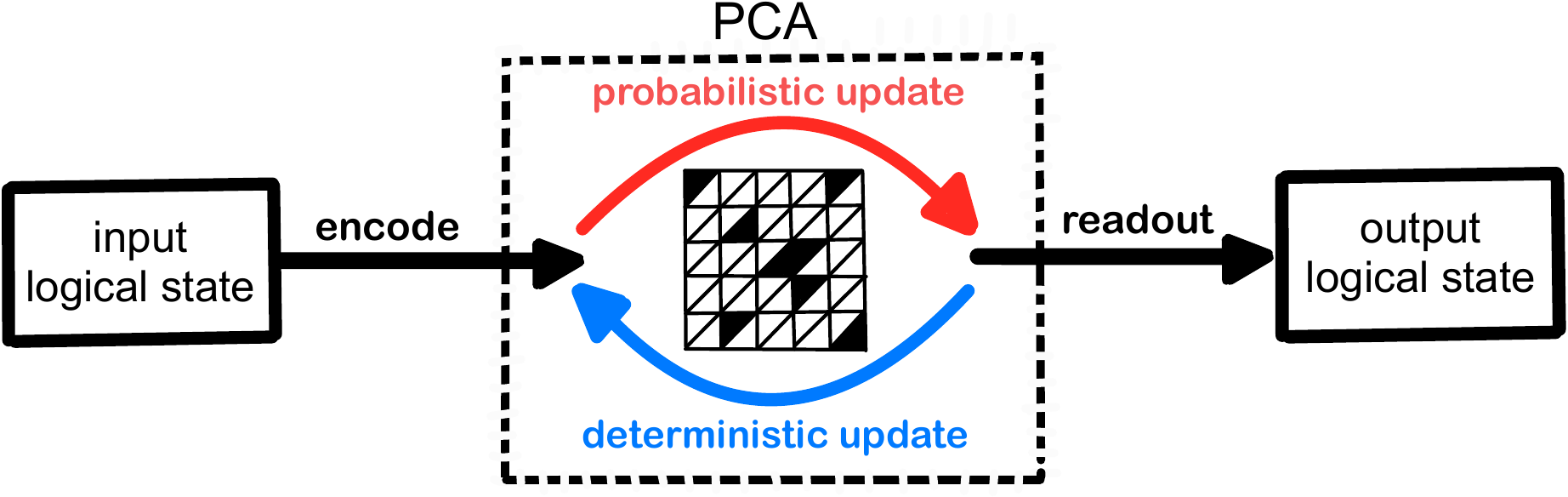}
	\caption{Schematic model of classical memory that protects logical information via a probabilistic cellular automaton. The deterministic update removes noise (random spin-flips) added to the system in the probabilistic update.
    }
	\label{AEMModel}
\end{figure}

At time $t$, the instantaneous magnetization $M(t)$ of the system is defined as
\begin{equation}
M(t) = \frac{1}{N}\sum_{i=1}^{N} s^{(t)}_{i}.
\label{Readout}
\end{equation} 
We can infer the state of the logical bit from the instantaneous magnetization $M(t)$ or from the coarse-grained magnetization, i.e., a simple moving average of the instantaneous magnetization; a positive value indicates logical state $1$, whereas a negative value indicates logical state $0$. In Sec.~\ref{Sim}, we discuss the evolution of the instantaneous magnetization and how coarse-graining can remove fluctuations in the instantaneous magnetization, making the coarse-grained readout a more reliable method of inferring the logical bit.

The evolution of this PCA simulates error correction in the system, where the deterministic rule may be viewed as a protocol that attempts to remove noise, i.e., random spin-flips introduced by the probabilistic rule.
From this perspective, 
\begin{align}
    T = p + q, \quad h = \frac{p-q}{p+q}
\end{align}
are analogous to temperature and magnetic field~\cite{bennettRoleIrreversibilityStabilizing1985}; however, for clarity, we will refer to $h$ as bias. By definition, $T\in[0,2]$, where $T = 1$ is analogous to the infinite temperature limit in thermodynamics; $T \in [0,1)$ and $T \in (1,2]$ are analogous to the positive and negative temperatures, respectively.
By definition, $h\in[-1,1]$.
The non-zero bias $h$ indicates whether one of the two possible spin-flips described in Eqs.~\eqref{noise1}-\eqref{noise2} is more likely than the other.
In particular, at $h=-1$ the probabilistic update only affects $+1$ spins, whereas only $-1$ spins are affected at $h=+1$.

\begin{figure*}[ht!]
    \includegraphics[width=0.3\textwidth]{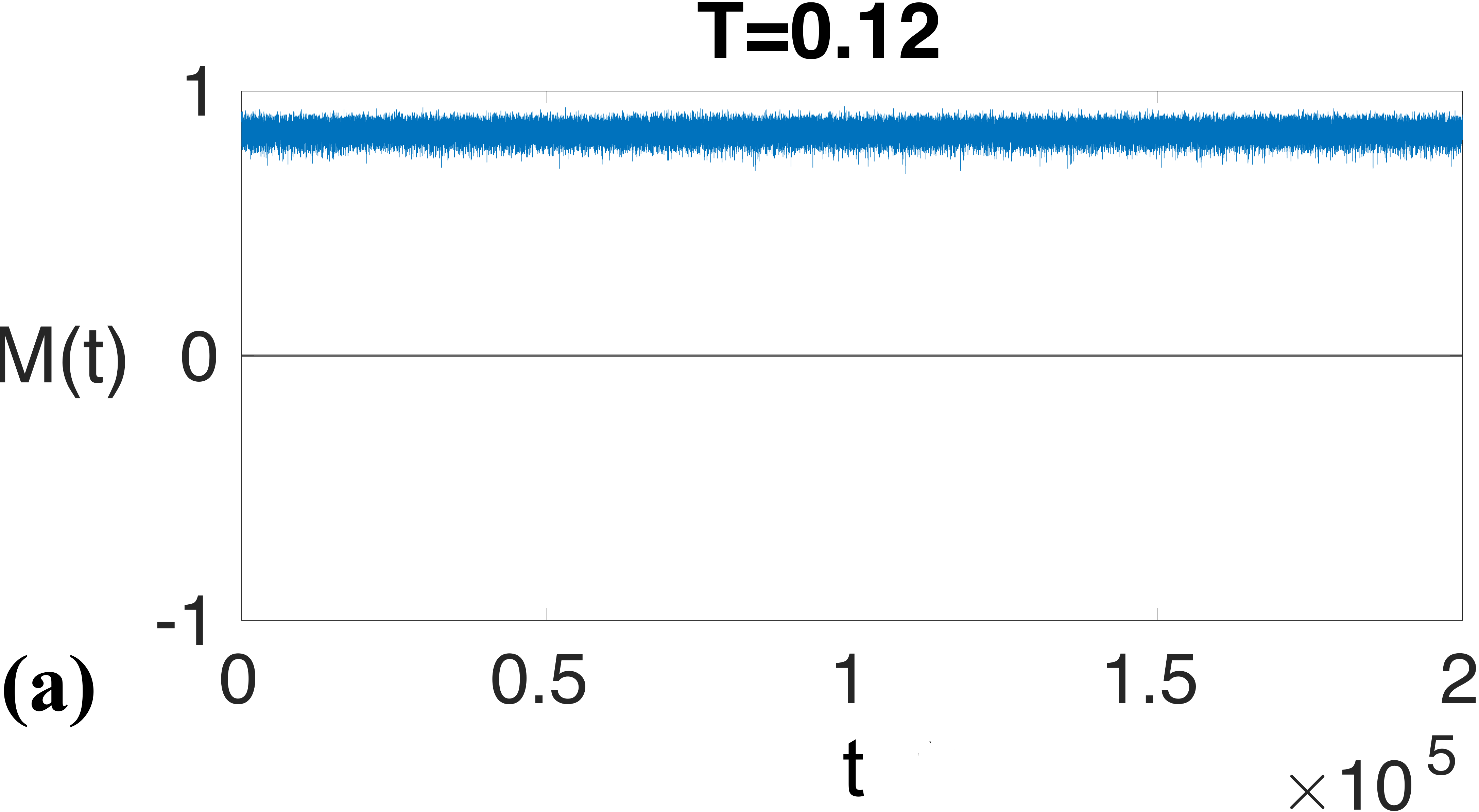} \qquad
    \includegraphics[width=0.3\textwidth]{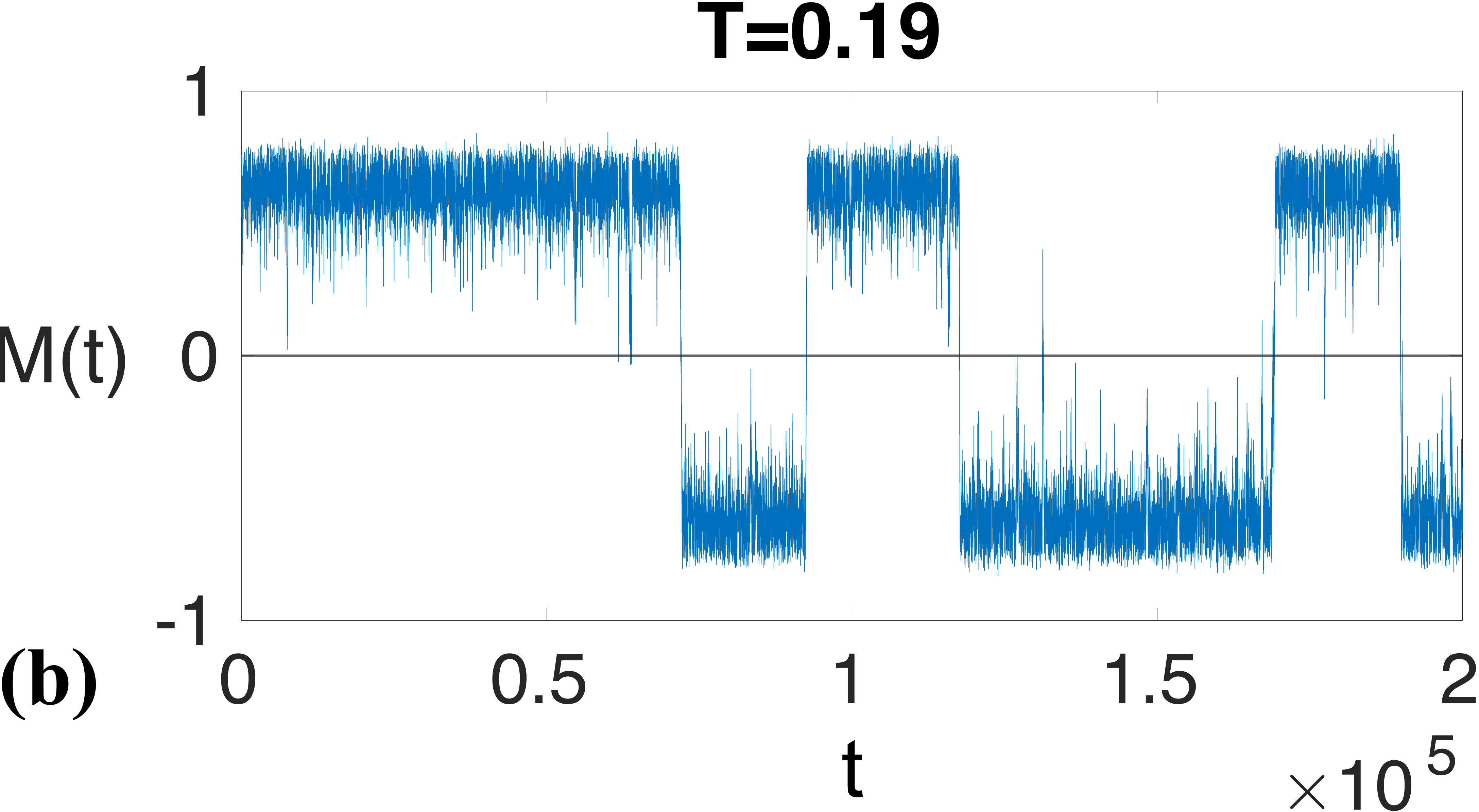} \qquad
    \includegraphics[width=0.3\textwidth]{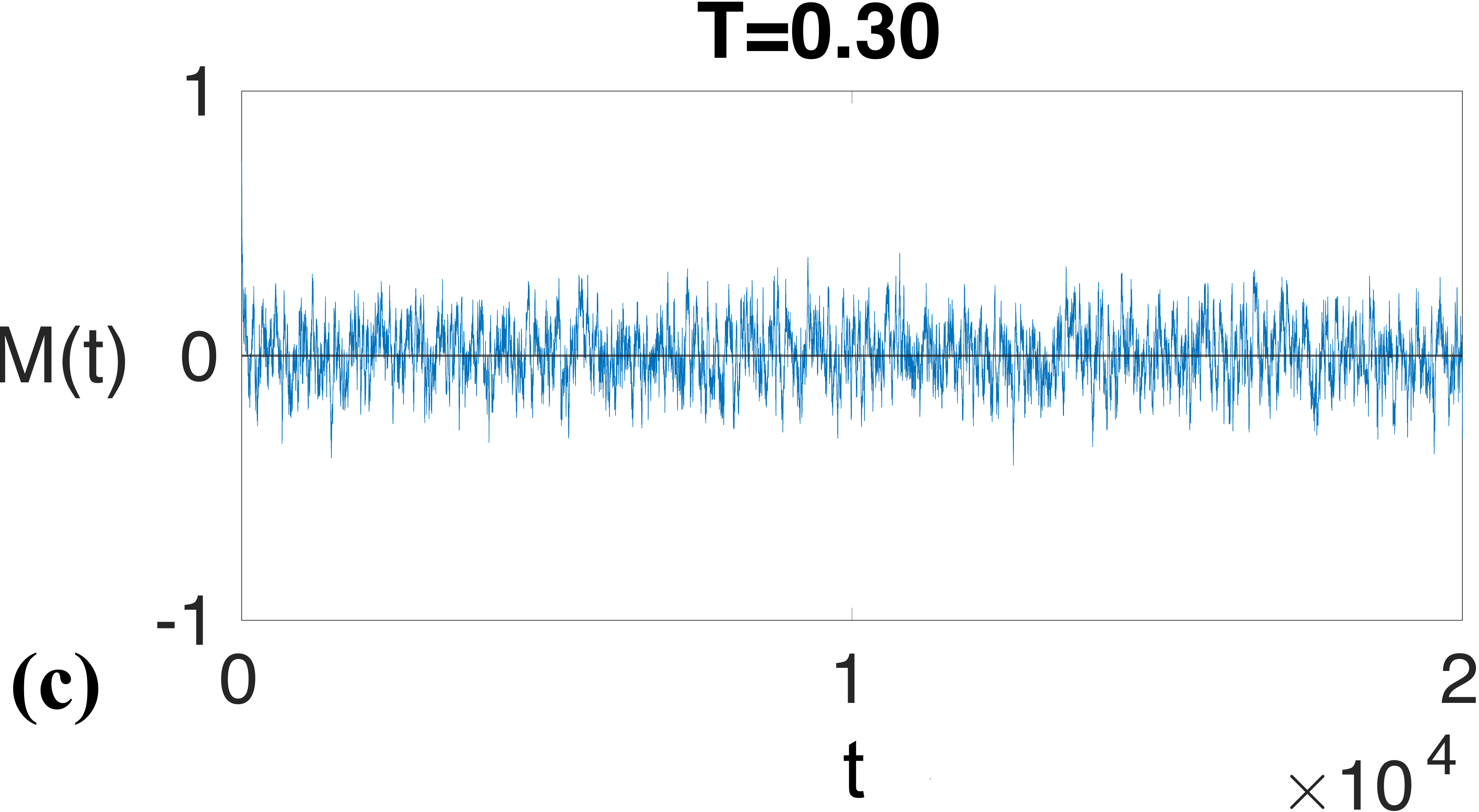} \\[1.5ex]
    \includegraphics[width=0.3\textwidth]{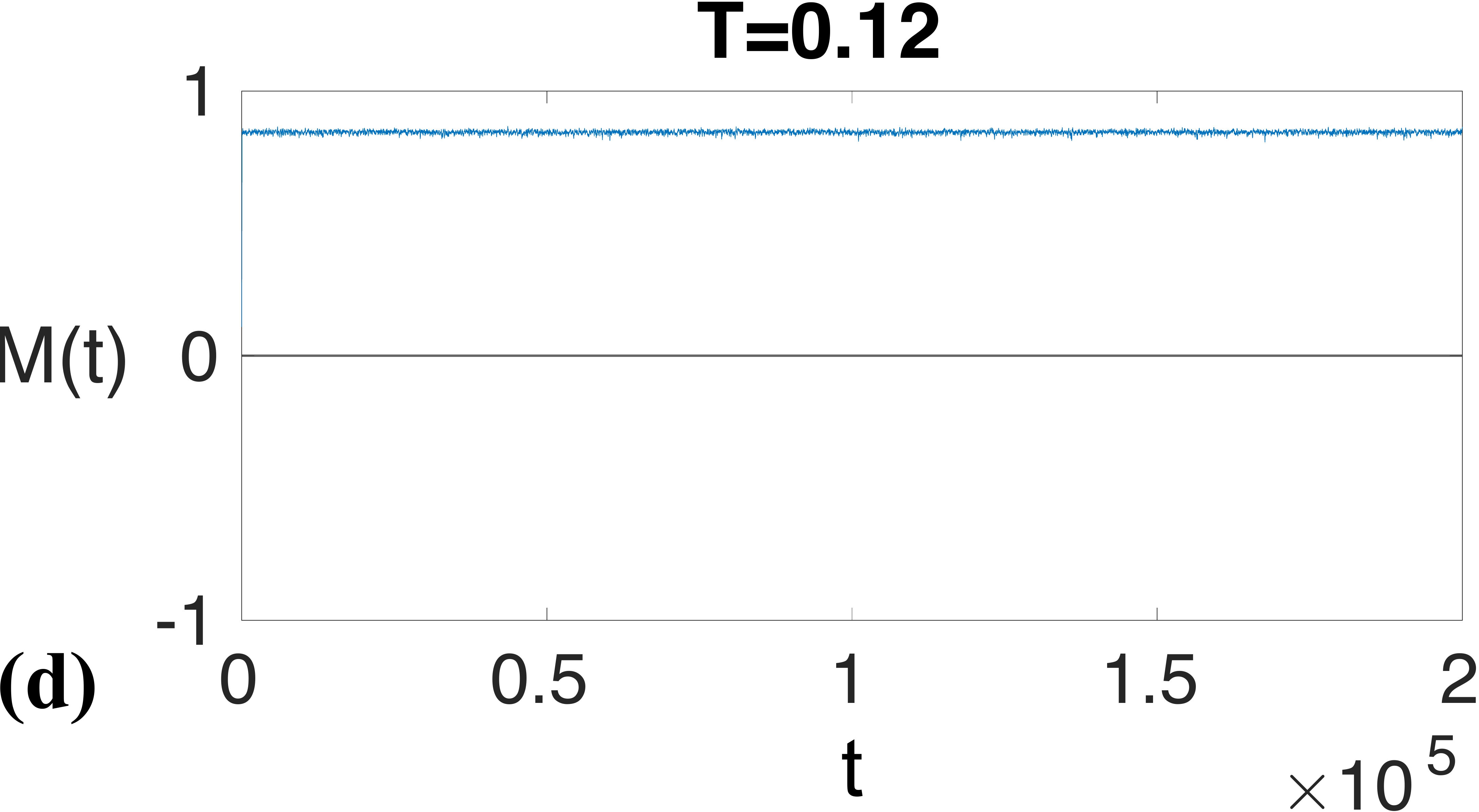} \qquad
    \includegraphics[width=0.3\textwidth]{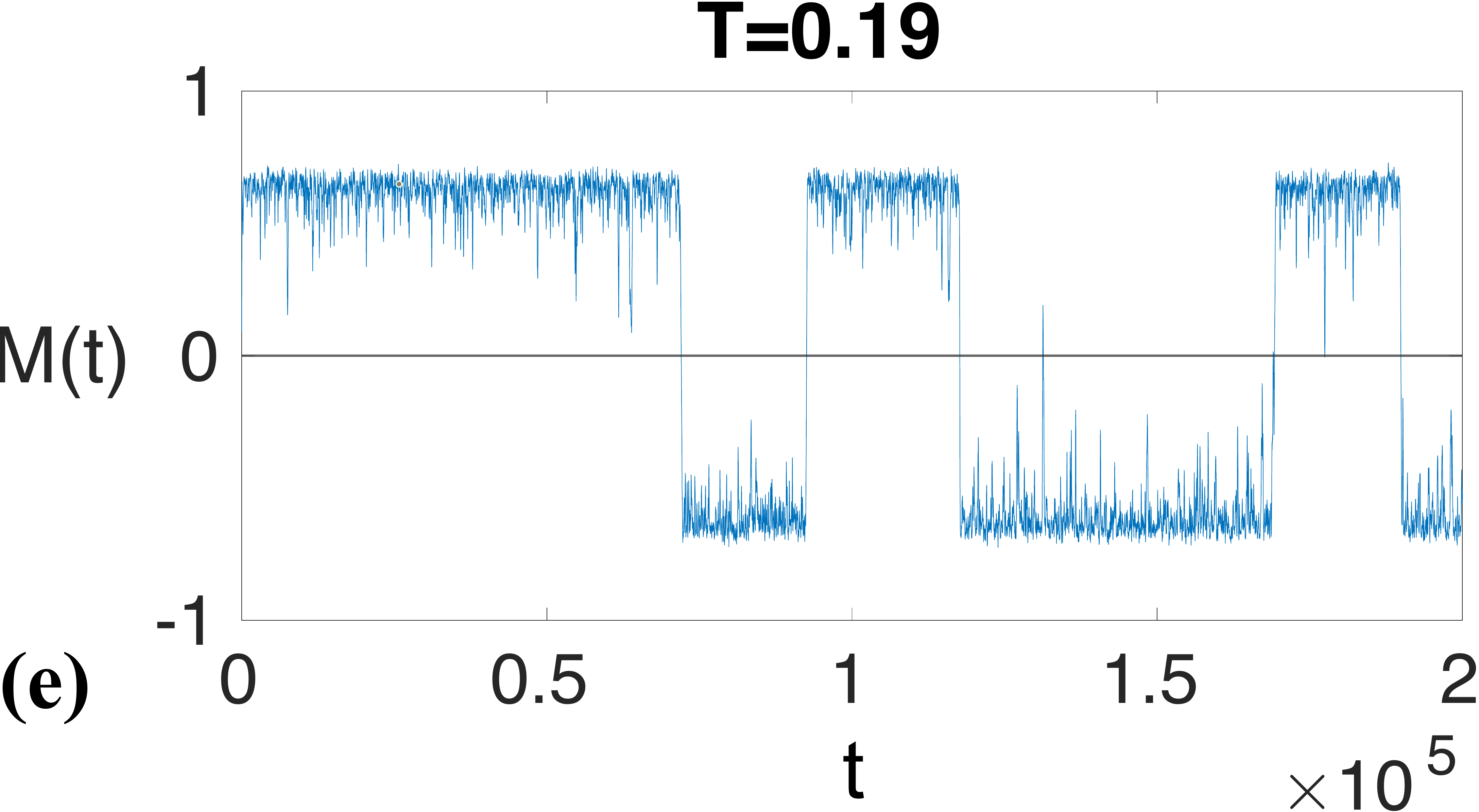} \qquad
    \includegraphics[width=0.3\textwidth]{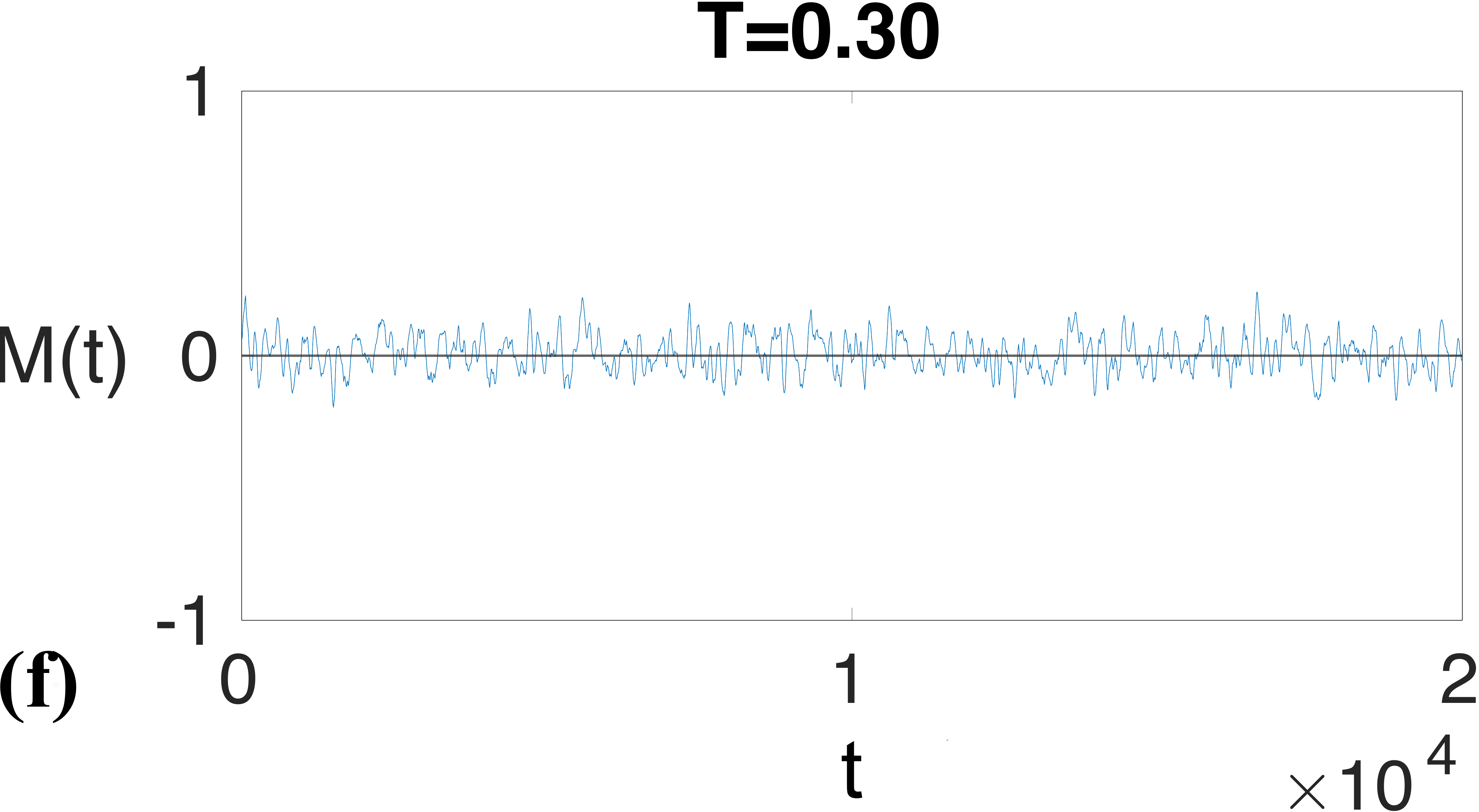}
    \caption{Evolution of magnetization of the PCA with $N=512$ spins on a $16 \times 16$ triangular lattice for $h=0$ at: (a)(d) low, (b)(e) moderate and (c)(f) high temperatures. The plots in (d)-(f) depict magnetization after it is smoothened (see Appendix~\ref{MemNumerics} for details). Logical bit-flips are noted as occurring at time steps when the value of smoothened magnetization is zero.
}
\label{MagEvolve}
\end{figure*}

\begin{figure*}
    \centering
    \includegraphics[width=0.3\textwidth]{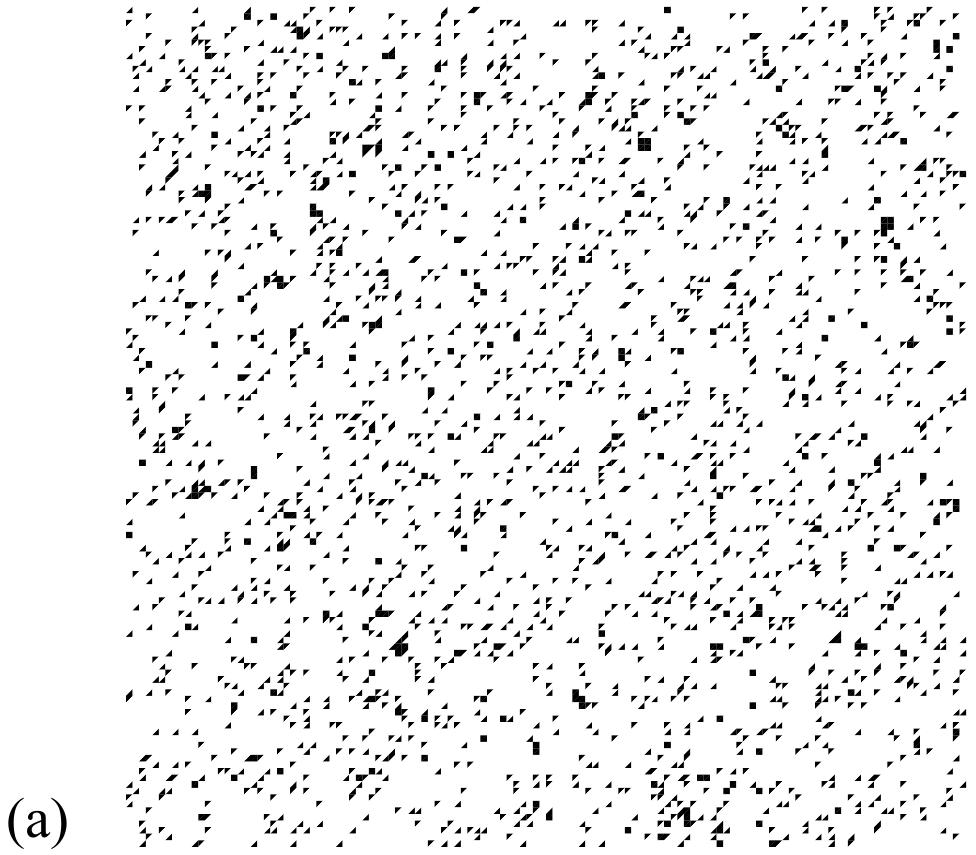} \qquad
    \includegraphics[width=0.3\textwidth]{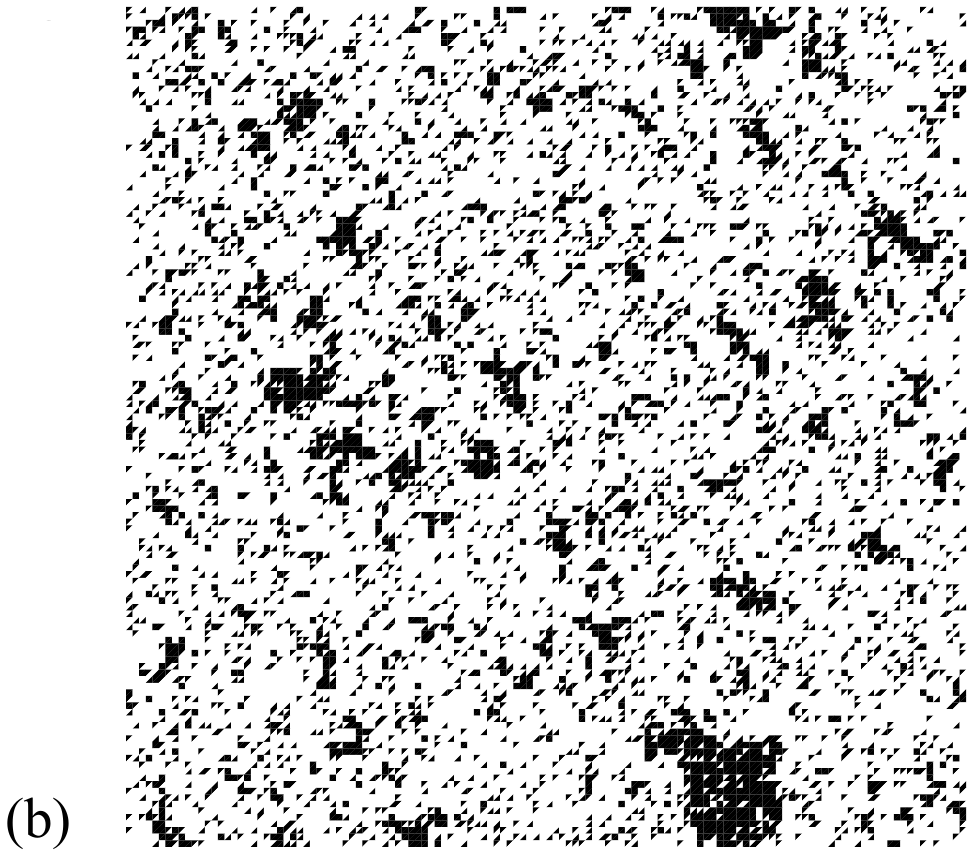} \qquad
    \includegraphics[width=0.3\textwidth]{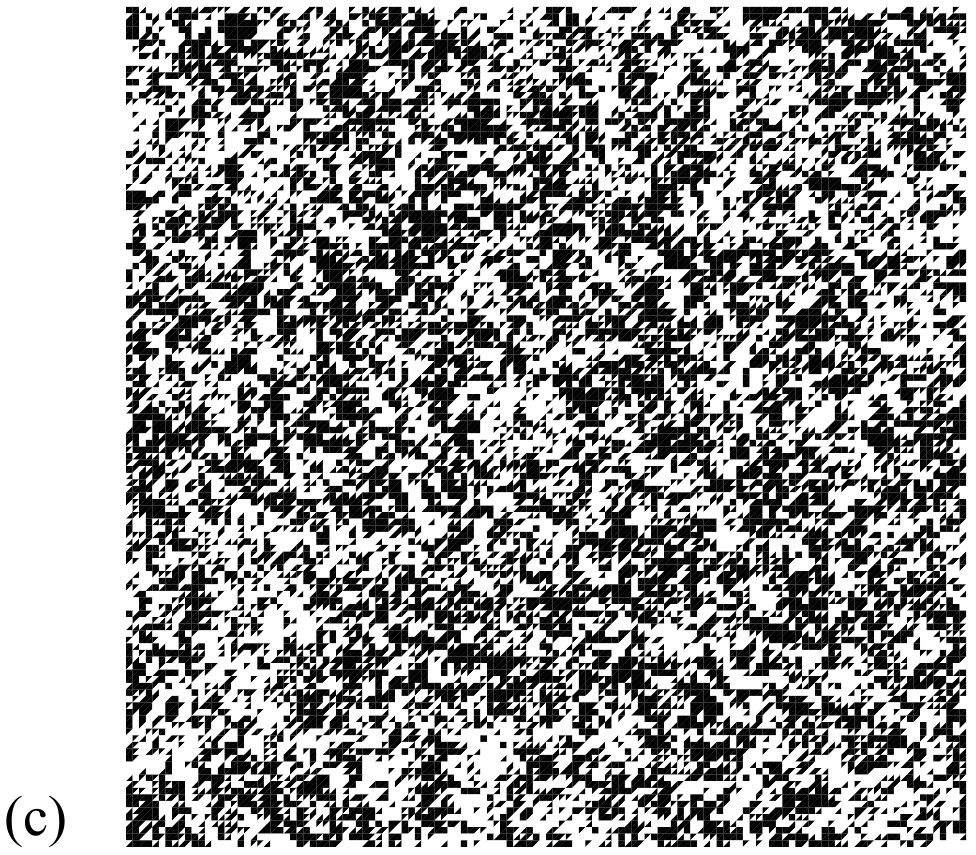} 
    \caption{Snapshots of spin configurations on a $128 \times 128$ triangular lattice with $N=32768$ spins at $h=0$.
 (a) At low temperature $T=0.12$, small clusters of $-1$ spins (black) are isolated from each other.
 (b) At $T=0.19$, large clusters of $-1$ spins of arbitrary size appear, indicating critical behavior. 
 As the system evolves, these clusters coalesce and cause logical bit-flips.
 (c) At high temperature $T=0.30$, the spin configuration is disordered, with roughly equal numbers of $+1$ and $-1$ spins.
 }
    \label{SweepSnaps}
\end{figure*}

We define the \emph{memory lifetime} of the PCA as the time it takes on average for a logical bit-flip in the system with dynamics described by the PCA.
The memory lifetime $\tau_{L,T}$ of the system of linear lattice size $L$ at temperature $T$ can be estimated as follows
\begin{equation}
\tau_{L,T} = \frac{1}{n}\sum_{j=0}^{n}t_{j}, 
\label{MemLyf}
\end{equation}
where $t_{j}$ is the time it takes for the first logical bit-flip to occur in the $j^{\text{th}}$ simulation of the PCA for $j \in \{1, 2,..., n\}$.
Our goal is to identify a region in the $(T,h)$ plane, where the memory lifetime of the system increases exponentially with system size, i.e., the system serves as a good memory. The boundary of this region tells us the maximum temperature (which we refer to as the critical temperature $T^h_{c}$) that can be tolerated at a given value of bias $h$.

\section{Simulating the Sweep Rule}
\label{Sim}

In this section, we study how a logical bit evolves to estimate the memory lifetime of the system.
Since we infer the state of the logical bit from the instantaneous magnetization $M(t)$, we numerically analyze the behavior of $M(t)$ and its moving average over time.

\subsection{Evolution of the instantaneous magnetization}
	
First, we study the system with no bias to identify the critical temperature $T^{0}_c$.
In this case, the spin-flip noise is symmetric with $p=q=T/2$.
Thus, without loss of generality, we focus on the evolution of the logical bit 1 encoded in the initial spin configuration $\forall i: s_i^{(0)}=+1$.
Over the course of simulation, we observe the magnetization of the system and detect three different kinds of behavior depending on the temperature $T$ and its relation to the critical temperature $T^0_c$ (see Fig.~\ref{MagEvolve}).

\begin{figure*}[ht!]
    \centering
    \includegraphics[width=0.3\textwidth]{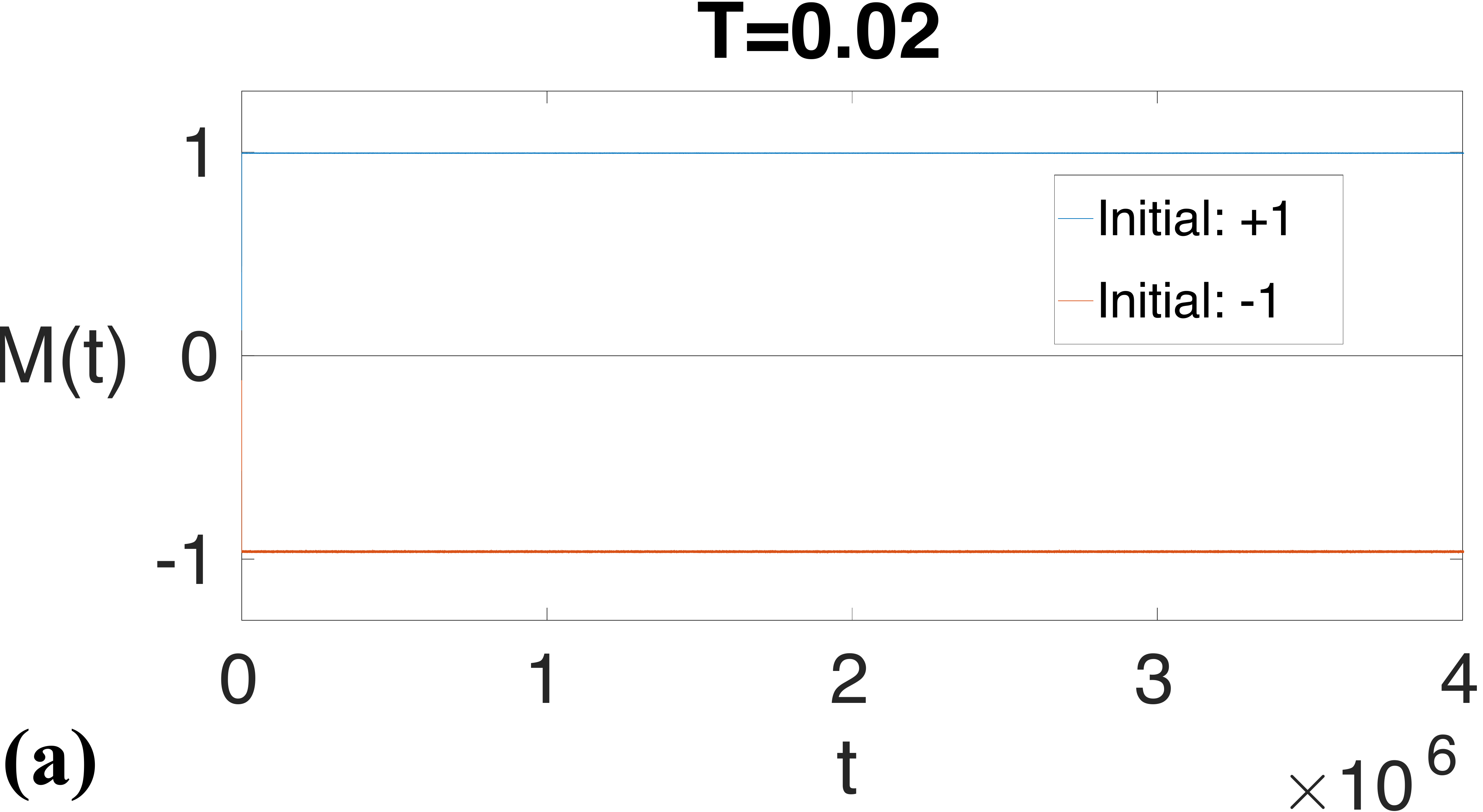} \qquad
    \includegraphics[width=0.3\textwidth]{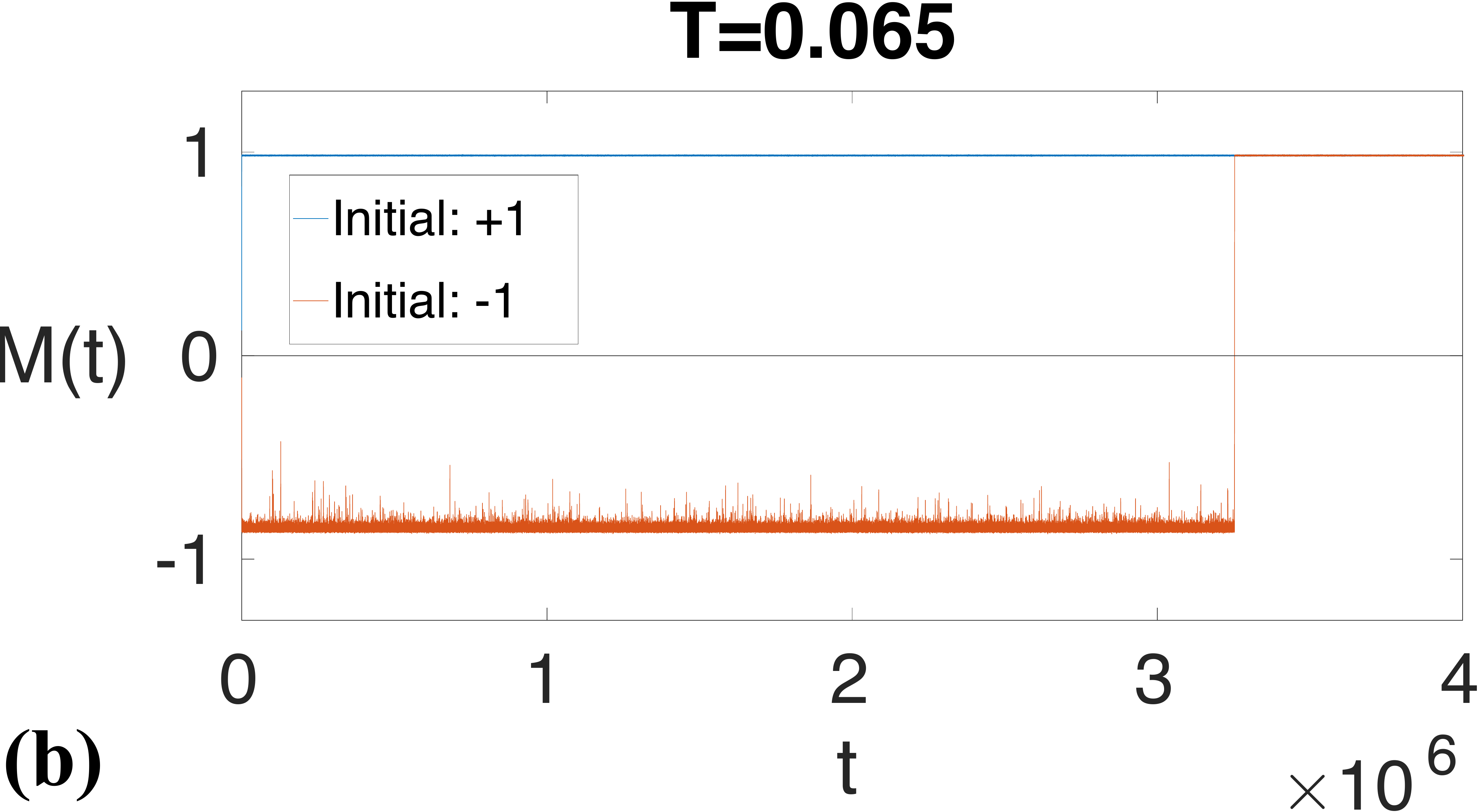} \qquad
    \includegraphics[width=0.298\textwidth]{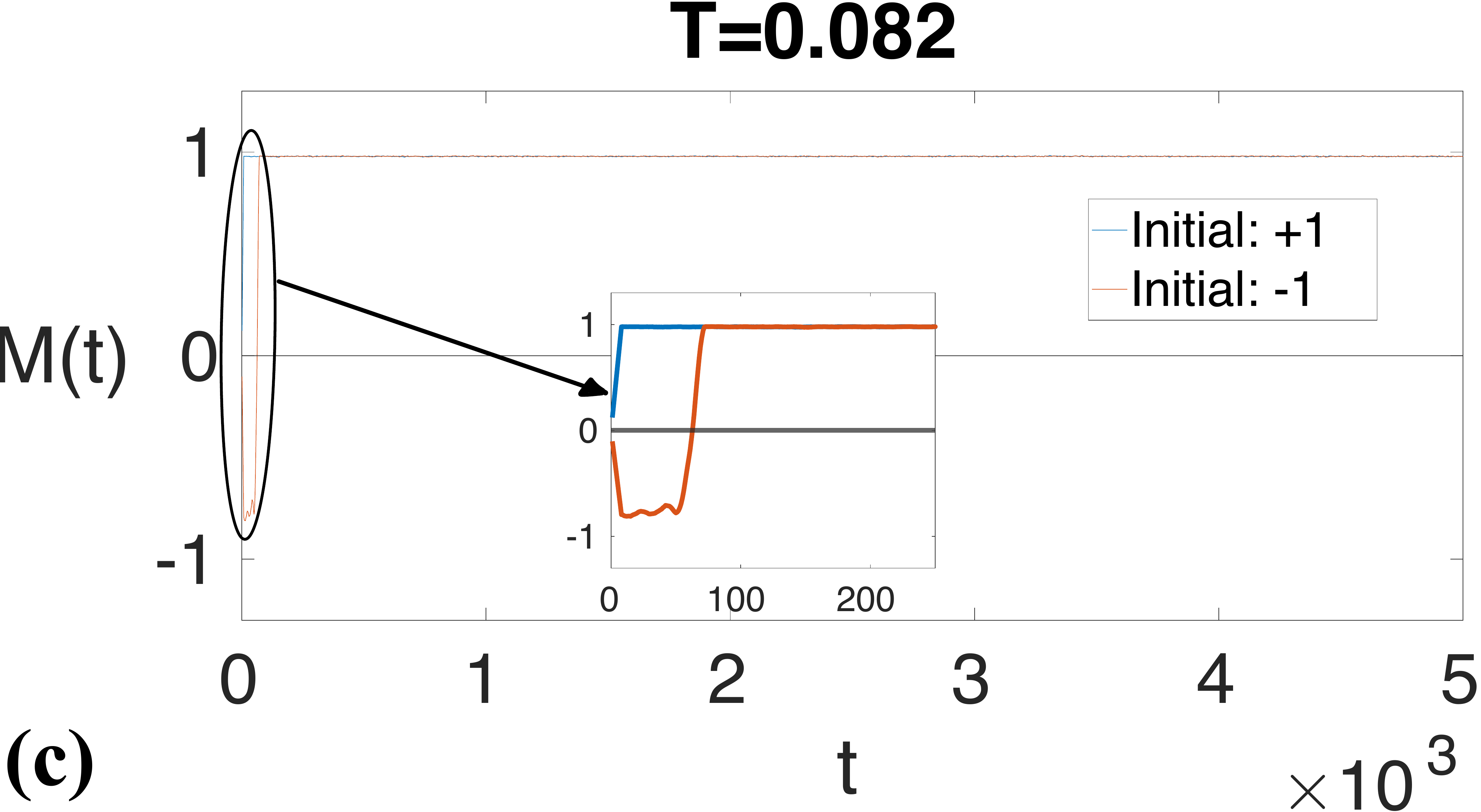} 
    \caption{Evolution of smoothened magnetization of the PCA with $N=512$ spins on a $16 \times 16$ triangular lattice for $h=0.75$ at: (a)low, (b) moderate and (c) high temperatures.
    Since $h>0$, the memory lifetime is limited by the lifetime of the logical state 0 (orange) that was initially encoded with $-1$ spins; the logical state 1 (blue) does not undergo logical bit-flips.
    }
    \label{MagEvolveh}
\end{figure*}

At low temperatures, i.e., $T < T^0_{c}$, the magnetization of the system exhibits fluctuations around a value $M_{\text{low}} \approx 1-T$ (see Fig.~\ref{MagEvolve}(a)) with standard deviation scaling linearly with $T$.
These fluctuations can be attributed to the addition of random spin-flips at each time step. At low temperatures, the probabilistic update creates small clusters of flipped spins (see Fig.~\ref{SweepSnaps}(a)) that are removed by the deterministic rule in the next few time steps ensuring that the value of magnetization remains close to $M_{\text{low}}$ for the entire duration of the simulation. Thus, at low temperatures, we observe that for a long time we are able to correctly infer the logical bit-state.
 
At moderate temperatures, i.e., $T \lesssim T^0_{c}$, at the beginning, the magnetization of the system fluctuates around some positive value $M_{\text{mod}} < 1- T$. The variance of the magnetization is greater than in the low temperature regime and diverges at the critical temperature $T_c$ as indicated by the plot of magnetic susceptibility $\chi$ (defined in Eq.~\eqref{MagVar}) in Fig.~\ref{DatCol}(b).
In this temperature range, at each time step, the probabilistic update creates clusters of flipped spins that are likely to be larger than those in the low-temperature regime (see Fig.~\ref{SweepSnaps}(b)).
The larger the cluster of flipped spins, the more time steps (and applications of the deterministic rule implementing error correction) are required to completely remove it from the system.
Large clusters have the potential to eventually coalesce and cause the majority of the spins to flip leading to a logical bit-flip. Thus, unlike in the low temperature range, where the value of magnetization remains close to $M_\text{low}$ over the whole duration of the simulation, in this case, the magnetization changes from a positive value $M_{\text{mod}}$ to a negative value $-M_{\text{mod}}$ (see Fig.~\ref{MagEvolve}(b)), indicating a logical bit-flip. Over the course of the simulation, the magnetization continues to flip between $M_{\text{mid}}$ and $-M_{\text{mid}}$.

At high temperatures, i.e., $T  > T^0_{c}$, the magnetization fluctuates about $M_{\text{high}} = 0$ over the duration of the entire simulation.
The variance of the magnetization in this temperature range is smaller than in the moderate temperature regime (see Fig.~\ref{DatCol}(b)). The system appears to lose the initially encoded state of the logical bit within a very short time (compared to the system size $N$).

The plots in Fig.~\ref{MagEvolve}(a)-(c) show that fluctuations in magnetization occur at all three temperature regimes due to the random physical spin-flips introduced by the probabilistic update.
At moderate and high temperatures, these fluctuations are likely to build up within a short time interval (proportional to the system size $N$) and cause logical bit-flips, while at low temperatures these fluctuations disappear and do not affect the logical bit state.
We thus choose to smoothen the plots of instantaneous magnetization (see Fig.~\ref{MagEvolve}(d)-(f)) by first coarse-graining it and then calculating its moving average (see Appendix~\ref{MemNumerics} for details).

For the PCA with non-zero bias, the spin-flip noise is not symmetric.
In particular, for $h>0$ the memory lifetime of the system is limited by the memory lifetime of the logical state $0$.
As in the case of $h=0$, we smoothen the magnetization (see Fig.~\ref{MagEvolveh}) to remove fluctuations that do not result in logical bit-flips.

\subsection{Numerical analysis of memory lifetime}\label{SimMem}

	\begin{figure*}[htp!]
		\centering
			\includegraphics[width=0.47\textwidth]{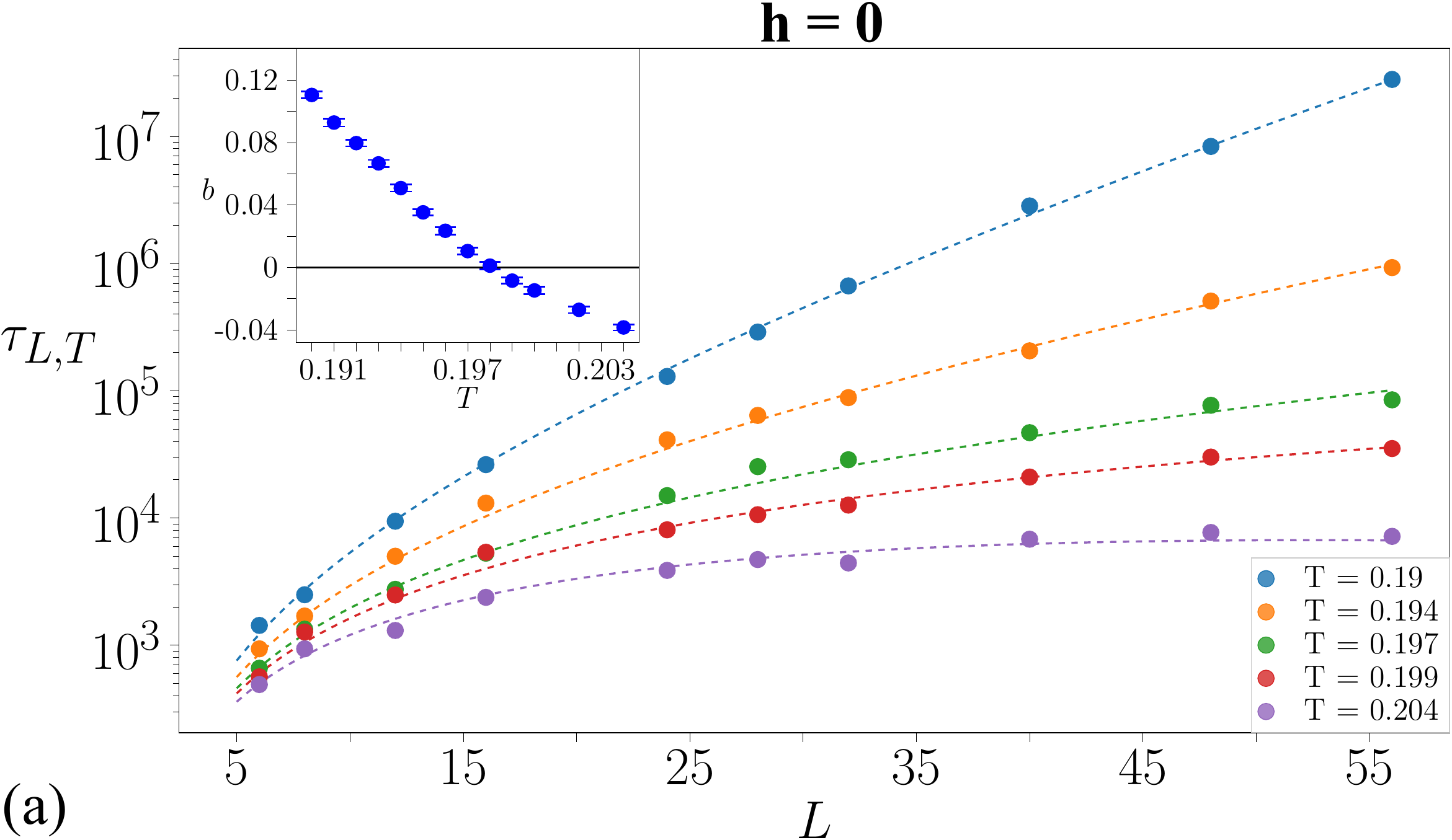}\qquad
			\includegraphics[width=0.47\textwidth]{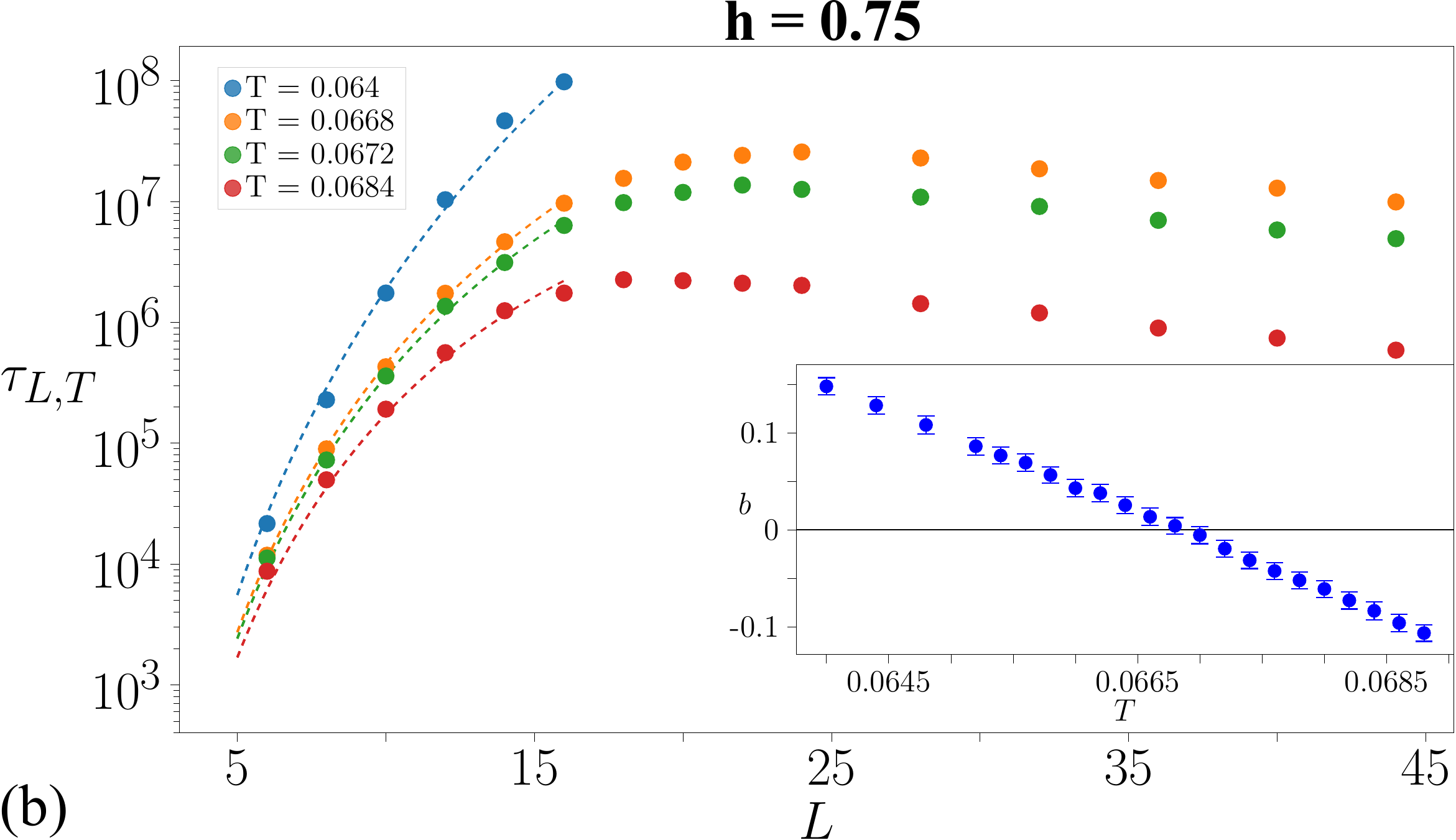}
		\caption{Plots of memory lifetime $\tau_{L,T}$ of the sweep rule as a function of the linear lattice size $L$ at various temperatures $T$. This data is fit into the ansatz in Eq.~\eqref{ansatz}. (a) At $h=0$, we identify the critical temperature $T^0_{c}=0.197\pm0.002$ as the point where $b=0$ (inset). The system behaves as a good memory below $T^0_{c}$. (b) At $h=0.75$, we similarly identify the critical temperature $T^{0.75}_{c}=0.0670\pm0.0004$. Additionally, for $T > T^{0.75}_{c}$ we observe that $\tau_{L, T}$ increases with $L$ as long as $L<L_T^*$, where $L_T^*$ is the linear lattice size maximizing $\tau_{L,T}$; further increase of $L$ decreases $\tau_{L,T}$.
    }
		\label{TMem}
	\end{figure*}

We estimate the memory lifetime at various values of temperature $T$ and bias $h$ (see Appendix~\ref{MemNumerics} for detailed data) and analyze how it scales with the linear lattice size $L$. We plot $\tau_{L,T}$ as a function of $L$ for various $T$ (see Fig.~\ref{TMem}). We fit this data in the numerical ansatz 
 \begin{align}
    \log\tau_{L,T} = a\log L + bL + c,  \label{ansatz}   
 \end{align}
where  $a$, $b$ and $c$ are fitting parameters. For $T \approx T^{h}_{c}$, we observe that the parameters $a, c$ do not exhibit a strong dependence on temperature. Therefore, we constrain $a, c$ to be fixed over the temperature range (see Fig.~\ref{abcFits} in Appendix~\ref{MemNumerics}) and report the behavior of $b$ as a function of $T$ (see Fig.~\ref{TriangTMembvT} in Appendix~\ref{MemNumerics}).

For $h=0$, we observe that at low temperatures, the memory lifetime scales exponentially with linear lattice size $L$ as indicated by the parameter $b$ being positive (see inset of Fig.~\ref{TMem}(a)).
At $T \approx 0.197$, $b=0$ and the memory lifetime $\tau_{L,T}$ scales polynomially in $L$, i.e., $\tau_{L,T} \propto L^a$ with  $a\approx 2$.
Thus, we estimate the critical temperature to be $T^{0}_{c}=0.197 \pm 0.002$.
Similarly, we perform the numerical analysis of memory lifetime of the PCA for multiple values of $h$.
For instance, when $h=0.75$, the parameter $b$ (see inset of Fig.~\ref{TMem}(b)) intersects the $x$-axis at $T^{0.75}_{c}=0.0670 \pm 0.0004$ and the memory lifetime $\tau_{L,T}$ scales as $L^a$, where $a\approx 7$, indicating a phase transition.
Subsequently, we obtain the phase transition points $T^{h}_{c}$ in each case (see Figs.~\ref{TriangTMem}~and~\ref{TriangTMembvT} in Appendix~\ref{MemNumerics}), thus recovering the phase diagram shown in Fig.~\ref{PhaseDiag}.

\section{Phase Transition in the Spin Model}
\label{StatMech}

	\begin{figure*}[htp!]
	\centering
        \includegraphics[width=0.32\textwidth]{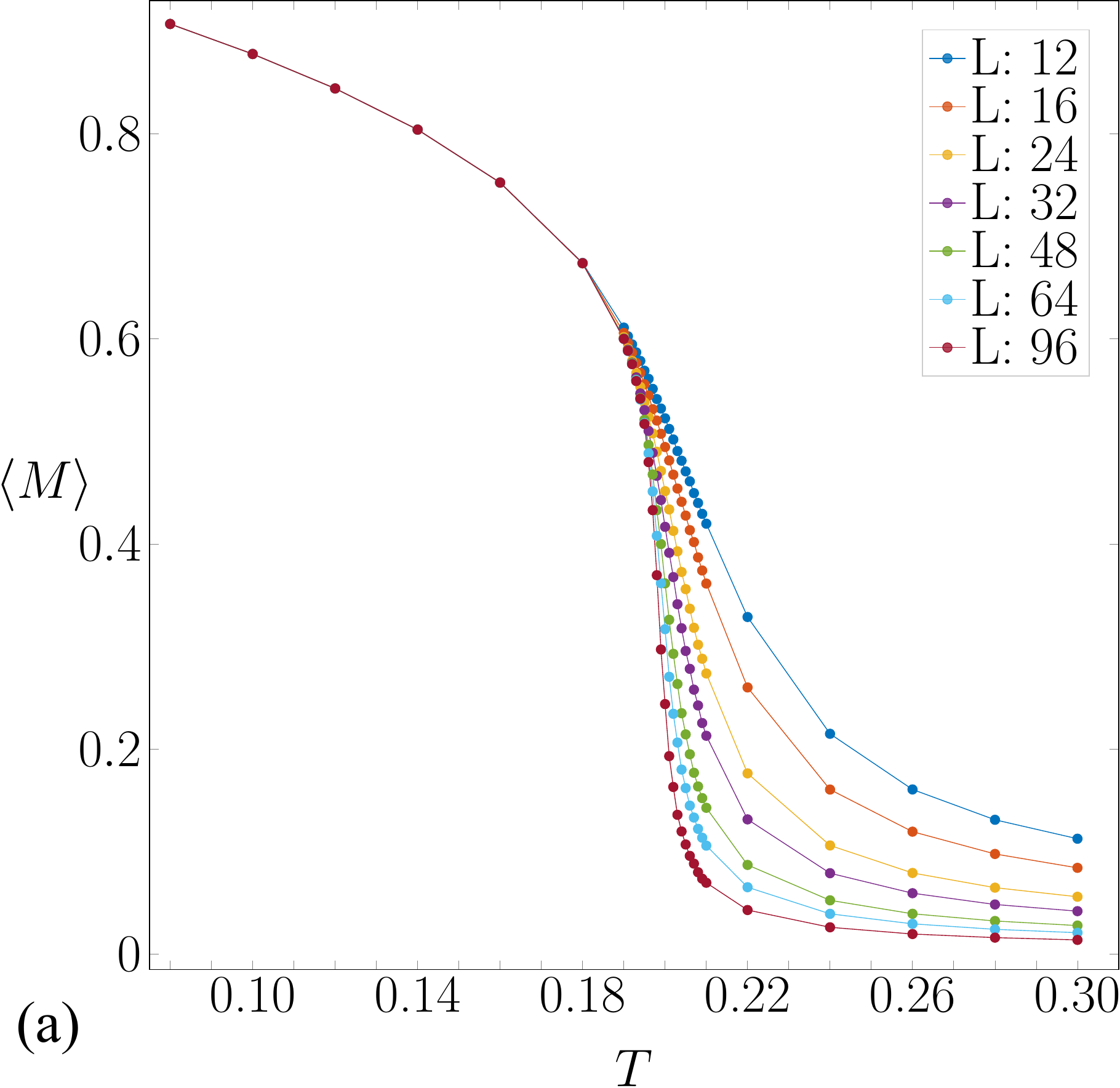}
        \includegraphics[width=0.32\textwidth]{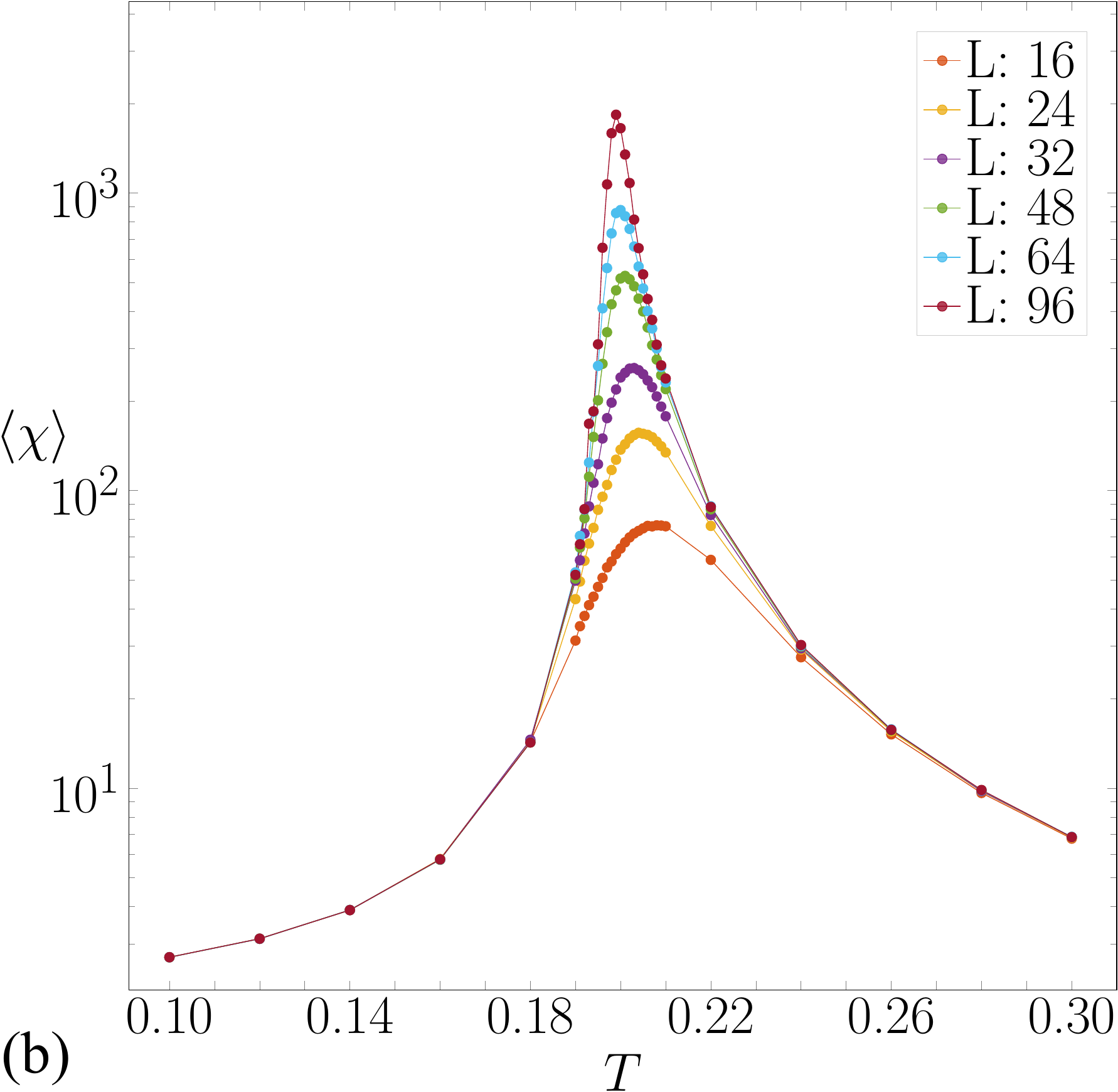}
        \includegraphics[width=0.322\textwidth]{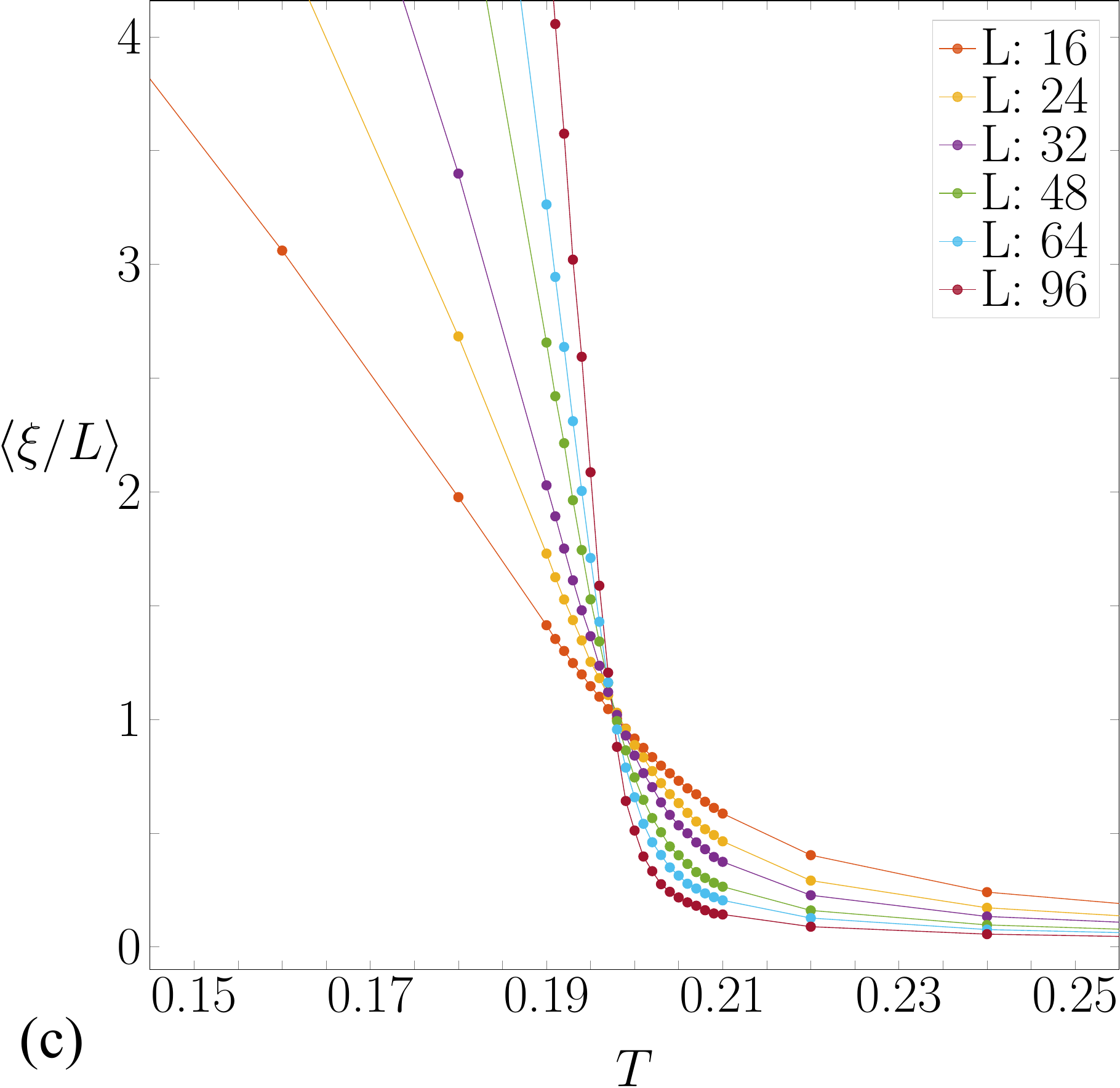}\\ [2ex]
        \includegraphics[width=0.32\textwidth]{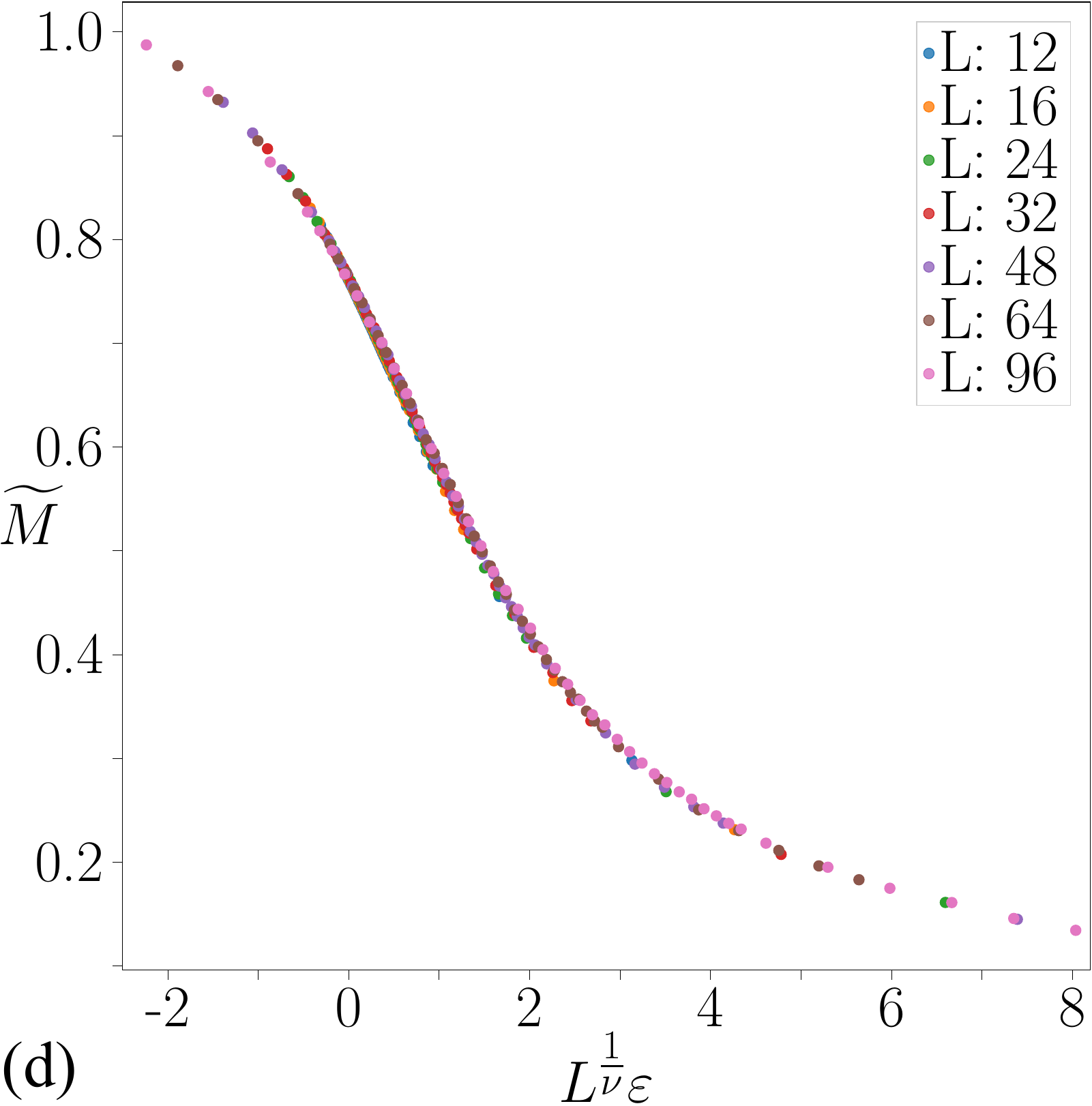}
        \includegraphics[width=0.32\textwidth]{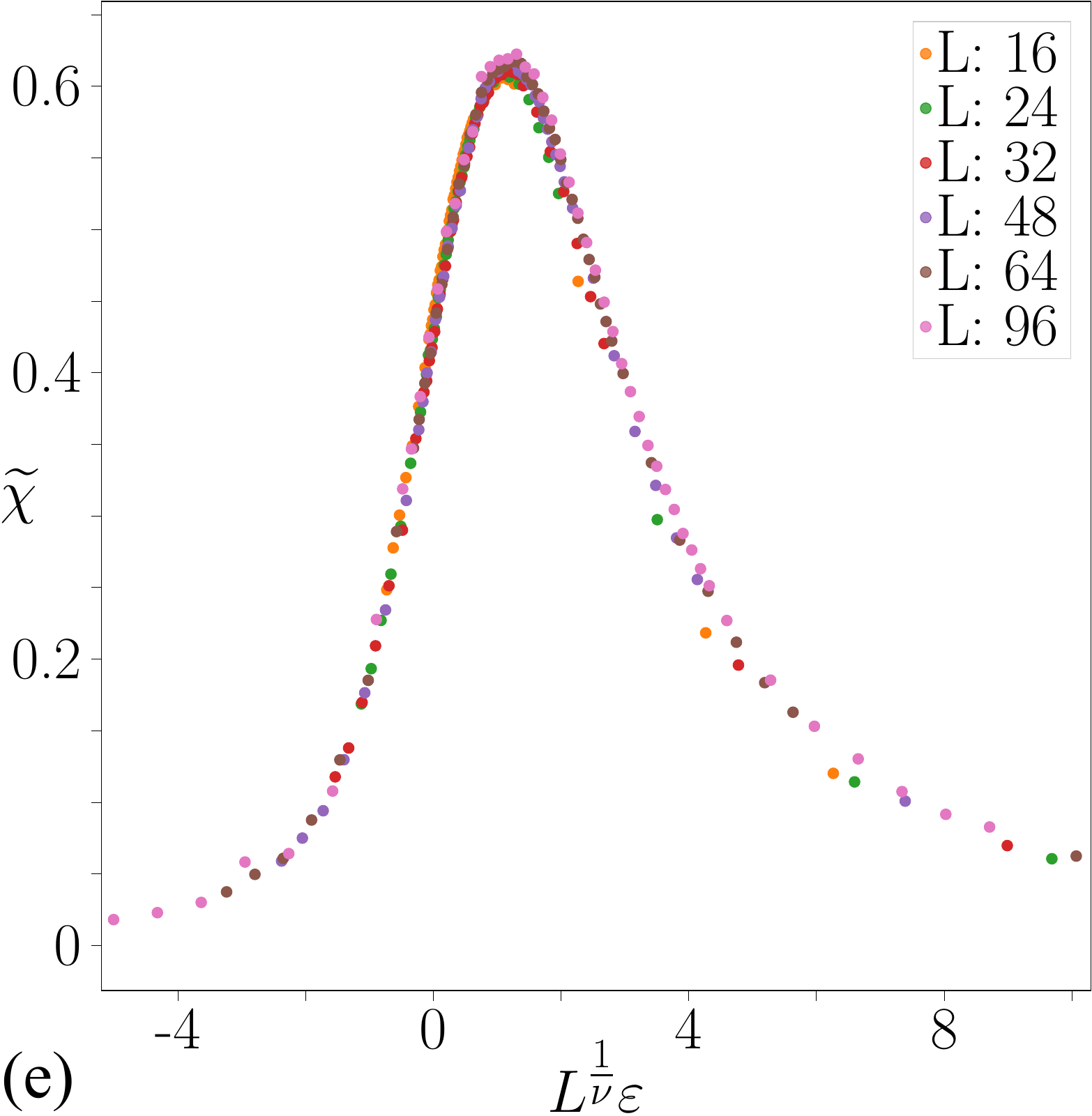}
        \includegraphics[width=0.32\textwidth]{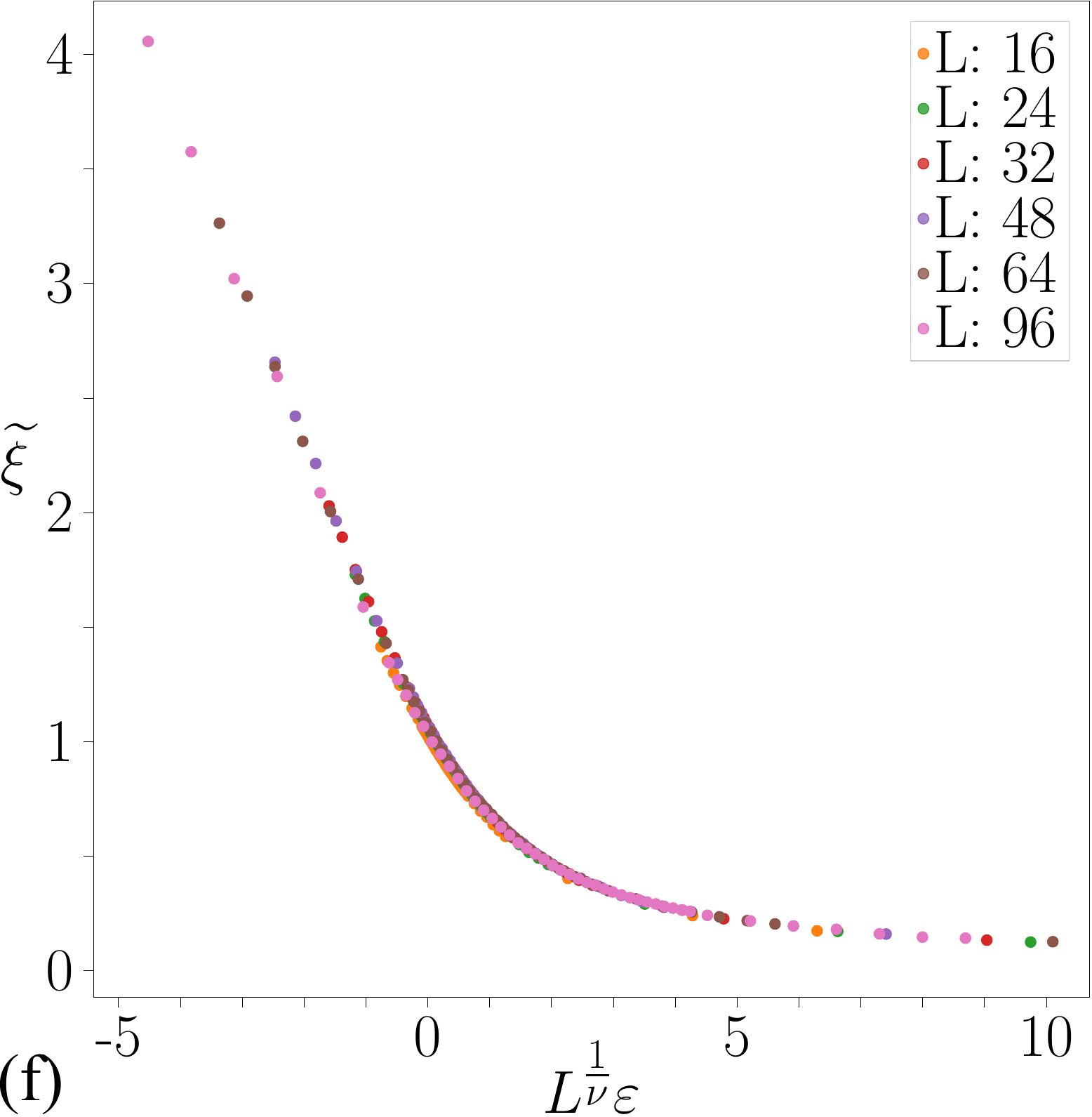}
		\caption{Critical behavior of the sweep rule at $h=0$. (a) The ensemble average of magnetization $\langle M \rangle$ drops continuously from a non-zero value to zero at the critical temperature, indicating a second-order phase transition. (b) The ensemble average of the magnetic susceptibility $\langle \chi \rangle$ captures the variance of the magnetization, which diverges at the critical temperature. (c) We can identify the critical temperature from the intersection of the plots of the ensemble average of two-point correlation functions $\langle \xi/L \rangle$ for different $L$.
  (d)-(f) The data collapse of $\langle M \rangle $, $\langle \chi \rangle$ and $\langle \xi/L \rangle $ into $L$-independent functions $\widetilde{M}$, $\widetilde{\chi}$ and $\widetilde{\xi}$ respectively, is used to obtain the critical exponents $\beta$, $\gamma$, $\nu$ (see Table~\ref{table:CritExps}) with critical temperature $T^{0}_{c}=0.1973 \pm 0.0001$.
  }
		\label{DatCol}
	\end{figure*}
	
 In this section, we use standard statistical-mechanical techniques to identify a phase transition in the sweep rule with no bias. In particular, we study the scaling of the (ensemble) averages of the system's magnetization $M$, magnetic susceptibility $\chi$, and two-point correlation function $\xi$ (see Appendix~\ref{StatMechApp} for details). In the neighborhood of $T^{0}_{c}$, this phase transition is characterized by critical exponents $\nu$, $\beta$ and $\gamma$ corresponding to the following scaling relations \cite{newmanMonteCarloMethods1999} 
	\begin{align}
	    \widetilde{\xi}(L^{\frac{1}{\nu}}\lvert \varepsilon \rvert) &= L^{-1} \langle \xi \rangle,\\
		\widetilde{\chi}(L^{\frac{1}{\nu}}\lvert \varepsilon\rvert) &= L^{-\frac{\gamma}{\nu}} \langle \chi \rangle,\\
	    \widetilde{M}(L^{\frac{1}{\nu}}\lvert \varepsilon \rvert) &= L^{\frac{\beta}{\nu}}\langle M \rangle,
	\end{align}
	where $\varepsilon=(T-T^{0}_{c})/T^{0}_{c}$ is the reduced temperature of the system.
 The numerical data for different linear lattice sizes $L$ is collapsed onto scaling functions $\widetilde{\xi}$, $\widetilde{M}$, and $\widetilde{\chi}$, which are independent of the lattice size $L$ (see Fig.~\ref{DatCol}).
This method yields the critical exponents $\beta=0.122 \pm 0.007$, $\gamma=1.63 \pm 0.02$ and $\nu=0.93 \pm 0.01$ with $T_{c}=0.197 \pm 0.001$. These values of critical exponents are also found to satisfy the hyperscaling relation~\cite{newmanMonteCarloMethods1999, makowiecUniversalityClassProbabilistic2002}
	\begin{equation}
		2 \beta + \gamma = \nu d,
	\end{equation}
where $d=2$ since we are studying 2D lattices. Moreover, the ratios $\gamma/\nu = 1.75 \pm 0.04$ and $\beta/\nu = 0.13 \pm 0.01$ obtained are equal to the corresponding ratios for the 2D Ising model indicating that the sweep rule belongs to the \emph{weak universality class}~\cite{suzukiNewUniversalityCritical1974} of the Ising model. This is also in agreement with the universality class of Toom's rule reported in Ref.~\cite{makowiecUniversalityClassProbabilistic2002}.
The statistical-mechanical analysis of the sweep rule independently identifies a phase transition at $T^{0}_{c}=0.1973 \pm 0.0001$ from ferromagnetic behavior at lower temperatures to paramagnetic behavior at higher temperatures. In the absence of bias, i.e., $h=0$, the ferromagnetic behavior of a system permits the stability of two different phases corresponding to average magnetization $+M$ and $-M$, which can be used to encode the two states of a logical bit. Such ferromagnetic behavior is an indicator that the system behaves as a good memory as in the case of the 2D Ising model at $h=0$. The use of two independent techniques, numerical analysis of memory lifetime scaling and statistical mechanical analysis of the sweep rule spin model, to identify the same phase transition suggests that the scaling of memory lifetime with linear lattice size is a good indicator of the sweep rule's phase transition. 

\section{Analytical model of sweep rule}
\label{Analytical}

In this section, following Ref.~\cite{dayIsingFerromagnetSelfcorrecting2012} we discuss a simple analytical model of the PCA at $h=0$. We note that in the infinite temperature limit, i.e., $T=1$, the probabilistic update flips each spin with probability $p = q = 0.5$, i.e., each spin has a $50 \%$ chance of being flipped irrespective of the outcome of the deterministic update. Thus, we can treat the system as a collection of $N$ independent spins with a mean flipping rate $\lambda=0.5$ per time step. At $T \lesssim 1$, the probability of spin-flips in one time step is less than 0.5 and depends on the outcome of the deterministic rule, which in turn depends on the state of the neighboring spins. As temperature decreases, the effects of the deterministic update compete with those of the probabilistic update. Nevertheless, for sufficiently high temperatures, the system is still predominantly affected by random spin-flips and  the system can be modelled by a smaller collection of $N_{\text{eff}} < N$ effective spins that independently flip at some mean flipping rate $\lambda < 0.5$ per time step.
Since we only take into account the effects of random spin-flips in this simple model, we can expect that the model will break down for temperatures close to the critical temperature, where the effect of the deterministic updates cannot be neglected.

 Using the same reasoning as presented in Ref.~\cite{dayIsingFerromagnetSelfcorrecting2012}, the fidelity $F(t)$ of the logical bit encoded in the PCA at time step $t$ can be written as
	\begin{align}
		F(t) & = \frac{1}{2}\left( 1 + \text{erf}\left(\frac{N_{\text{eff}}-2\mu}{2\sqrt{2}\sigma}\right) \right), \label{eq:F(t)_triang}\\ 
		\mu  & = \frac{1}{2}N_{\text{eff}}(1-e^{-2\lambda t}),\\
		\sigma & = \frac{1}{2}[N_{\text{eff}}(1-e^{-2\lambda t})(1+e^{-2\lambda t})]^{1/2}.
	\end{align}  
Here, $F(t)$ is the probability that at time step $t$ a logical bit-flip does not occur in the system, i.e., fewer than half of the spins in the system have flipped an odd number of times since initialization, $\text{erf}(\cdot)$ denotes the error function, and $N_{\text{eff}}, \lambda$ are the two parameters of the model that vary with temperature.
As per this model, $F(t)$ approximates a binomial distribution with mean $\mu$ and standard deviation $\sigma$. By definition, the fidelity at initialization is $F(0) = 1$. As time progresses, the fidelity is expected to decrease from this initial value and 
$\lim_{t \to \infty} F(t) = 0.5$.

\begin{figure*}[ht!]
    \centering
	\includegraphics[width=0.45\textwidth]{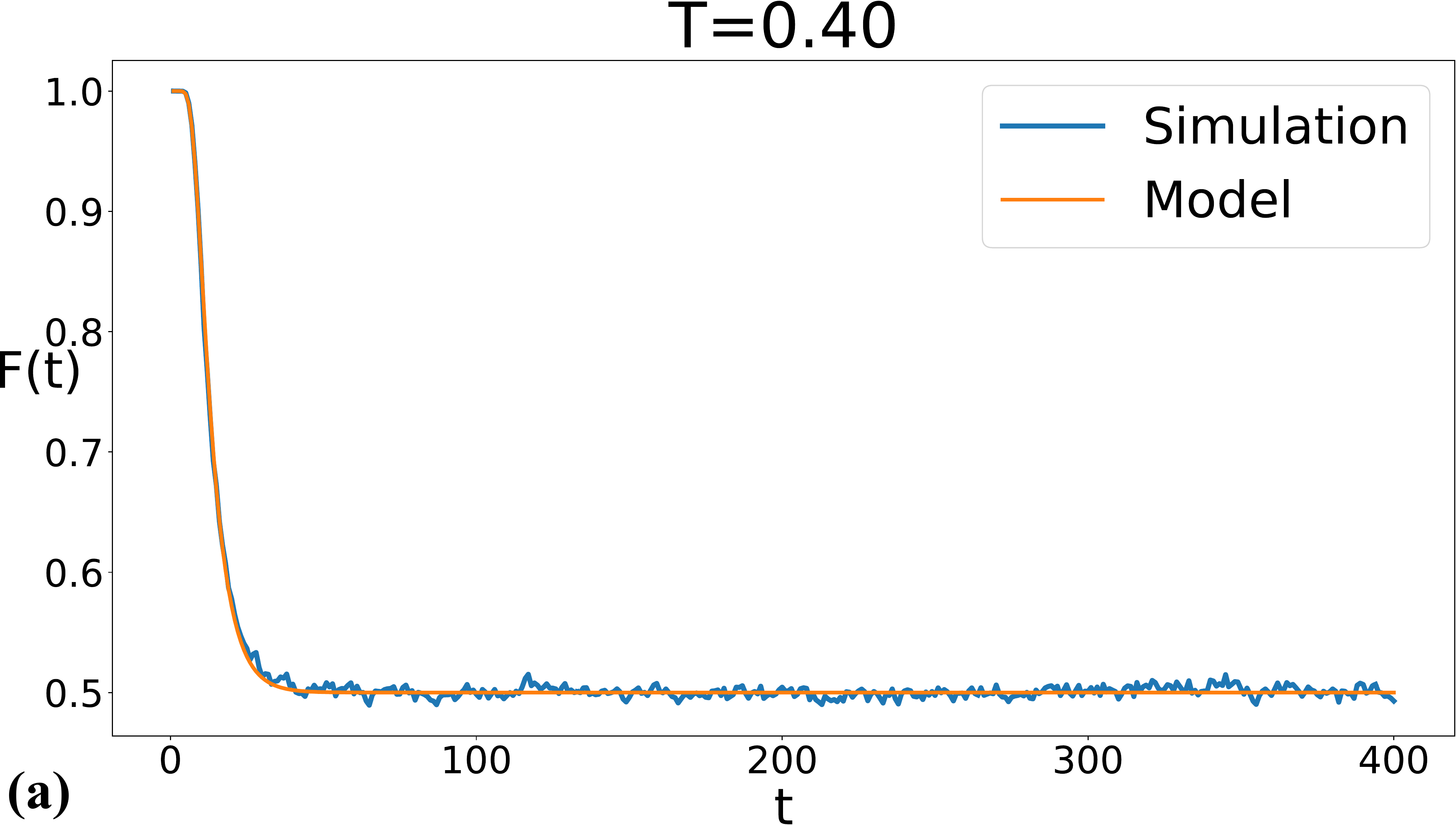}\qquad
	\includegraphics[width=0.45\textwidth]{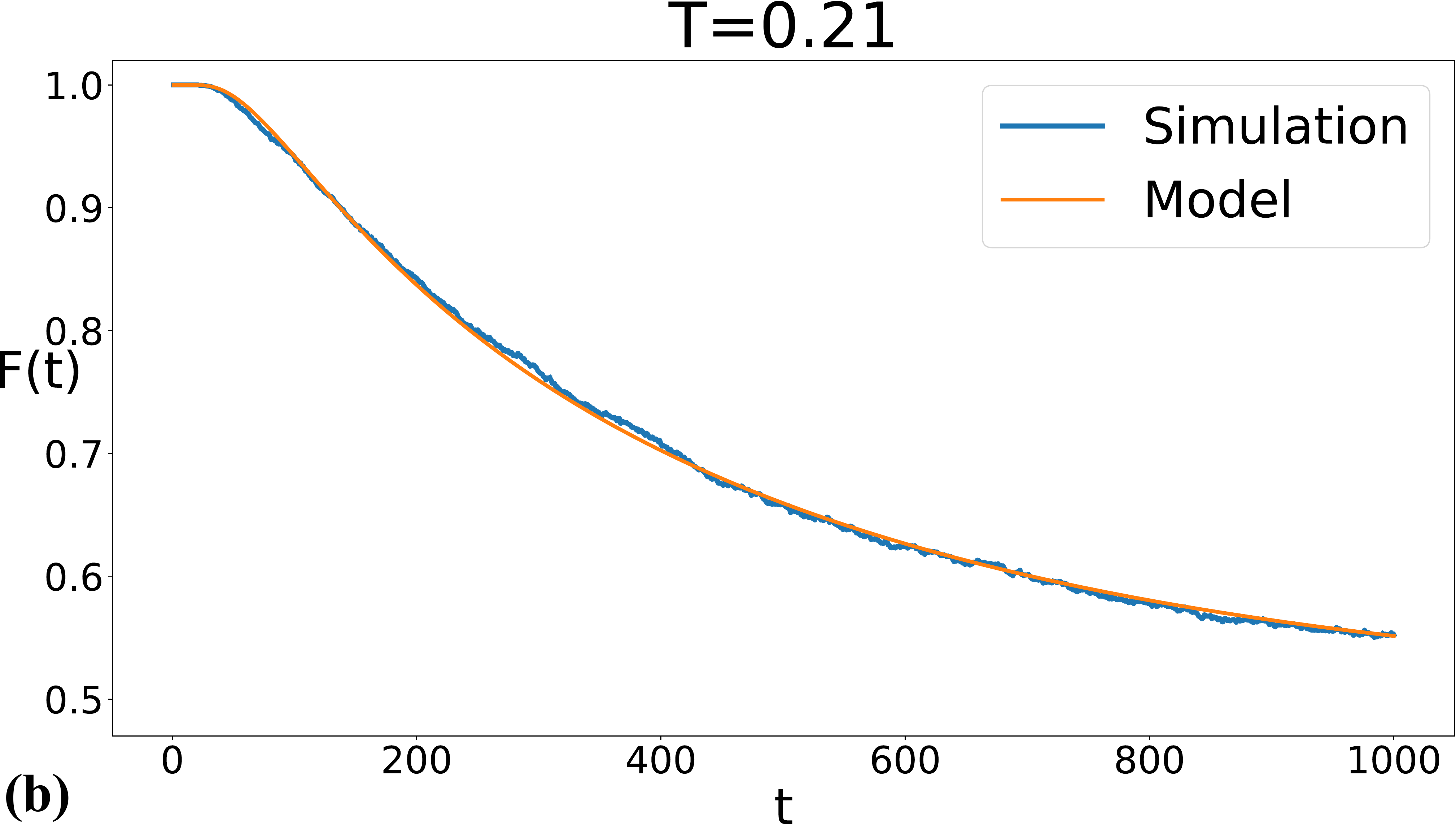}
	\caption{The plots of fidelity $F(t)$ of the logical bit encoded in the PCA at $h=0$. The width of the peak in the $F(t)$ plot captures the memory lifetime of the PCA at that temperature. (a) At $T=0.40$, the fidelity estimated from simulation agrees very well with the proposed model in Eq.~\eqref{eq:F(t)_triang}. For a system with $N=200$ spins, we obtain parameters $N_{\text{eff}}=38.7 \pm 1.2$, $\lambda= 0.0884 \pm 0.0001$ from the model, with goodness of fit parameters $\varchi^2 = 335$ and degrees of freedom $\text{DOF}=398$.
  (b) For the same system at a lower temperature $T=0.21$, we obtain the parameters to be $N_{\text{eff}}=1.373 \pm 0.006$, $\lambda= (1.105 \pm 0.003) \times 10^{-3}$ with $\varchi^2 = 1181$, $\text{DOF}=998$. The ratio $\varchi^2/\text{DOF}$ is found to increase as $T$ is lowered, indicating that the model best captures the behavior of the system at high temperatures~\cite{dayIsingFerromagnetSelfcorrecting2012}.}
	\label{ModelTest_triang}    
\end{figure*}

\begin{figure*}[ht!]
 	\centering
    \includegraphics[width=0.269\textwidth]{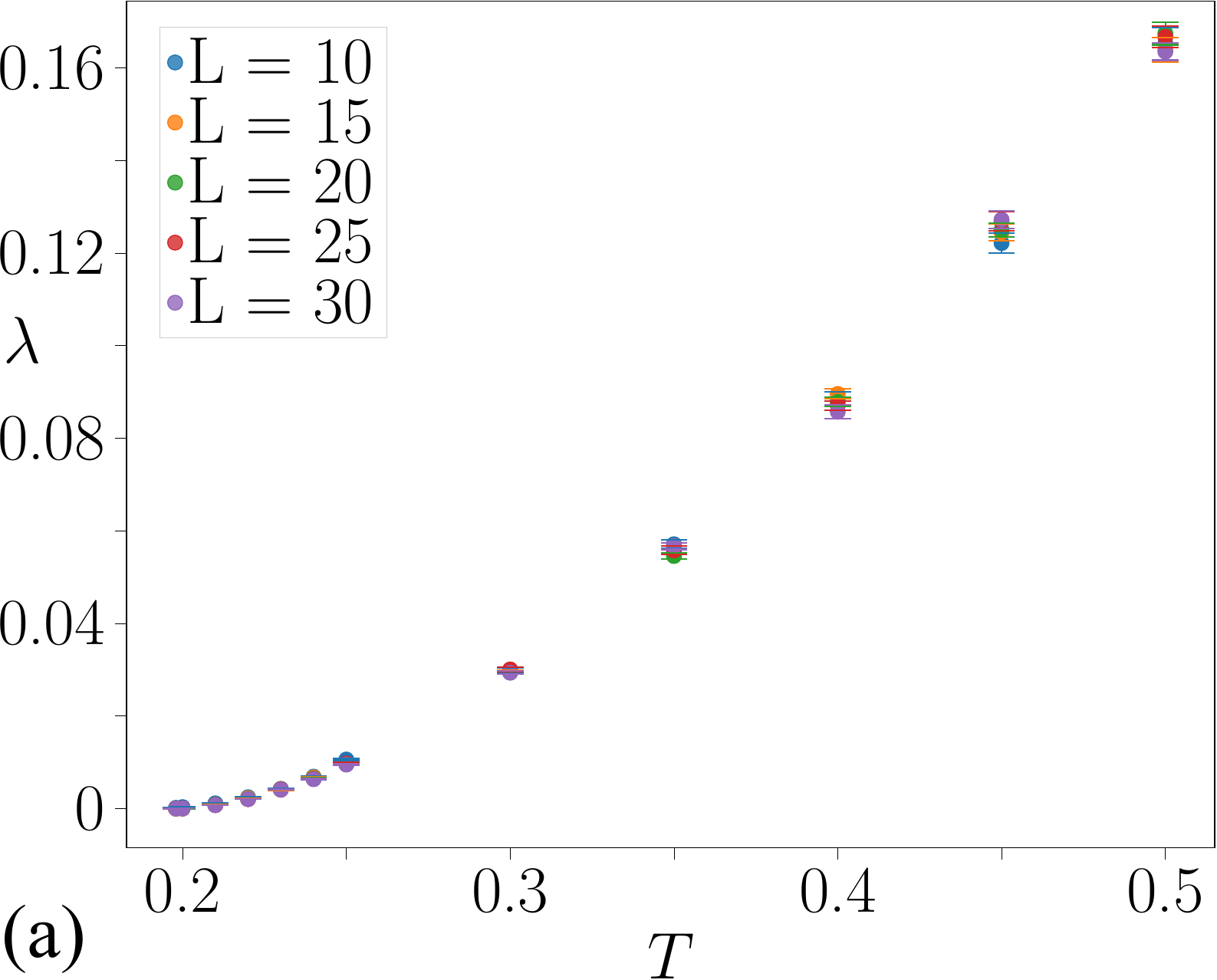}\qquad
    \includegraphics[width=0.33\textwidth]{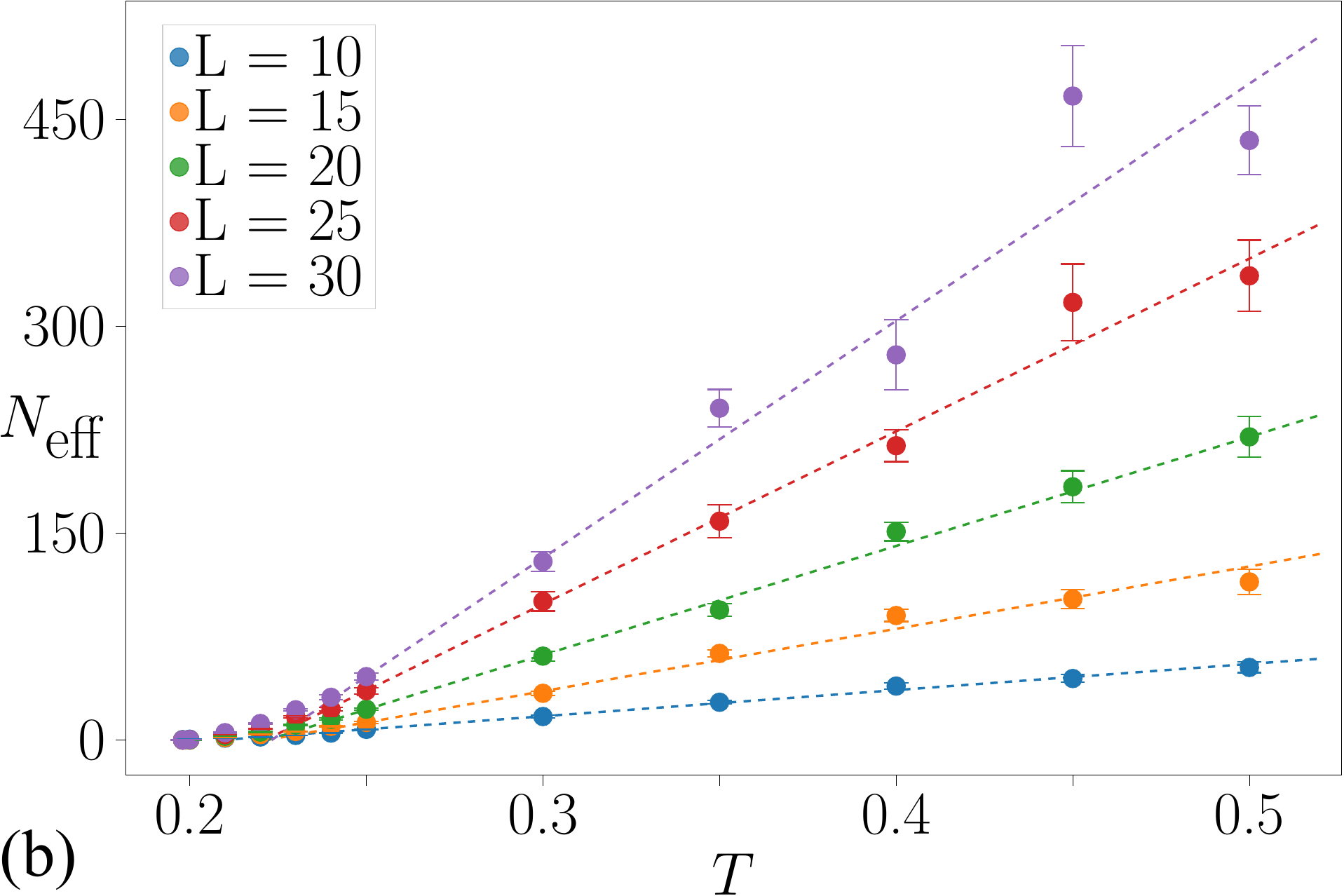}\qquad
    \includegraphics[width=0.245\textwidth]{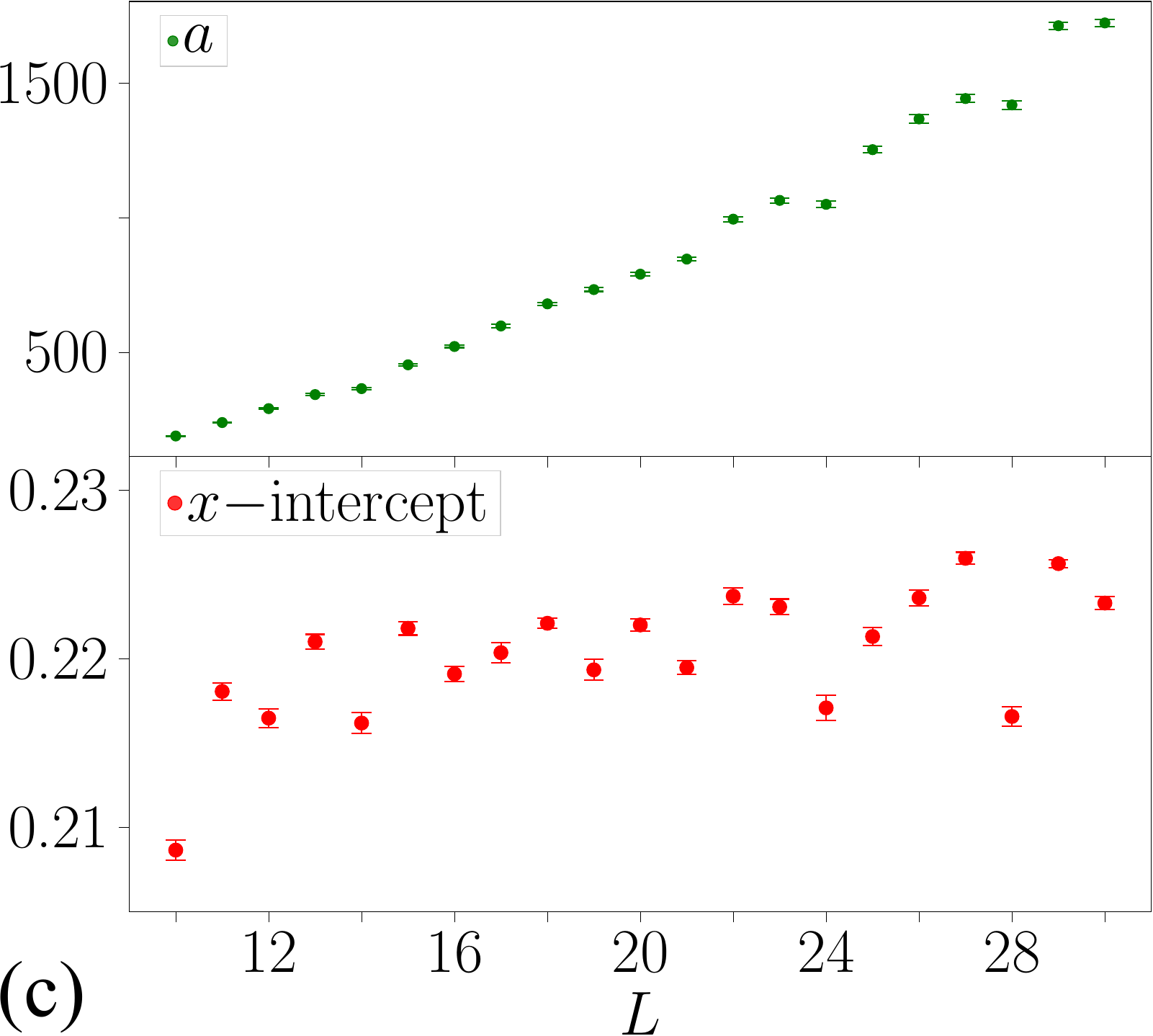}
    \caption{
    (a) The plot of the mean flipping rate $\lambda$ as a function of $T$. $\lambda$ does not change noticeably with $L$. (b) The plot of the effective number of spins $N_{\text{eff}}$ as a function of $T$. In the range $T\in[0.25, 0.5]$, this parameter scales linearly with temperature, as captured by Eq.~\eqref{eq_neff}. (c) We plot the slope $a$ and the $x$-intercept of this linear fit as a function of lattice size $L$. The $x$-intercept gives the temperature at which $N_{\text{eff}} = 0$, subsequently indicating the breakdown of the model.
    }
    \label{AnalytParamsvT}
\end{figure*}

We test the validity of the model in Eq.~\eqref{eq:F(t)_triang} by numerically estimating the fidelity of the system as a function of time and checking the goodness of fit of the model. 
We simulate the PCA initialized with all spins in state $+1$ for $n=10^4$ trials and at each time step $t$, we estimate $F(t)$ as the fraction of the trials for which $M(t) > 0$.
We observe that in the high temperature range, i.e., $T\in [0.25, 1]$, the model is a good description of the system. For instance at $T=0.4$ (see Fig.~\ref{ModelTest_triang}(a)), the goodness of fit ratio indicates agreement between numerical data and Eq.~\eqref{eq:F(t)_triang}. However, as the temperature is lowered, the quality of this fit diminishes. In particular, at $T=0.21$, the goodness of fit ratio (see Fig.~\ref{ModelTest_triang}(b)) suggests that the model starts to break down.

In Fig.~\ref{AnalytParamsvT}, we plot the parameters $\lambda$ and $N_{\text{eff}}$ against temperature $T$ for  different lattice sizes $L$.
We observe that $\lambda$ is predominantly dependent on $T$ and does not change noticeably with $L$.
In the range $T\in[0.25, 0.5]$, the parameter $N_{\text{eff}}$ can be approximated with the following ansatz
\begin{equation}
    N_{\text{eff}} = aT + b,
    \label{eq_neff}
\end{equation}
where $a,b$ are fitting parameters.
The slope $a$ scales linearly with $L$ and by finding the $x$-intercept of the numerical ansatz in Eq.~\eqref{eq_neff} we can identify that our simple analytical model breaks down at $T\approx 0.22$, as $N_{\text{eff}}$ (that corresponds to the effective number of independent spins) becomes zero (see Fig.~\ref{AnalytParamsvT}(c)).
This is indicative of a phase transition in the system and corroborates our numerical findings at $h=0$.

 \section{Discussion}\label{Conclude}

In this article, we studied a classical spin system evolving under the sweep rule, and used numerical simulations to find the region in the $(T,h)$ plane, where its memory lifetime scales exponentially with linear lattice size. 
For $h=0$, we supported our numerical analysis by an additional statistical-mechanical study of the sweep rule and a simple analytical model that captures the memory lifetime at high temperatures.
We also found  that the sweep rule belongs to the weak 2D Ising universality class.

In the main text, we focused on the synchronous sweep rule on triangular lattices.
However, our results do not change qualitatively if we consider square lattices instead (see Appendix~\ref{sec_Toomsqure}).
In fact, the sweep rule on square lattices is identical to Toom's rule.
From this perspective, our results recover and are consistent with the previous works on Toom's rule~\cite{toom1980stable, bennettRoleIrreversibilityStabilizing1985,makowiecUniversalityClassProbabilistic2002}.
We also remark that our findings are largely the same for the asynchronous sweep rule, with the main difference being that it belongs to the Ising universality class (see Appendix~\ref{Variants} for details).

PCAs such as the sweep rule not only stabilize classical information; they can also help to protect quantum information encoded into topological quantum error-correcting codes, such as the toric code~\cite{kitaevFaulttolerantQuantumComputation2003, dennisTopologicalQuantumMemory2002} and the color code~\cite{bombinTopologicalQuantumDistillation2006, bombinOptimalResourcesTopological2007, kubicaaleksandermarekABCsColorCode2018}. For instance, Toom's rule and the sweep rule are at the heart of many decoding algorithms~\cite{dennisTopologicalQuantumMemory2002, breuckmannLocalDecoders2D2016, kubicaEfficientColorCode2023, kubicaCellularAutomatonDecodersProvable2019, vasmerCellularAutomatonDecoders2021, breuckmannPhDThesisHomological2018, pastawskiQuantumMemoriesBased2011}.
In general, decoding algorithms based on PCAs are particularly appealing because they are, by definition, parallelizable and only use local information about errors affecting the system.
Lastly, we remark that studying classical spin models and their phase transitions give insights into the performance of quantum error-correcting codes and their optimal decoding algorithms~\cite{dennisTopologicalQuantumMemory2002, bombinTopologicalSubsystemCodes2010, bombinUniversalTopologicalPhase2012, kubicaThreeDimensionalColorCode2018, chubbStatisticalMechanicalModels2021, duaClifforddeformedSurfaceCodes2022}. 
We hope that studying PCAs will bring us closer to realizing (possibly self-correcting) quantum memories~\cite{brownQuantumMemoriesFinite2016, terhalQuantumErrorCorrection2015}.

\acknowledgements

A.R. thanks R. Melko and M. Vasmer for providing valuable feedback during the writing of this article.
A.R. acknowledges the use of Compute Canada resources for numerical simulations reported in this article.
A.K. thanks C. H. Bennett and J. Smolin for exciting discussions about Toom's rule.
This research was undertaken thanks in part to funding from the Government of Canada through the Natural Sciences and Engineering Research Council of Canada (NSERC).

\onecolumngrid
\appendix
\section{Details of numerical analysis of memory lifetime}\label{MemNumerics}
\subsection{Obtaining data}
	
	For parameters $T, h$ and linear lattice size $L$, we simulate the sweep rule $n=1000$ times, each simulation starting with all spins aligned. For each simulation, we smoothen the magnetization $M(t)$ as follows.
 	\begin{itemize}
		\item Coarse graining the magnetization $M(t)$ by a factor of $L/2$ to obtain the coarse grained magnetization
        \begin{align}
            M_\text{CG}(t') = \frac{2}{L} \sum_{t = t'+1}^{t'+ L/2} M(t),
        \end{align}
        where $t'$ is a multiplicity of $L/2$.
  		\item Convolving $M_{\text{CG}}(t')$ by a rectangular filter (moving average filter) of width $w = L$ (at low temperatures) or $w = L/2$ (at high temperatures) to obtain the smoothened magnetization
        \begin{align}
        \label{smooth}
          M_{\text{smooth}}(t') = \frac{1}{w} \sum_{i=0}^{w-1} M_\text{CG}\left(t'+ \tfrac{1}{2}i L\right). 
        \end{align}

	\end{itemize}
	For the $i^{\text{th}}$ simulation, where $i \in \{1, 2,..., n\}$, we record the smallest value of $t_{i}$ for which $M_{\text{smooth}}(t_{i}) = 0$. This is noted as the time of the first logical bit-flip in the $i^{\text{th}}$ simulation. The memory lifetime $\tau_{L,T}$ of the system of lattice size $L$ at temperature $T$ is then estimated by the mean of the recorded values $t_{i}$.
	
	We remark that in our numerical analysis we also compared the rectangular filter in Eq.~\eqref{smooth} with Gaussian filters of the same width. We observed that the smoothened magnetization $M_{\text{smooth}}$ obtained with the rectangular filter enabled a clearer identification of logical bit-flips than that obtained with the Gaussian filter.
 
	\subsection{Identifying phase transitions}
	
\begin{figure}[ht!]
	\centering
		\includegraphics[width=0.47\textwidth]{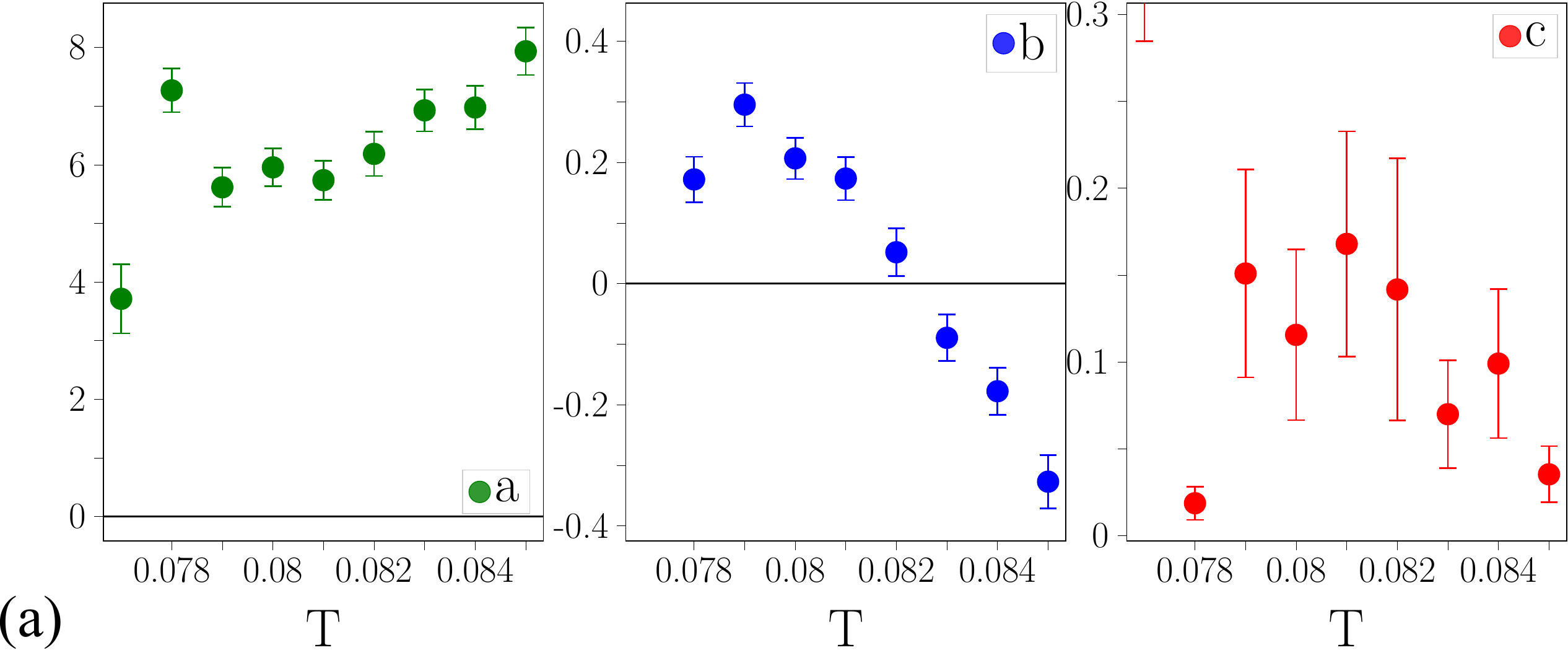}\qquad
	    \includegraphics[width=0.47\textwidth]{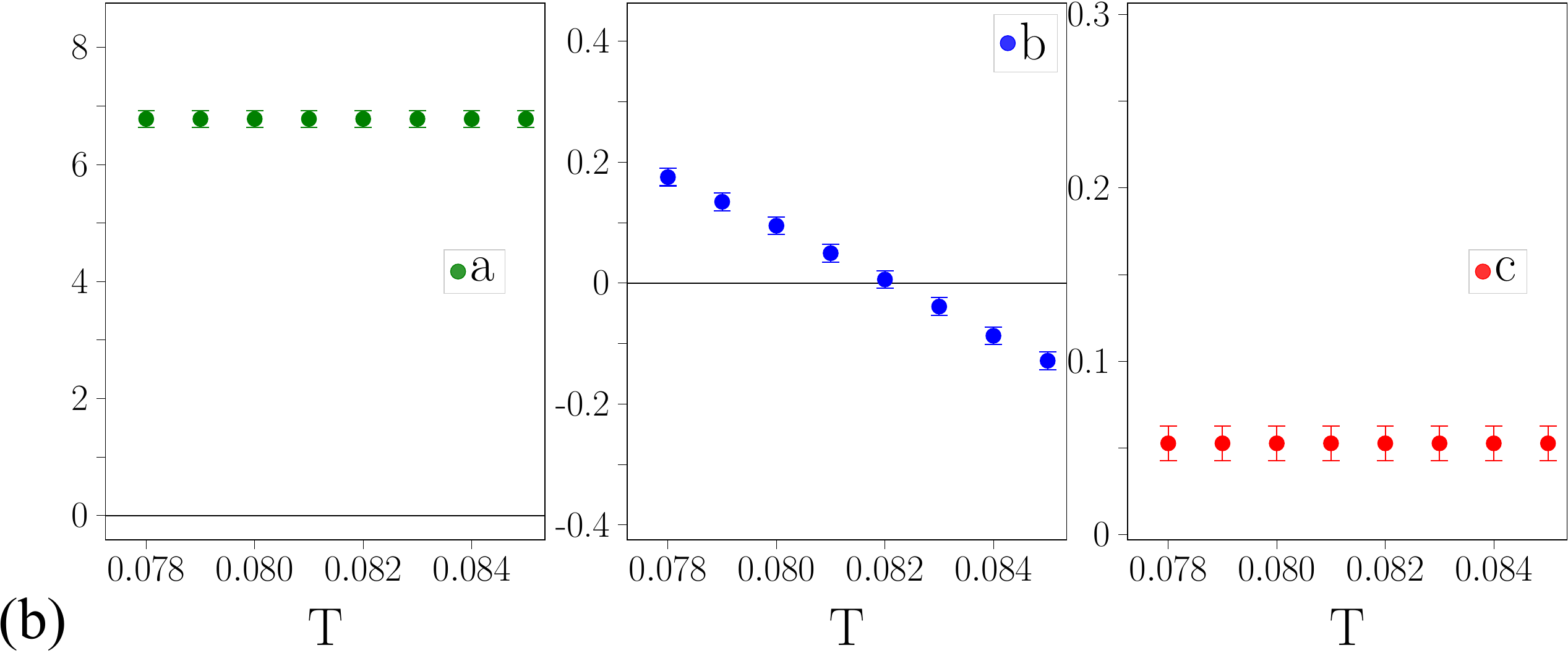}
		\caption{(a) Estimating parameters $a,b,c$ from the values obtained by fitting resampled data sets at $h=0.5$ into the ansatz Eq.~\eqref{ansatz}.
  (b) By constraining $a,c$ to be constant over the range of temperatures we obtain a smoother plot of $b$ as a function of $T$ and, subsequently, can identify the phase transition as the point where $b=0$, i.e., $T^{0.50}_{c} = 0.082 \pm 0.001$.} 
		\label{abcFits}
	\end{figure}

	For a given value of $h$, we fit the $\tau_{L,T}$ data obtained for various $L$ and $T$ into the fitting ansatz in Eq.~\eqref{ansatz}. We use the bootstrapping method to estimate the variance of the fitting parameters $(a,b,c)$, i.e., for each $L$, $T$, we resample (with replacement) from the data set $t_{i}$, $i \in \{1, 2,..., n\}$ to obtain data sets $t^{1}_{i}$, $t^{2}_{i}$,..., $t^{\text{B}}_{i}$, where $B=50$. For each of the $B$ data sets we calculate $\tau_{L,T}$ and fit into the ansatz to obtain the value of the fit parameters $a,b,c$. Each parameter is then estimated by its mean and variance over the $B$ samples (see Fig.~\ref{abcFits}).
	
	We identify the phase transition as the temperature where the parameter $b$ becomes zero. In our analysis, we observed that in the neighbourhood of the phase transition, the parameters $a, c$ do not change significantly with temperature $T$. So, in this neighbourhood, we fit the resampled memory lifetime data into the ansatz while constraining parameters $a, c$ to be constant over the range of temperatures. Constraining parameters $a,c$ in this manner (see Fig.~\ref{abcFits}(b)) gives a smooth plot of $b$ as a function of $T$ allowing us to clearly identify the point where $b=0$.

	\subsection{Analysis of the sweep rule on triangular lattices}
	\begin{figure}[ht!]
			\includegraphics[width=0.32\textwidth]{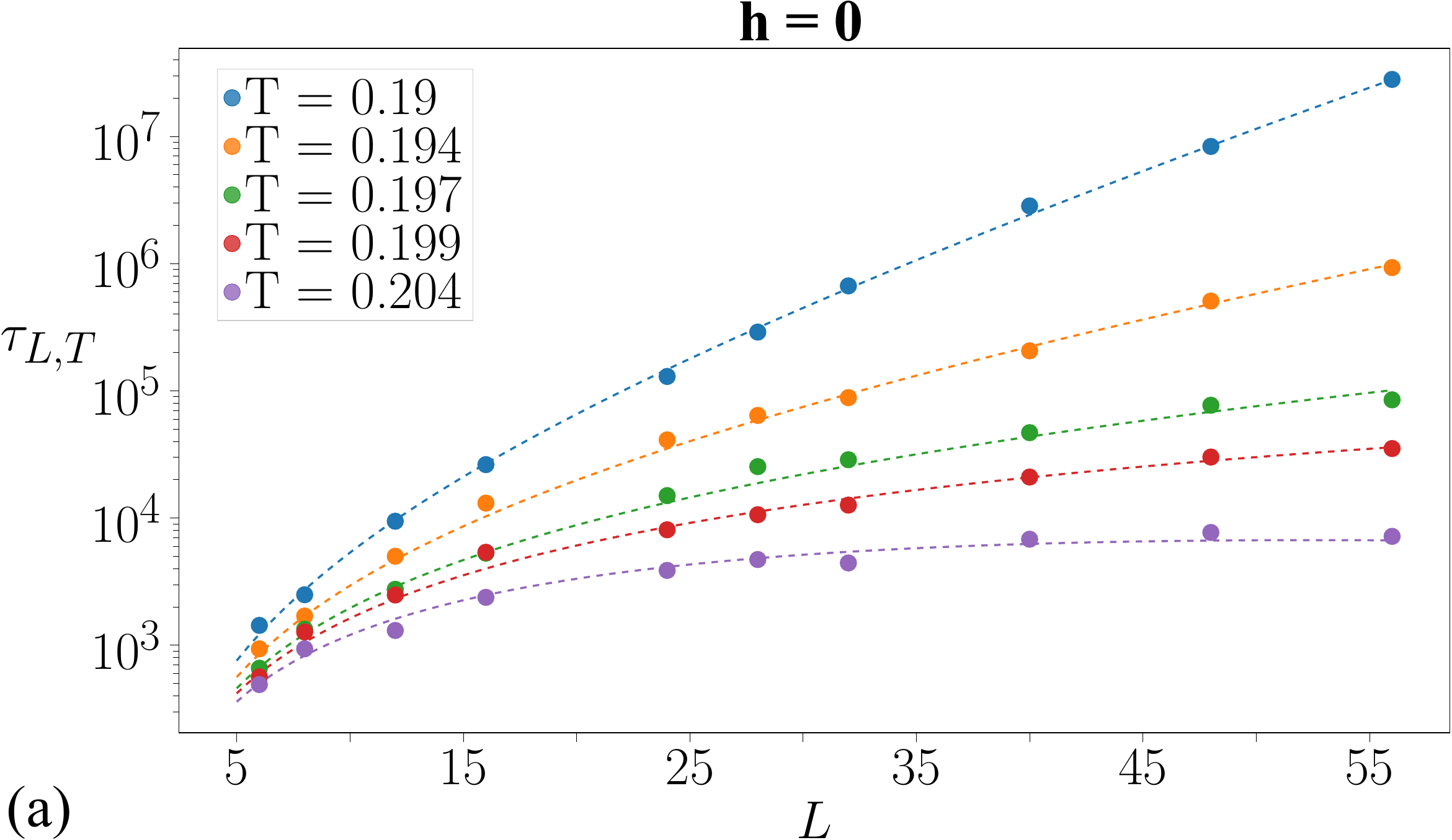} 
			\includegraphics[width=0.32\textwidth]{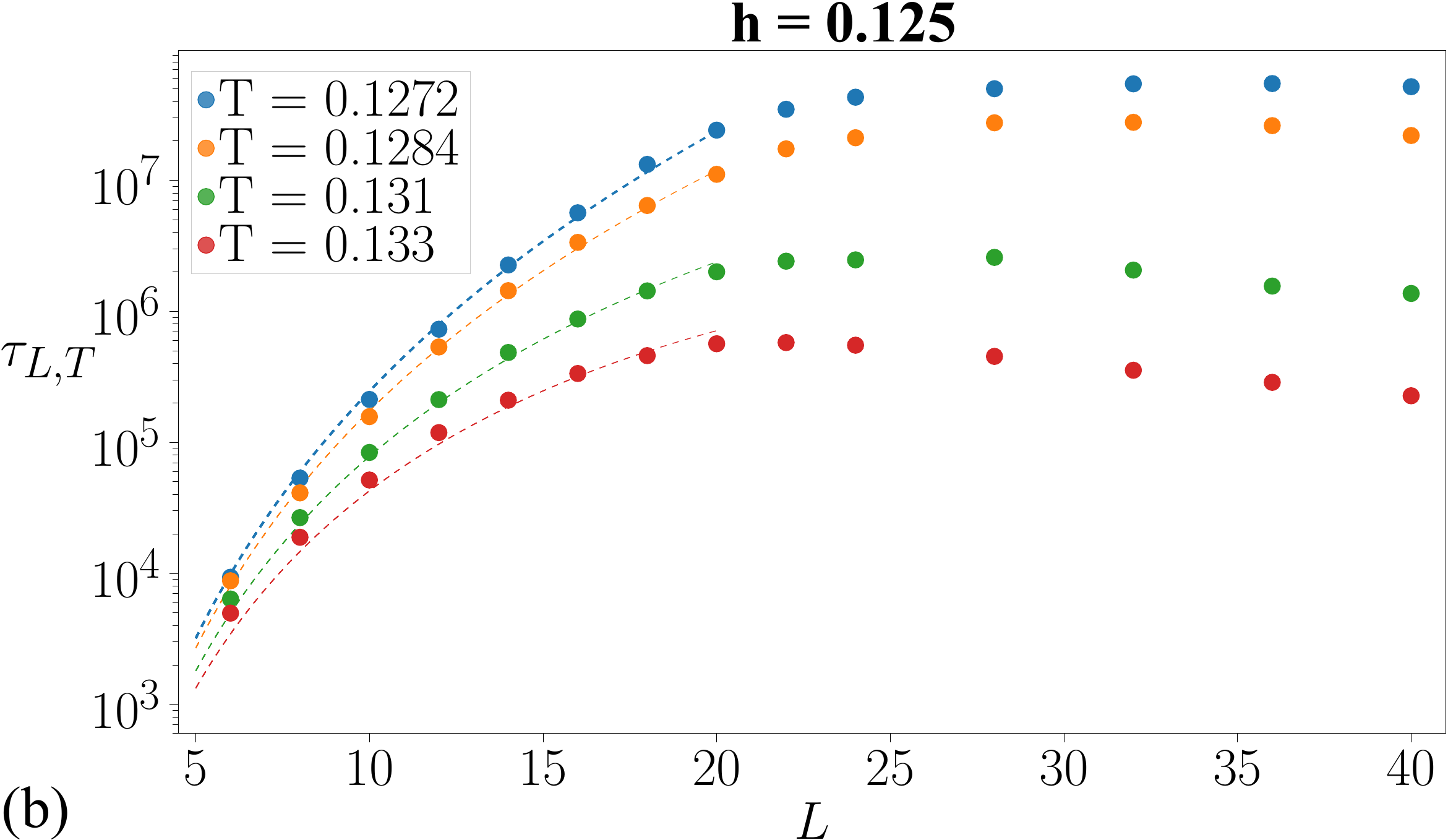} 
			\includegraphics[width=0.32\textwidth]{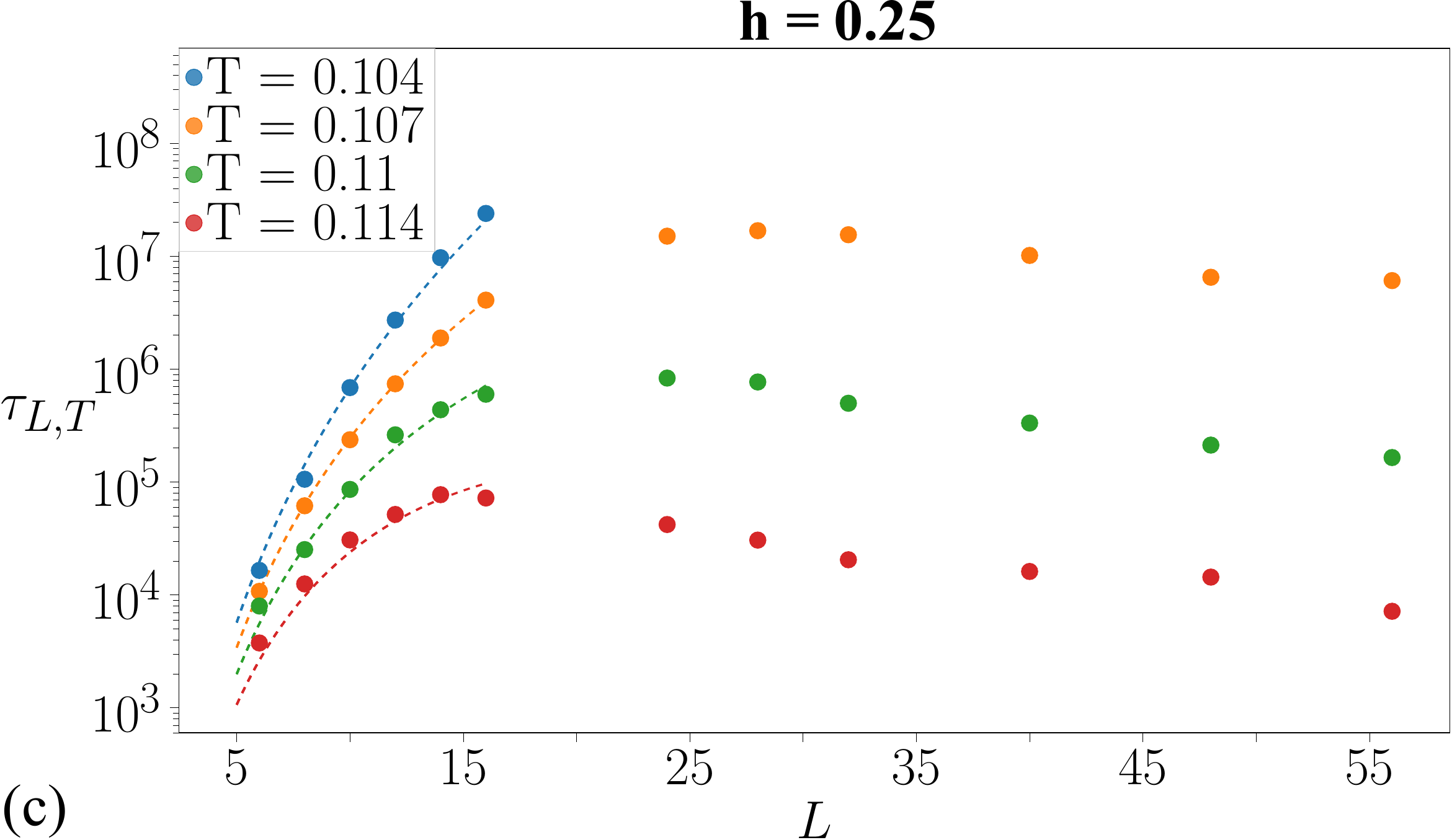} \\ [1.5ex]
			\includegraphics[width=0.32\textwidth]{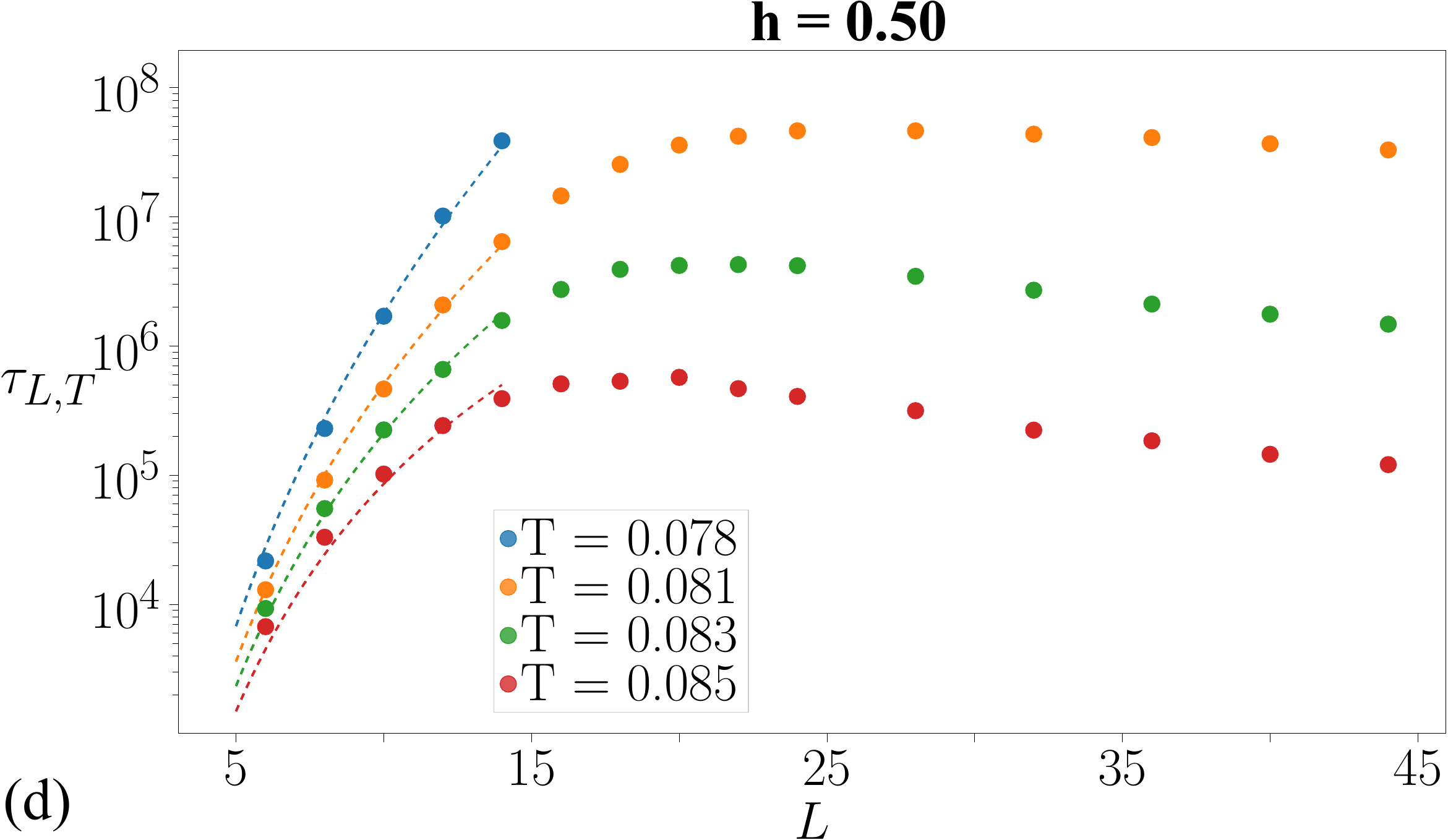} 
			\includegraphics[width=0.32\textwidth]{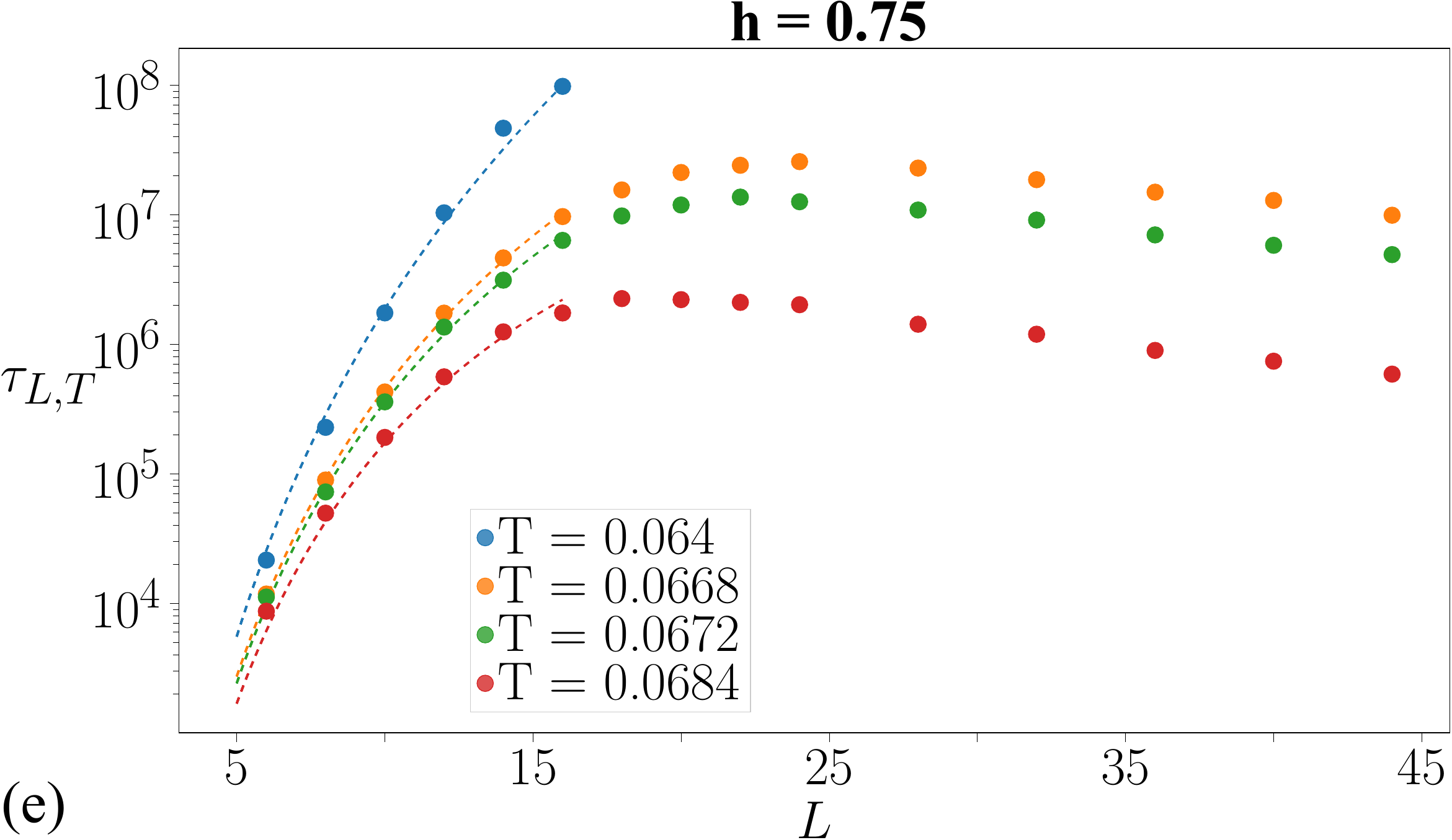} 
			\includegraphics[width=0.32\textwidth]{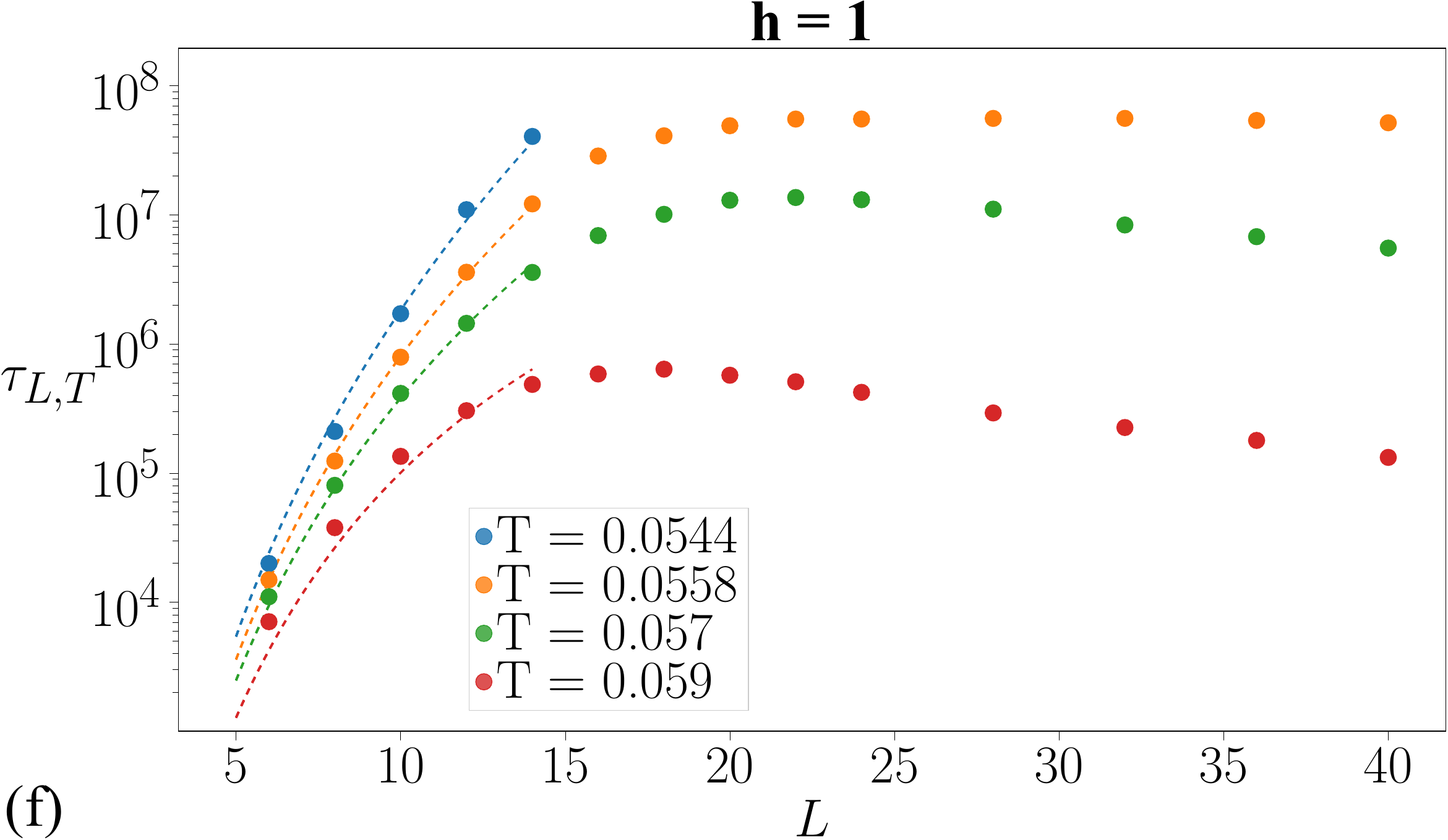} 
		\caption{Plots of memory lifetime $\tau_{L,T}$ of the sweep rule as a function of linear lattice size $L$. The dashed lines indicate the best fit into the ansatz in Eq.~\ref{ansatz}. The data in these plots have been fit by taking into account the statistical errors in the numerical estimation of memory lifetime. To avoid clutter, we do not display the corresponding asymmetric error bars.
  }
		\label{TriangTMem}
	\end{figure}
	\begin{figure}[ht!]
		\includegraphics[width=0.32\textwidth]{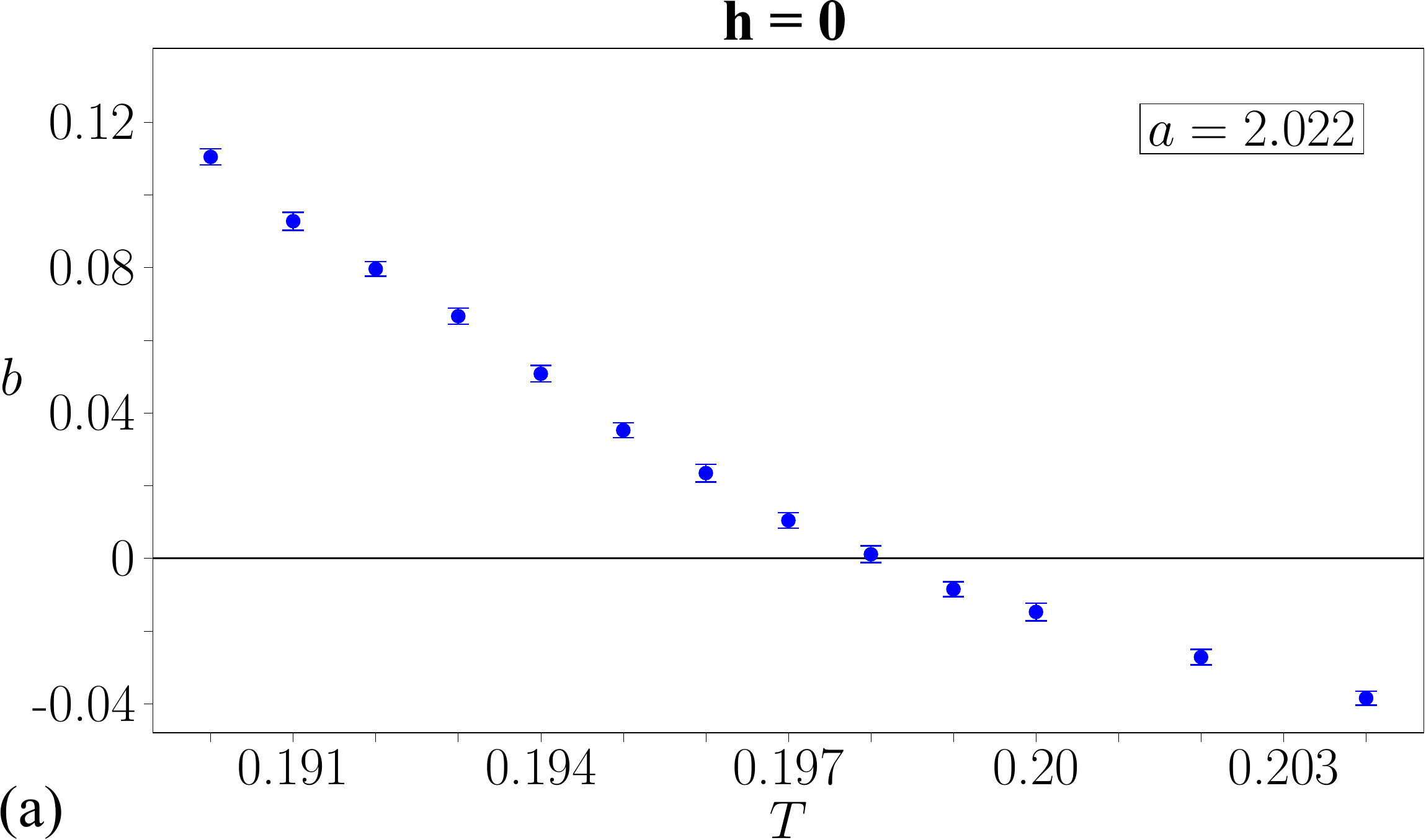} 
		\includegraphics[width=0.32\textwidth]{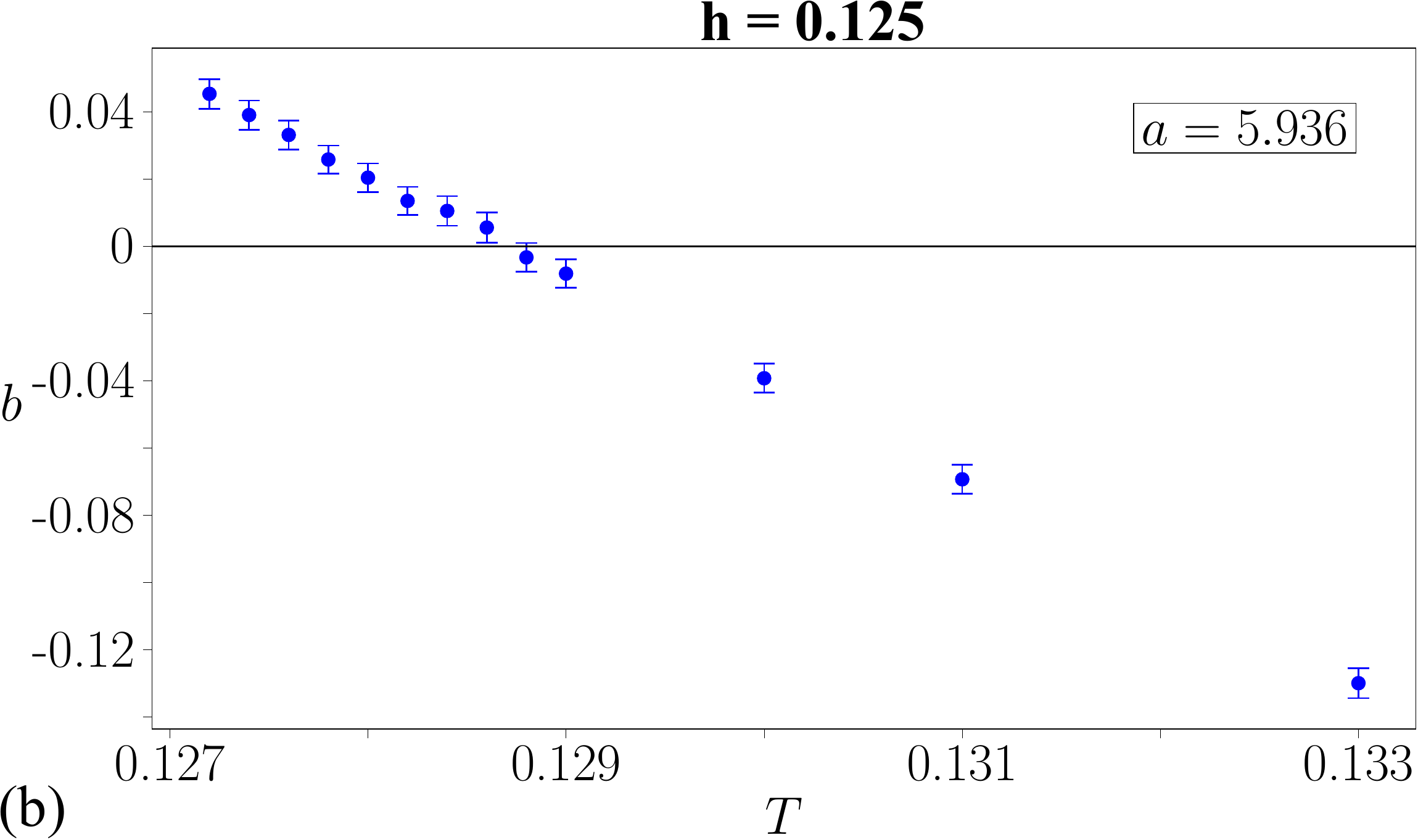} 
		\includegraphics[width=0.32\textwidth]{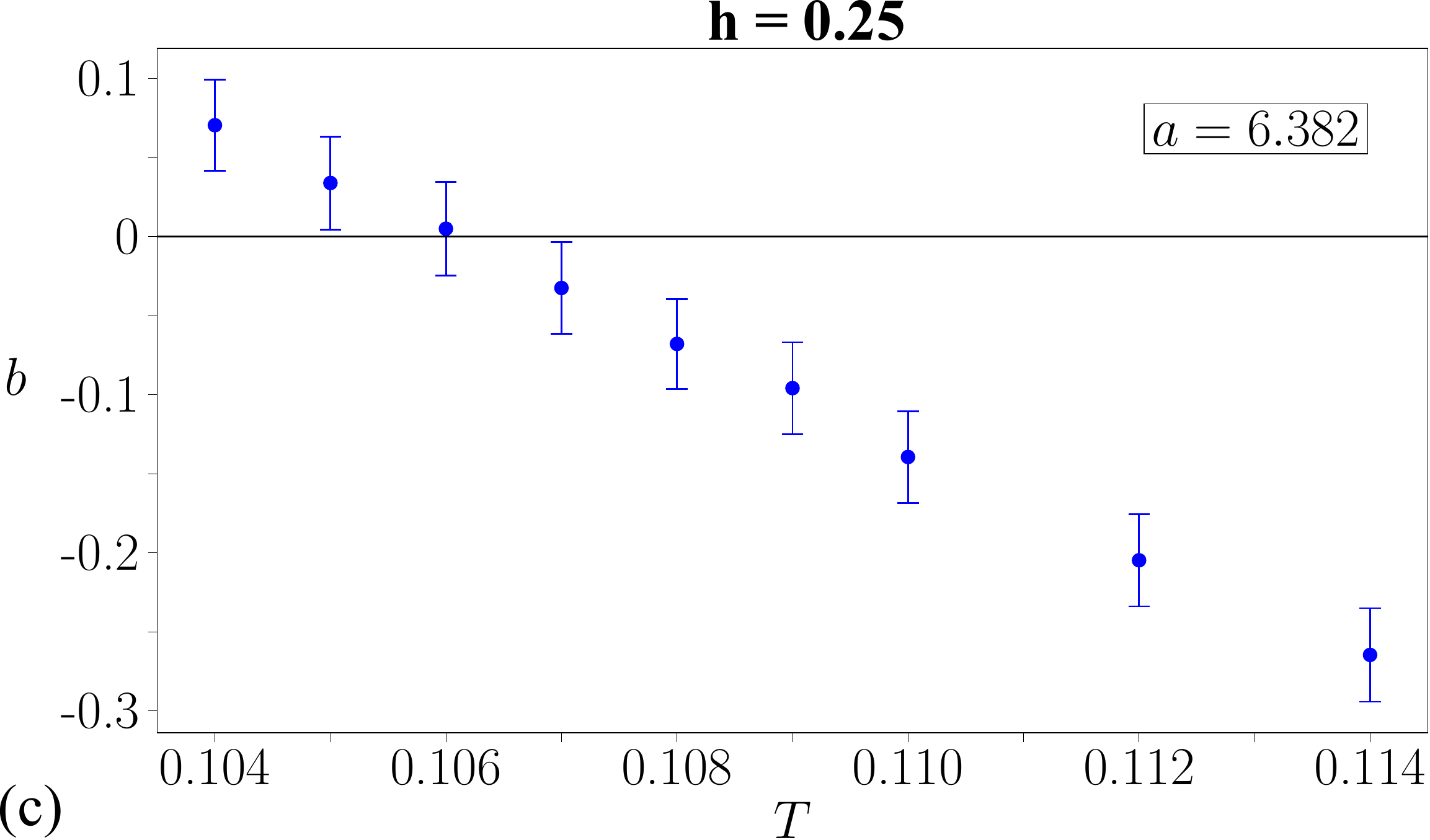} \\ [1ex]
		\includegraphics[width=0.32\textwidth]{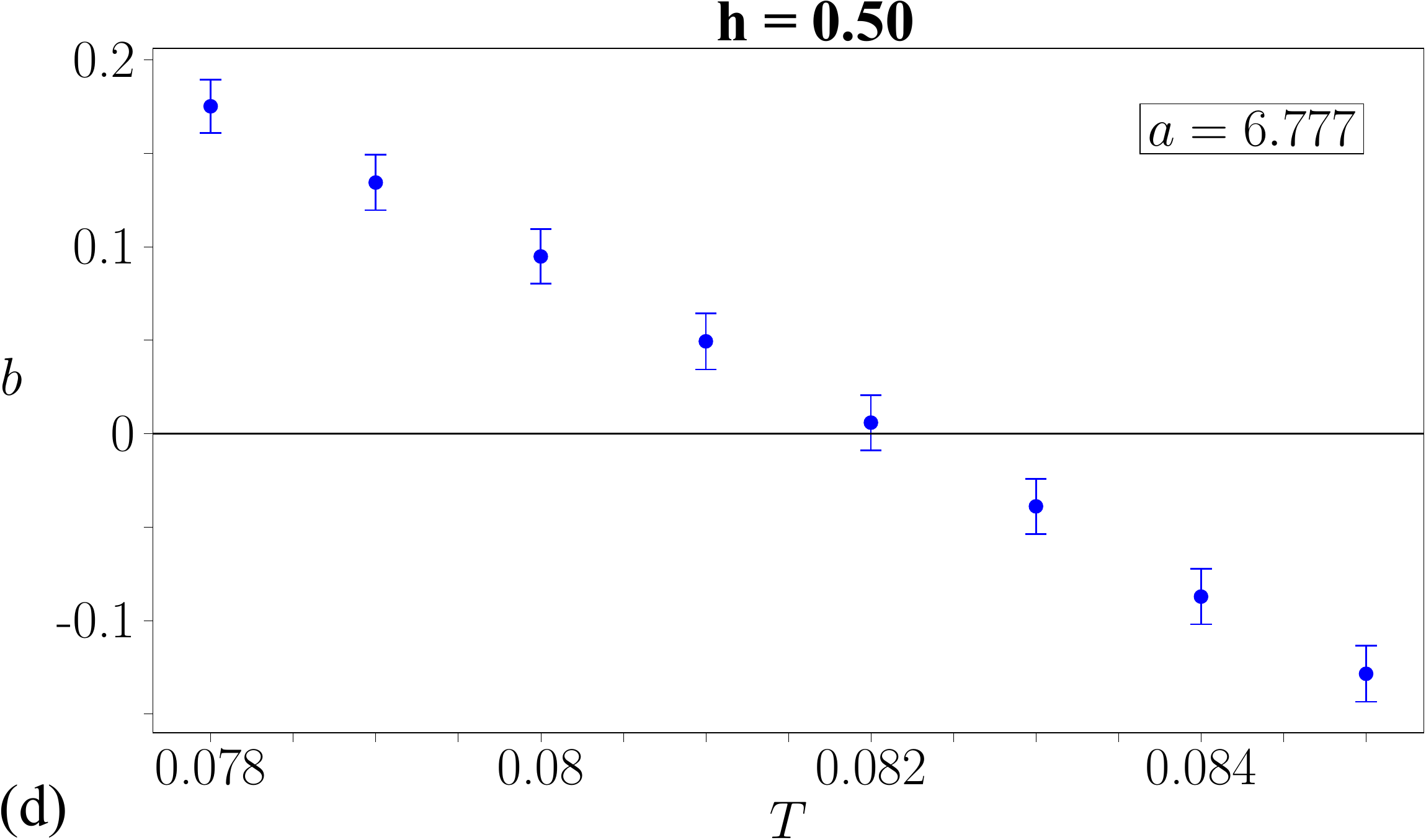} 
		\includegraphics[width=0.32\textwidth]{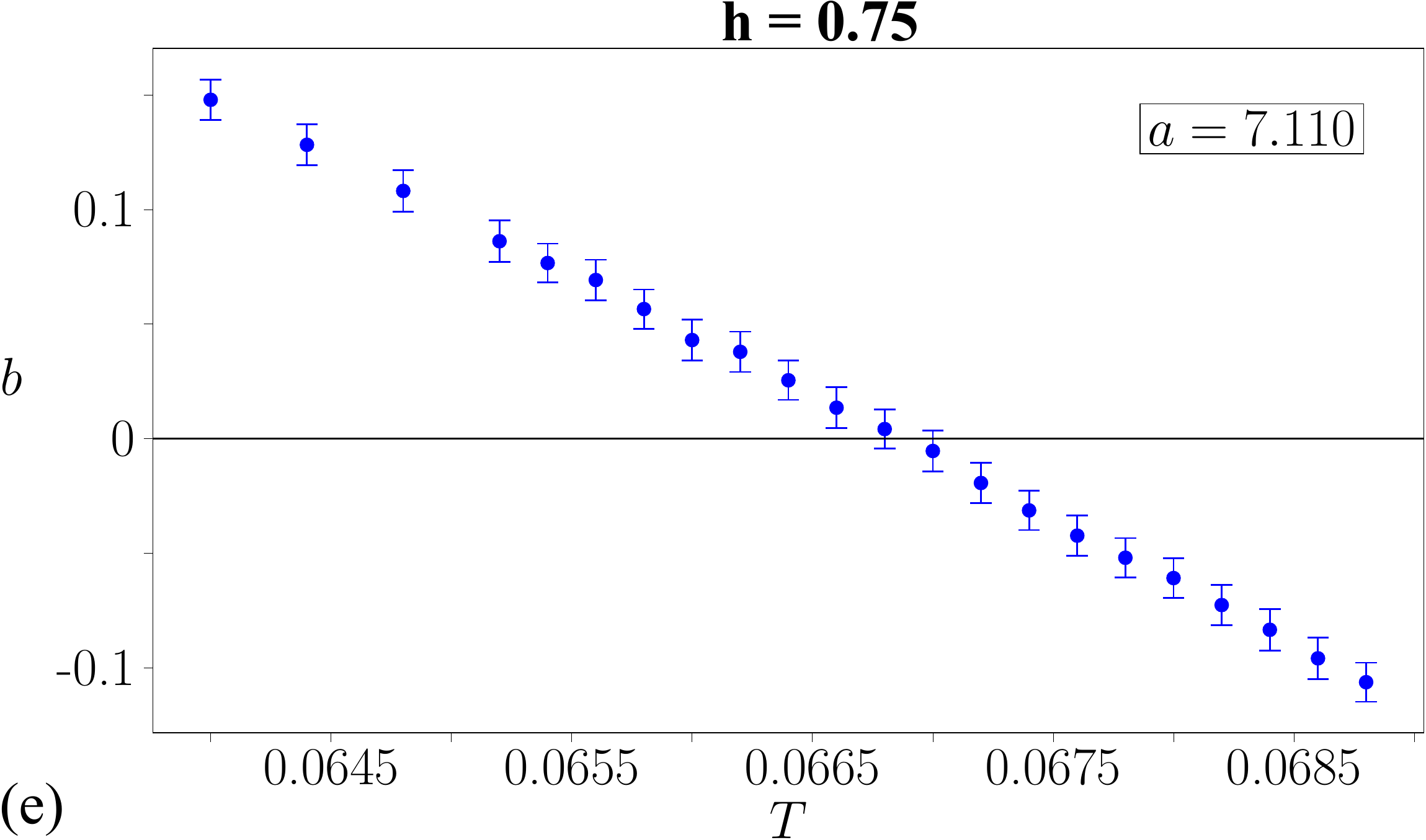} 
		\includegraphics[width=0.32\textwidth]{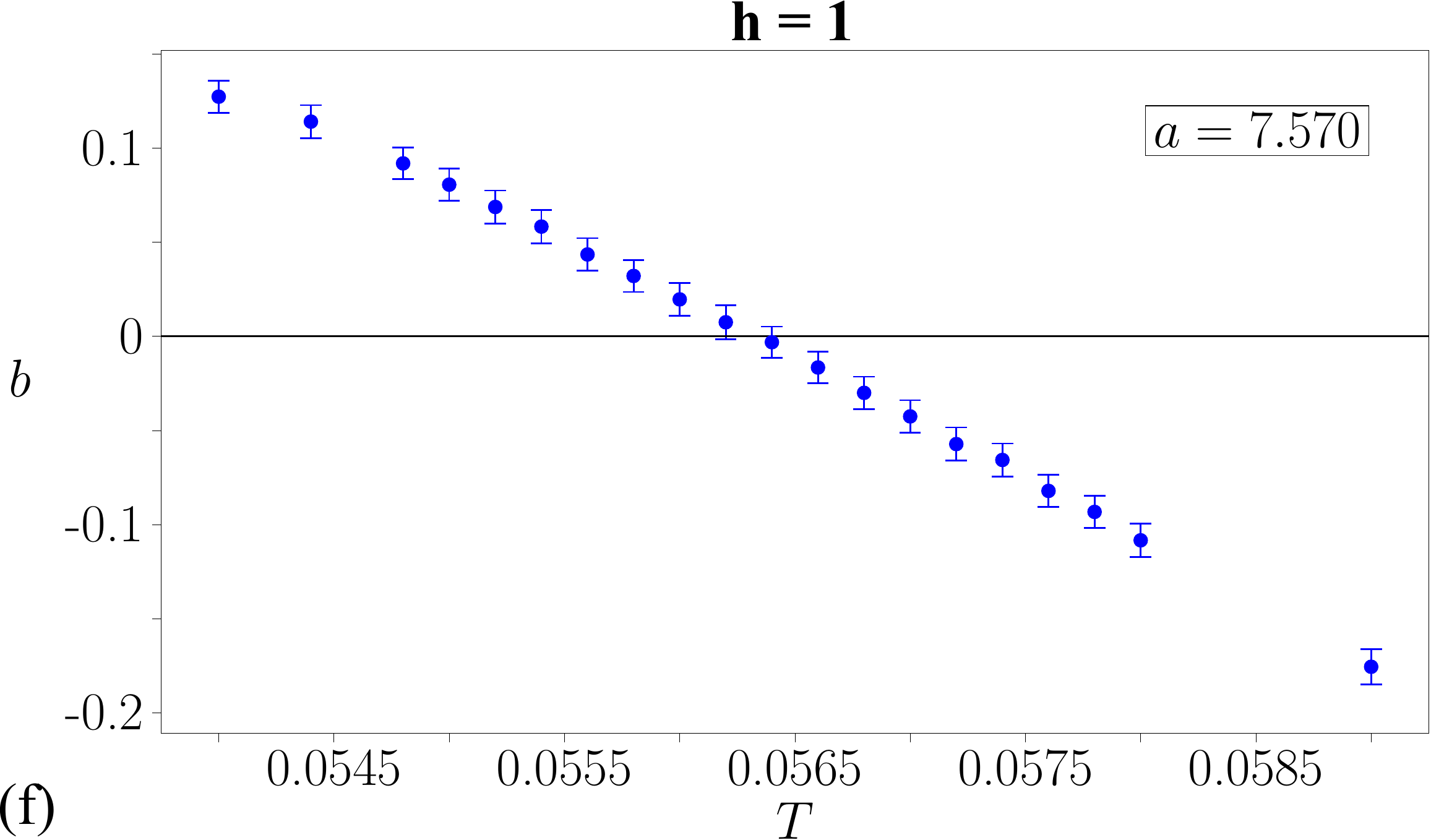} 
		\caption{Plot of parameter $b$ as a function of temperature $T$ for respective values of $h$ corresponding to the fit obtained in Fig.~\ref{TriangTMem}. We identify the critical temperature $T^{h}_{c}$ at each value of $h$ as the point where $b=0$. In each case, we constrain the parameters $a, c$ to be fixed in the neighborhood of $T^{h}_c$ and report the values of $a$ (polynomial scaling). The critical temperatures thus obtained are $T^{0}_{c} = 0.198 \pm 0.001$, $T^{0.125}_{c} = 0.129 \pm 0.001$, $T^{0.25}_{c} = 0.106 \pm 0.002$, $T^{0.50}_{c} = 0.082 \pm 0.001$, $T^{0.75}_{c} = 0.0670 \pm 0.0004$ and $T^{1.00}_{c} = 0.0564 \pm 0.0004$.}
        \label{TriangTMembvT}
	\end{figure}
	We use the aforementioned techniques and identify the critical temperatures $T^h_{c}$ of the sweep rule at different values of bias $h = \{0, 0.125, 0.25, 0.5, 0.75, 1\}$. In Fig.~\ref{TriangTMem} we show the fit obtained at each value of $h$ for the memory lifetime $\tau_{L}$ estimated using the original data set (no resampling). Fig.~\ref{TriangTMembvT} shows the plots of $b$ as a function of $T$ for the corresponding values of $h$ obtained after resampling from the original data set. The points $(T^{h}_c, h)$ are use to construct the phase diagram of the sweep rule shown in Fig.~\ref{PhaseDiag}.

	\subsection{Analysis of Toom's rule on square lattices}
    \label{sec_Toomsqure}
	\begin{figure}[ht!]
		\centering
		\includegraphics[width=0.4\textwidth]{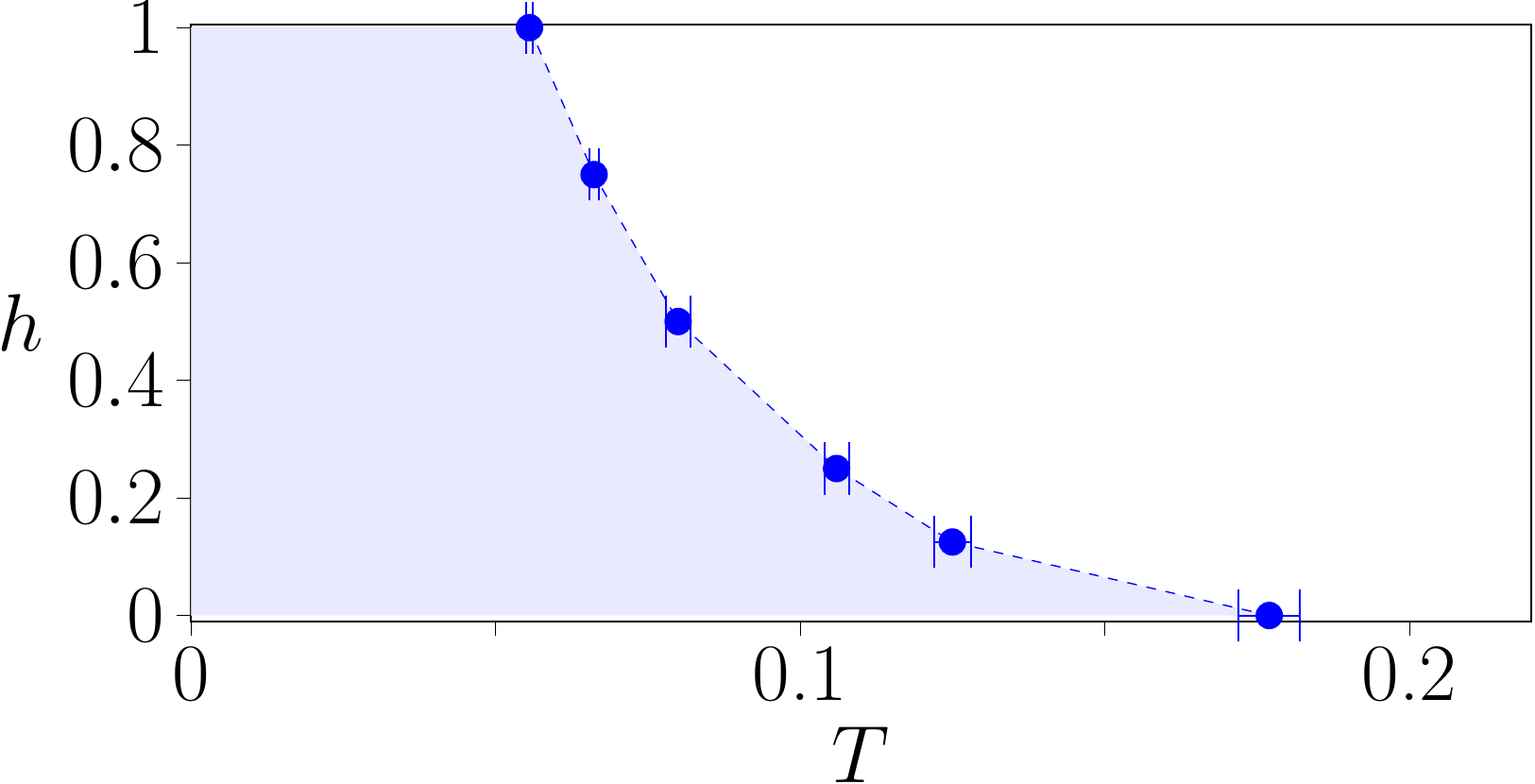}
		\caption{Phase diagram of the Toom's rule in the $(T,h)$ plane. The shaded region corresponds to a region where two stable phases coexist and the Toom's rule can work as memory. The diagram is symmetric with respect to the $h=0$ axis.}
		\label{PhaseDiagToom}
	\end{figure}
	\begin{figure}[ht!]
			\includegraphics[width=0.32\textwidth]{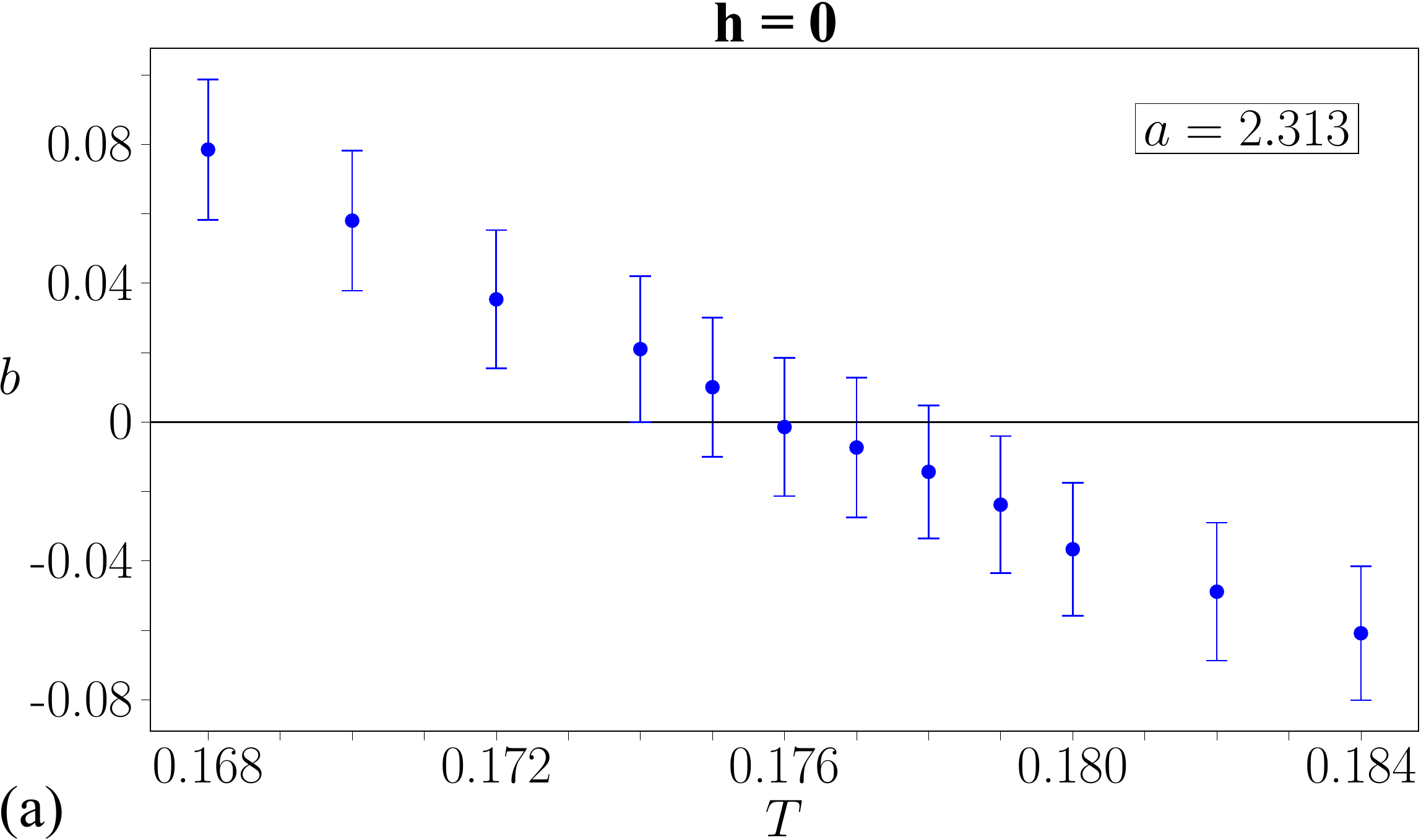} 
			\includegraphics[width=0.32\textwidth]{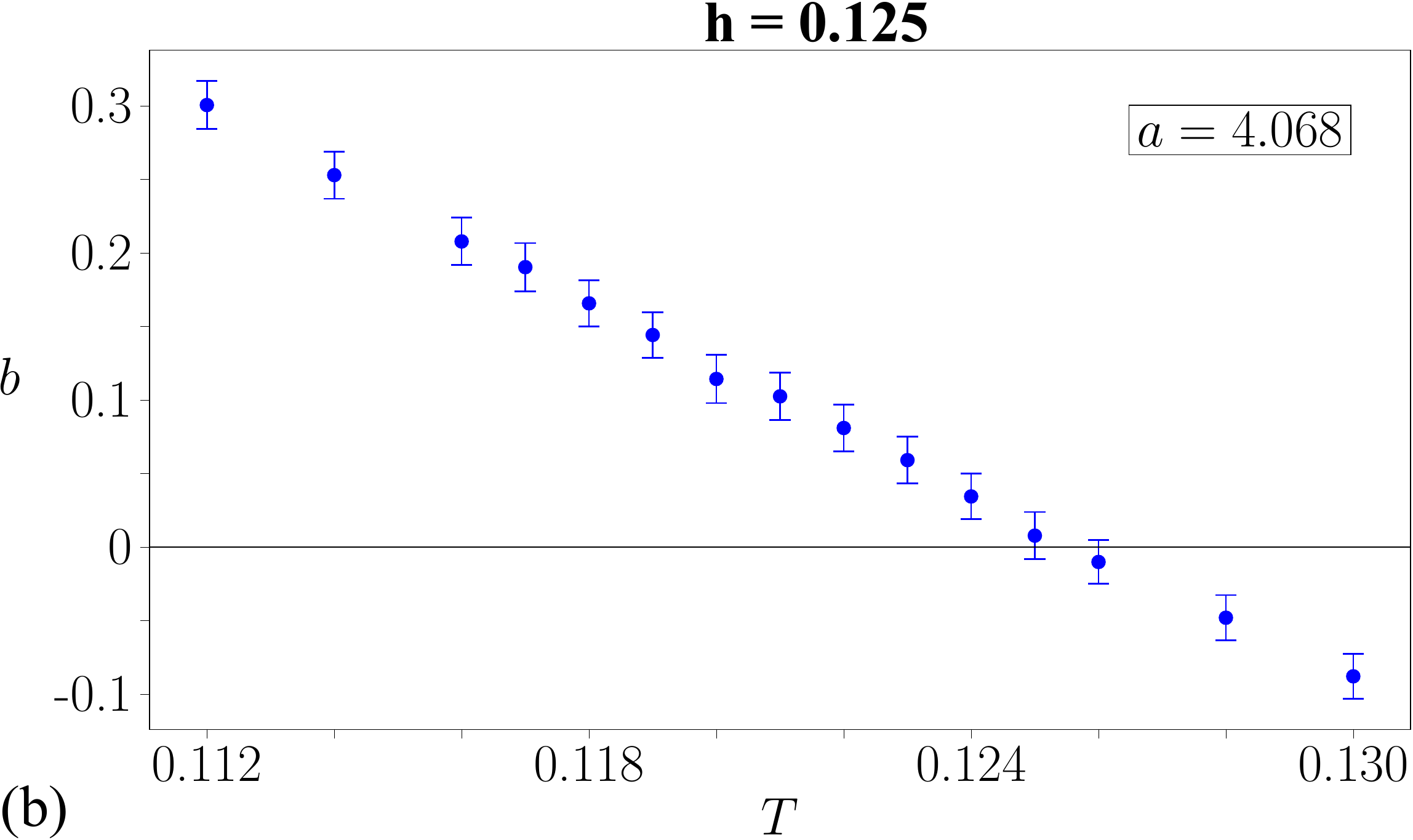} 
			\includegraphics[width=0.32\textwidth]{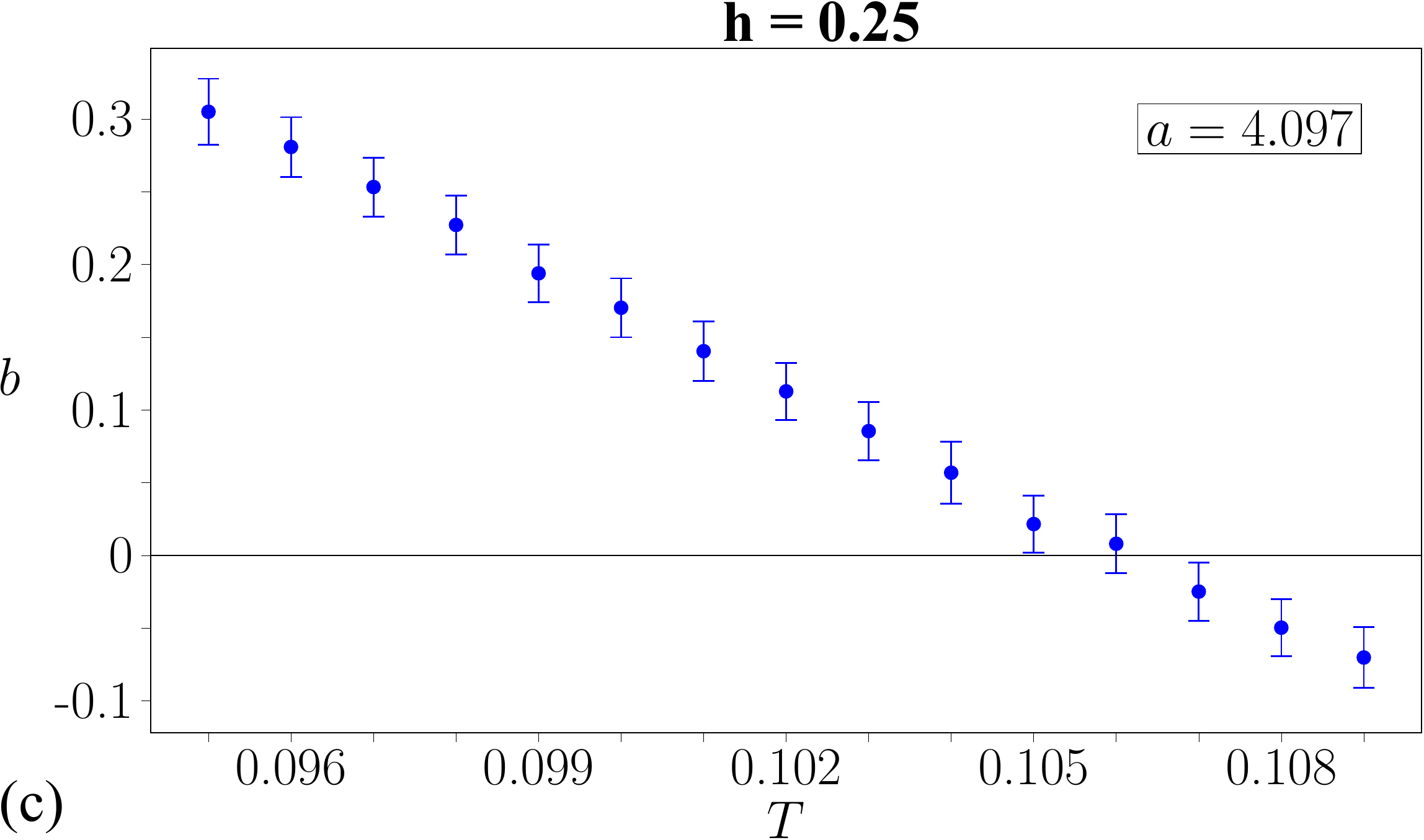} \\ [1.5ex]
			\includegraphics[width=0.32\textwidth]{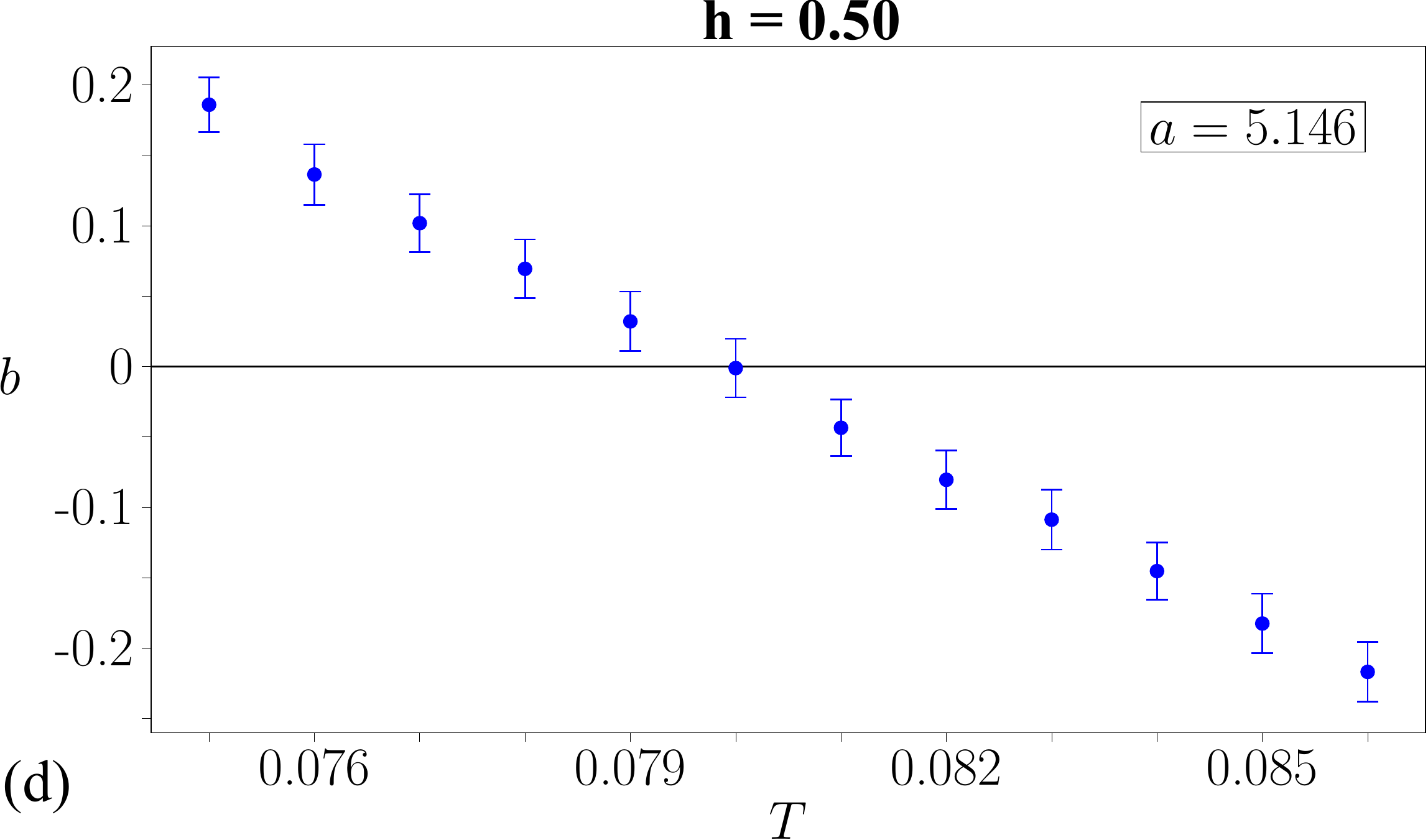} 
			\includegraphics[width=0.32\textwidth]{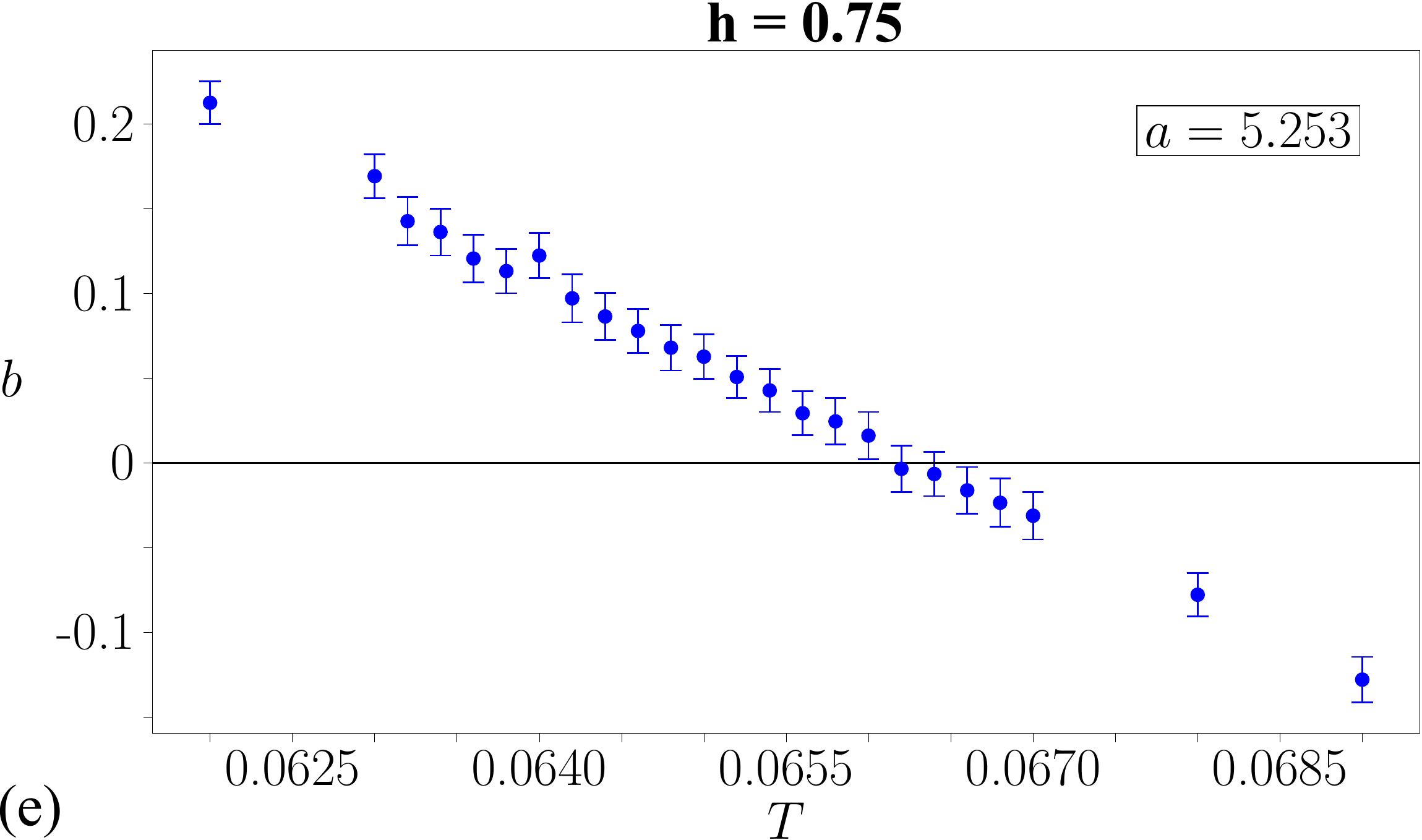} 
			\includegraphics[width=0.32\textwidth]{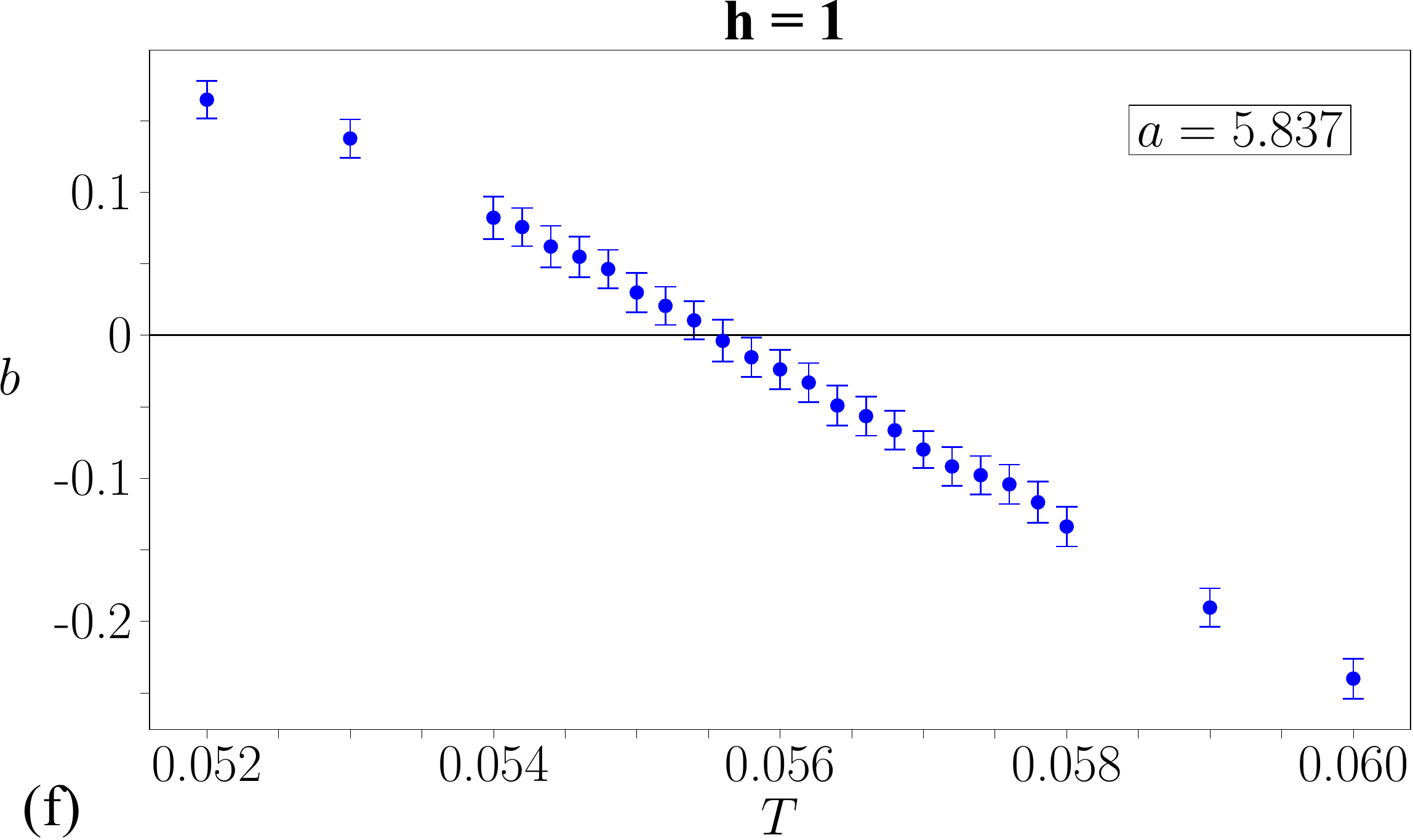} 
		\caption{Plots of parameter $b$  as a function of temperature $T$ obtained from fitting memory lifetime of Toom's rule into the ansatz in Eq.~\eqref{ansatz}. The critical temperatures obtained are $T^{0}_{c} = 0.177 \pm 0.005$, $T^{0.125}_{c} = 0.125 \pm 0.003$, $T^{0.25}_{c} = 0.106 \pm 0.002$, $T^{0.50}_{c} = 0.080 \pm 0.002$, $T^{0.75}_{c} = 0.0662 \pm 0.0008$ and $T^{1.00}_{c} = 0.0556 \pm 0.0006$.} 
		\label{SqTMembvT}
	\end{figure}
	
	We also report the same numerical analysis (see Fig.~\ref{SqTMembvT}) of memory lifetime for Toom's rule on square lattices and obtain the corresponding phase diagram (see Fig.~\ref{PhaseDiagToom}). This phase diagram corroborates the prediction in Ref.~\cite{bennettRoleIrreversibilityStabilizing1985} of the existence of a region of non-zero measure in the $(T,h)$ plane where two stable phases coexist.

    \section{Details of statistical-mechanical analysis}\label{StatMechApp}
\begin{figure}[ht!]
		\centering
		\includegraphics[width=0.35\textwidth]{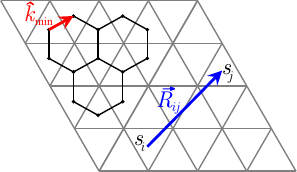}
		\caption{The sweep rule is implemented on a triangular lattice with periodic boundary conditions. A lattice of linear lattice size $L = 4$ is shown here. The $N=32$ spins situated at the face centres of the triangles form a honeycomb lattice (black edges) whose wave vector lies along the unit vector $\hat{k}_{\text{min}}$ (red). The displacement vector $\Vec{R}_{i,j}$ from $s_{i}$ to $s_{j}$ is depicted in blue.
        }
		\label{TriangLat}
	\end{figure}

    We simulate the sweep rule at $h=0$ for $2^{26}-1$ time steps, divided into consecutive windows of size
    $W \in \{2^0,2^1, \ldots, W_\mathrm{max} = 2^{25}\}$ for linear lattice sizes $L = \{12, 16, 24, 32, 48, 64, 96\}$. The states of the system at each time step in the last window of the simulation (from $t = 2^{25}$ to $t=2^{26}-1$) together constitute an ensemble of states sampled from the stationary distribution of the PCA. We estimate ensemble averages of a property $A$ of the PCA as
    \begin{equation}
        \langle A \rangle = \frac{1}{W_\mathrm{max}} \sum_{t=W_\mathrm{max}}^{2W_\mathrm{max}-1} A(t),
    \end{equation}
    where $A(t)$ is the value of the property $A$ of the system at time step $t$. In Sec.~\ref{StatMech}, we study the ensemble averages of the magnetization $M$, magnetic susceptibility $\chi$ and the two-point correlation function $\xi$ (see Fig.~\ref{DatCol}). These ensemble averages are estimated as follows
    \begin{align}
        \langle M \rangle &= \frac{1}{W_\mathrm{max}} \sum_{t=W_\mathrm{max}}^{2W_\mathrm{max}-1} |M(t)|, \label{AbsMag}\\
        \langle \chi \rangle &= \frac{1}{2L^{2}} \frac{\langle M^{2} \rangle - \langle M \rangle^{2}}{T}, \label{MagVar}\\
        \langle \xi \rangle &= \frac{1}{2\sin (k_\mathrm{min}/2)}\sqrt{\frac{\langle\chi_{m}(\Vec{0})\rangle}{\langle\chi_{m}(\Vec{k}_\mathrm{min})\rangle} - 1}, \label{2Correl}
    \end{align}
    where $\Vec{k}_\text{min}$ is the shortest wave vector of the lattice, 
    \begin{align}
        \chi_{m}(\Vec{k}) = \frac{1}{N}\left\langle \sum_{i,j} s_{i} s_{j} e^{i\Vec{k}\cdot\Vec{R_{i,j}}} \right\rangle,
    \end{align}
     and $\Vec{R}_{ij}$ is the displacement vector from the spin at the $i^{\text{th}}$ cell to the spin at the $j^{\text{th}}$ cell of the lattice. Note that although we represent the triangular lattice (see Fig.~\ref{SweepRule}) as a square lattice with double the number of faces for convenience, we are actually studying the triangular lattice of equilateral triangles (see Fig.~\ref{TriangLat}) with periodic boundaries. This is reflected in the way the displacement vector $\Vec{R}_{i,j}$ is calculated to estimate the ensemble average of the two-point correlation function $\xi$~\cite{andristUnderstandingTopologicalQuantum2012}.

    \section{Critical exponents of variants of the sweep rule}\label{Variants}
\begin{figure*}[hpt!]
	\centering
        \includegraphics[width=0.28\textwidth]{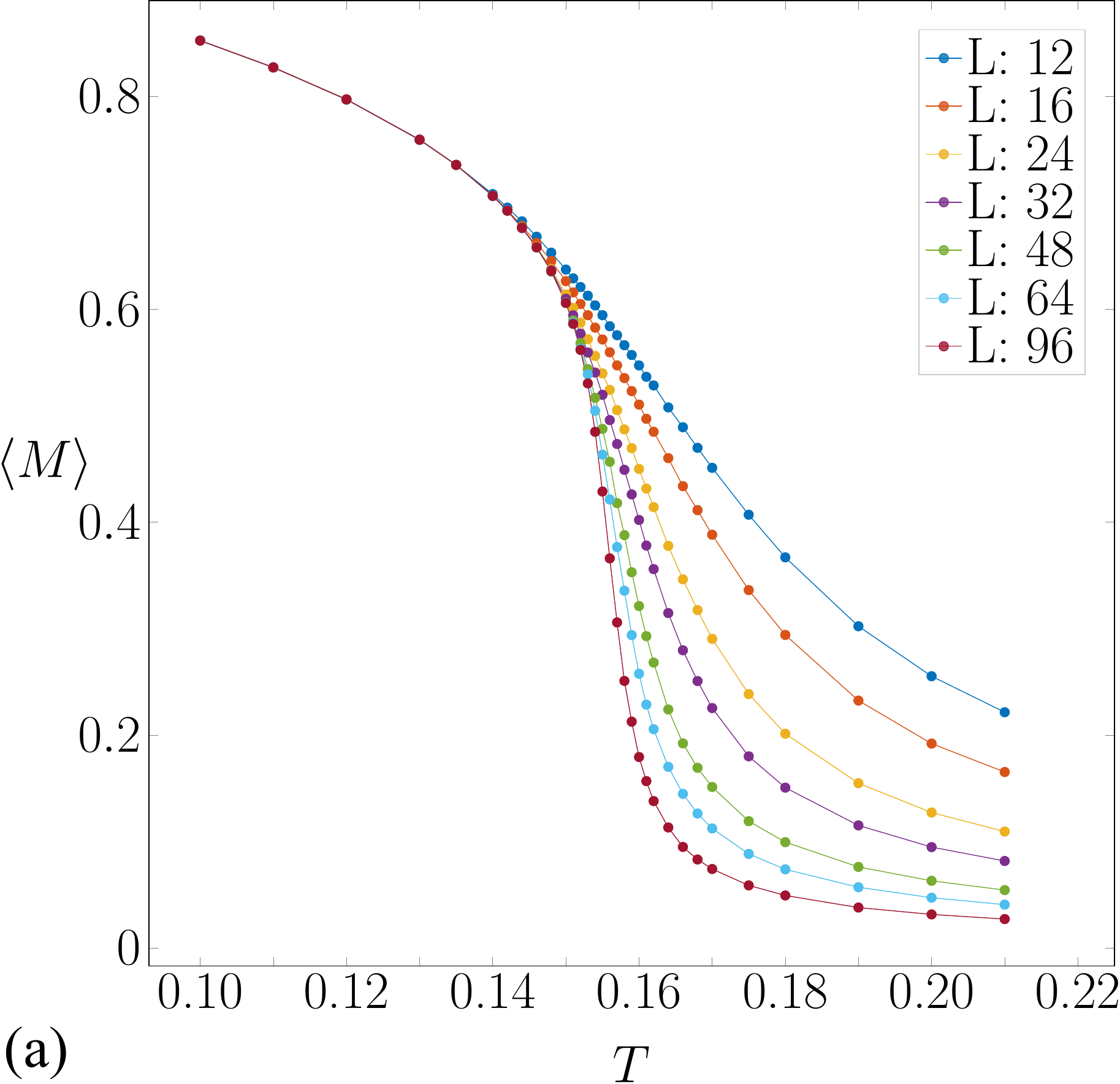}
        \includegraphics[width=0.28\textwidth]{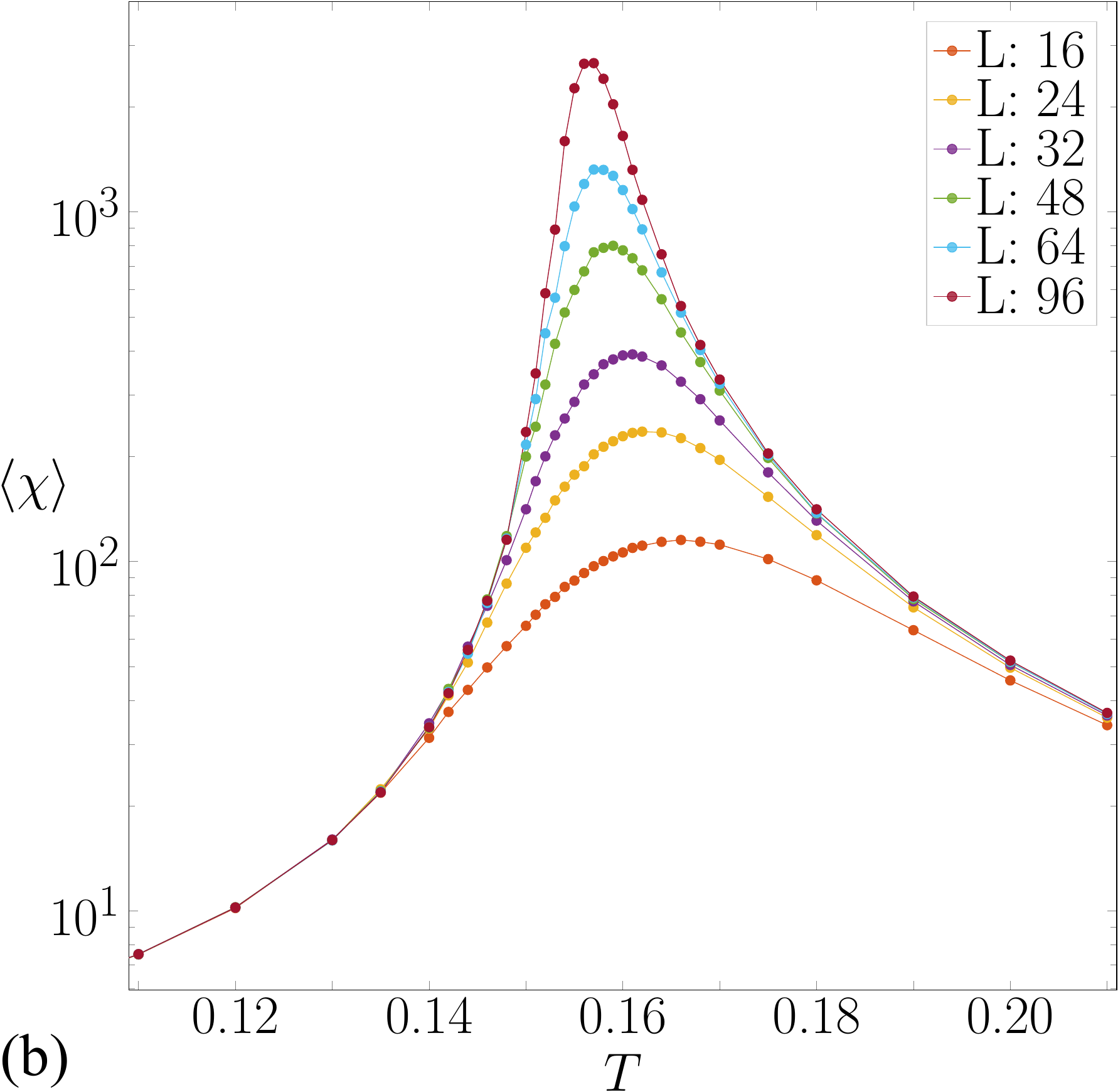}
        \includegraphics[width=0.279\textwidth]{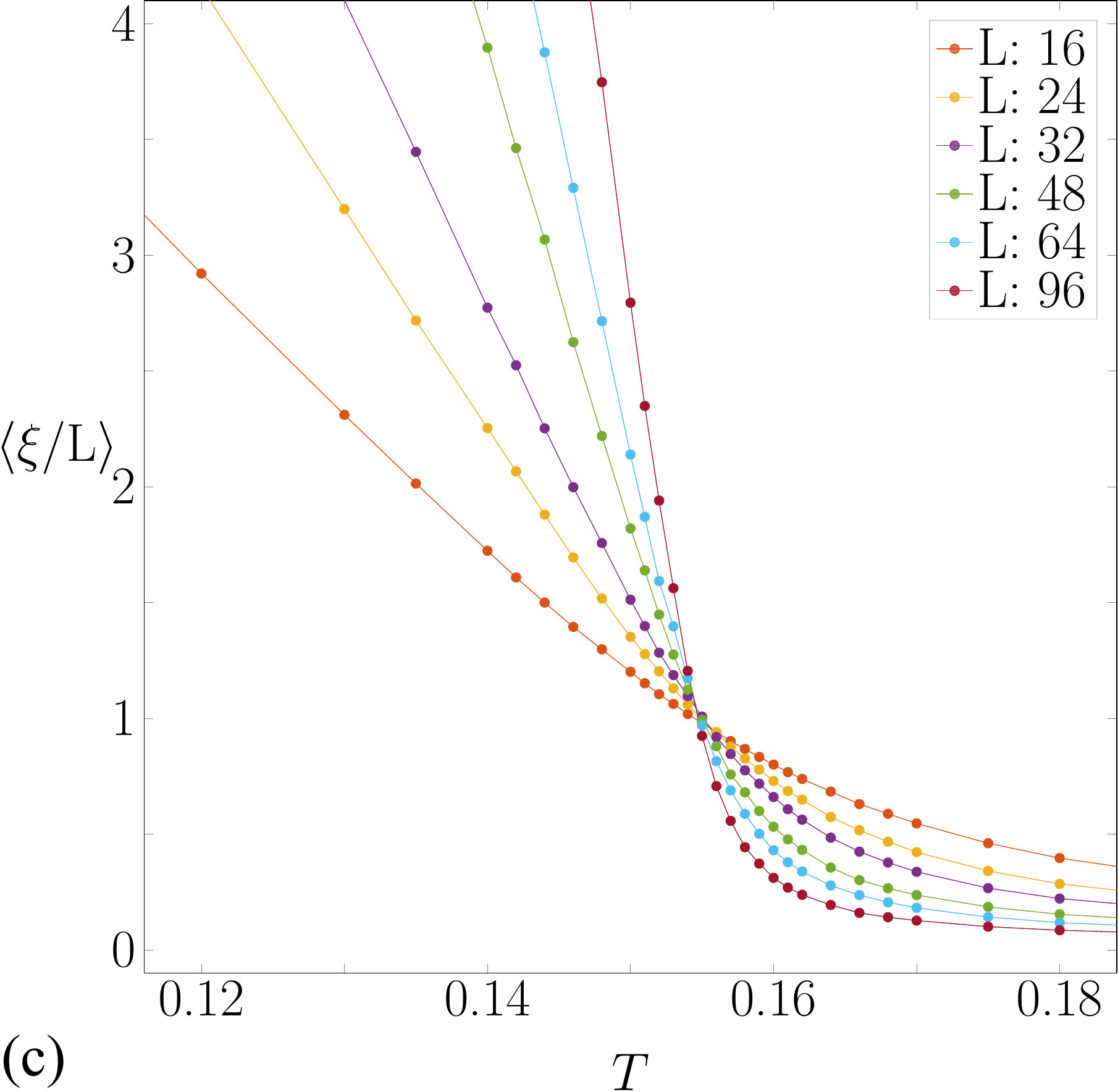}\\ [2ex]
        \includegraphics[width=0.277\textwidth]{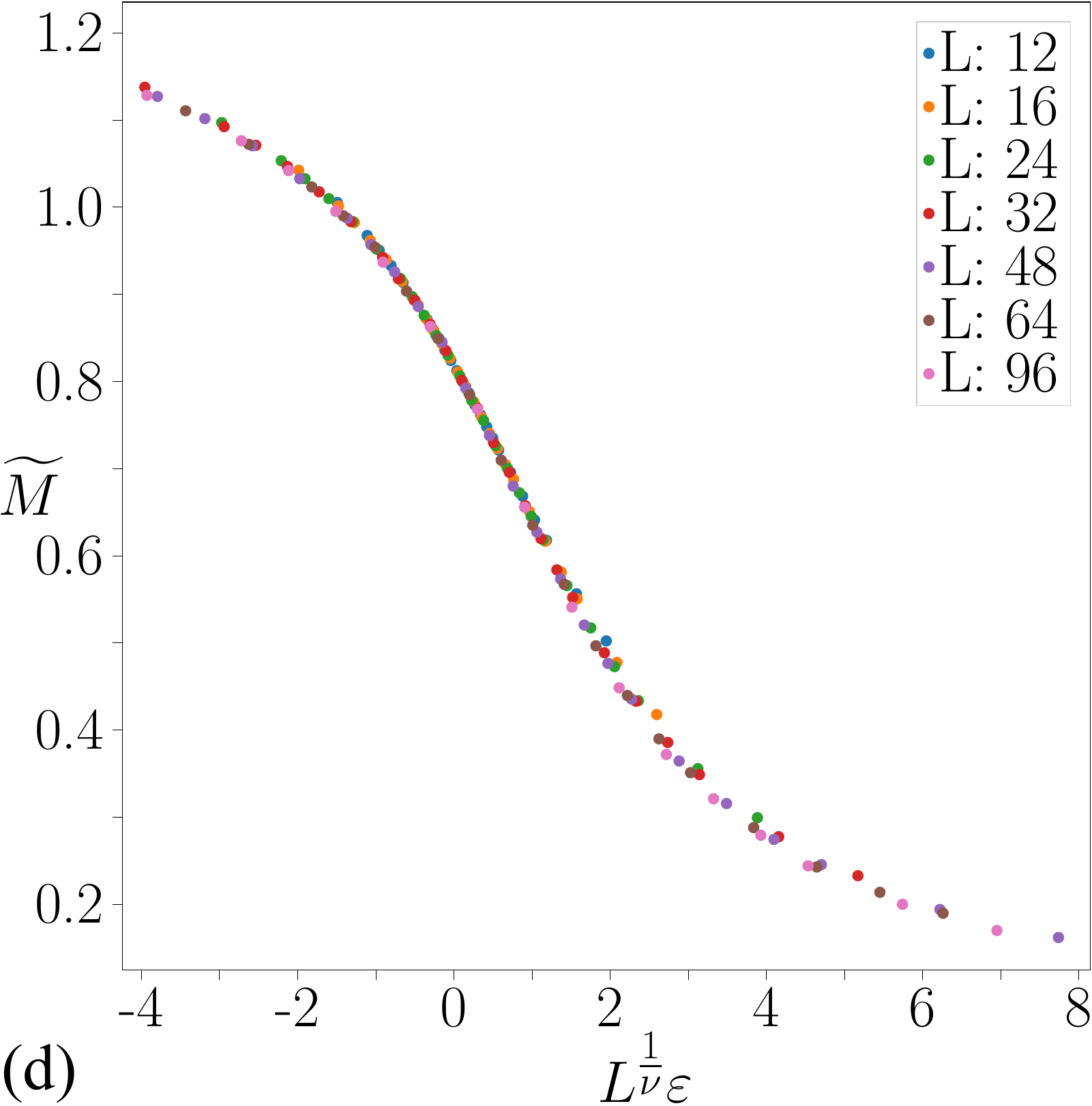}
        \includegraphics[width=0.277\textwidth]{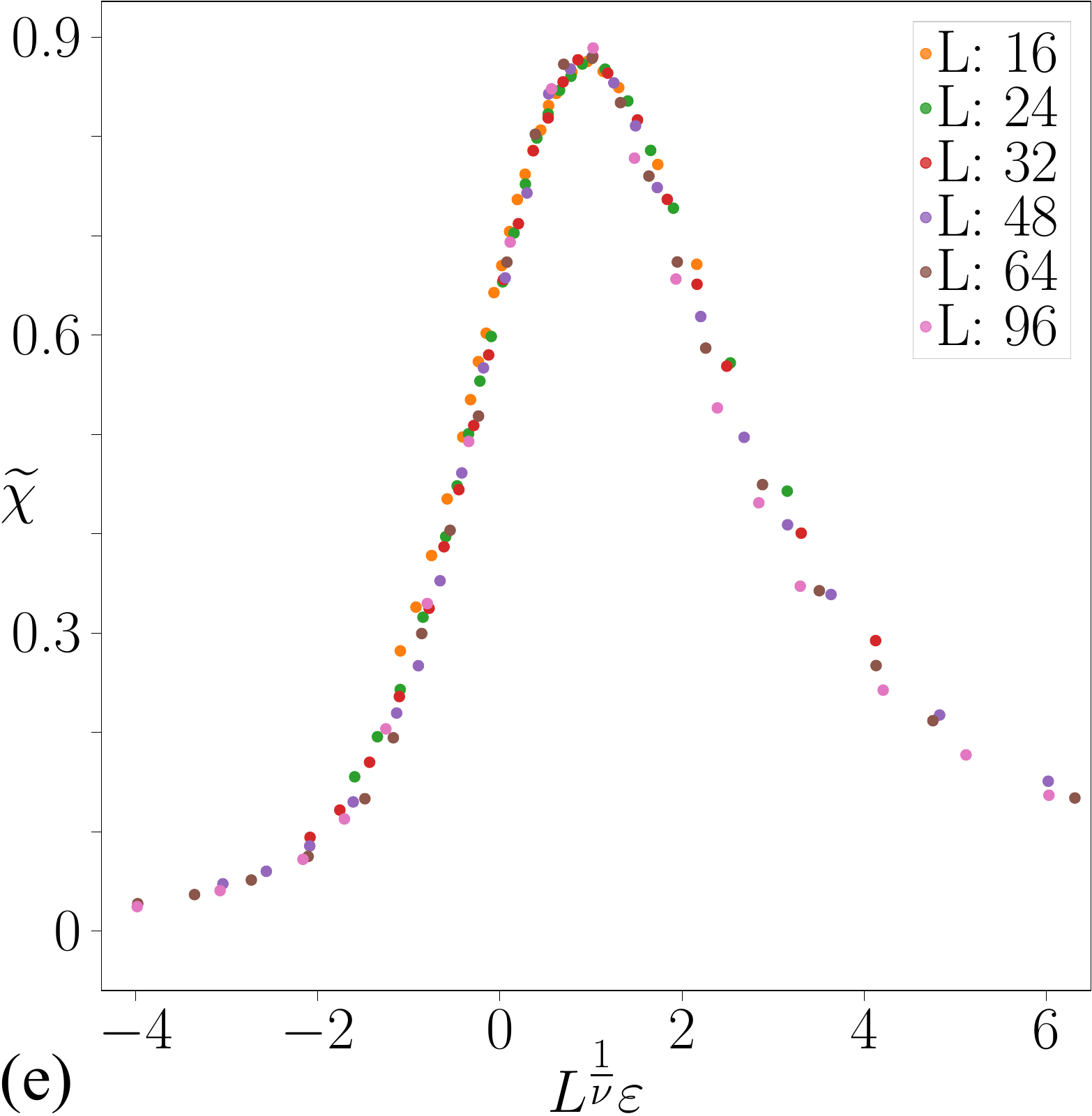}
        \includegraphics[width=0.277\textwidth]{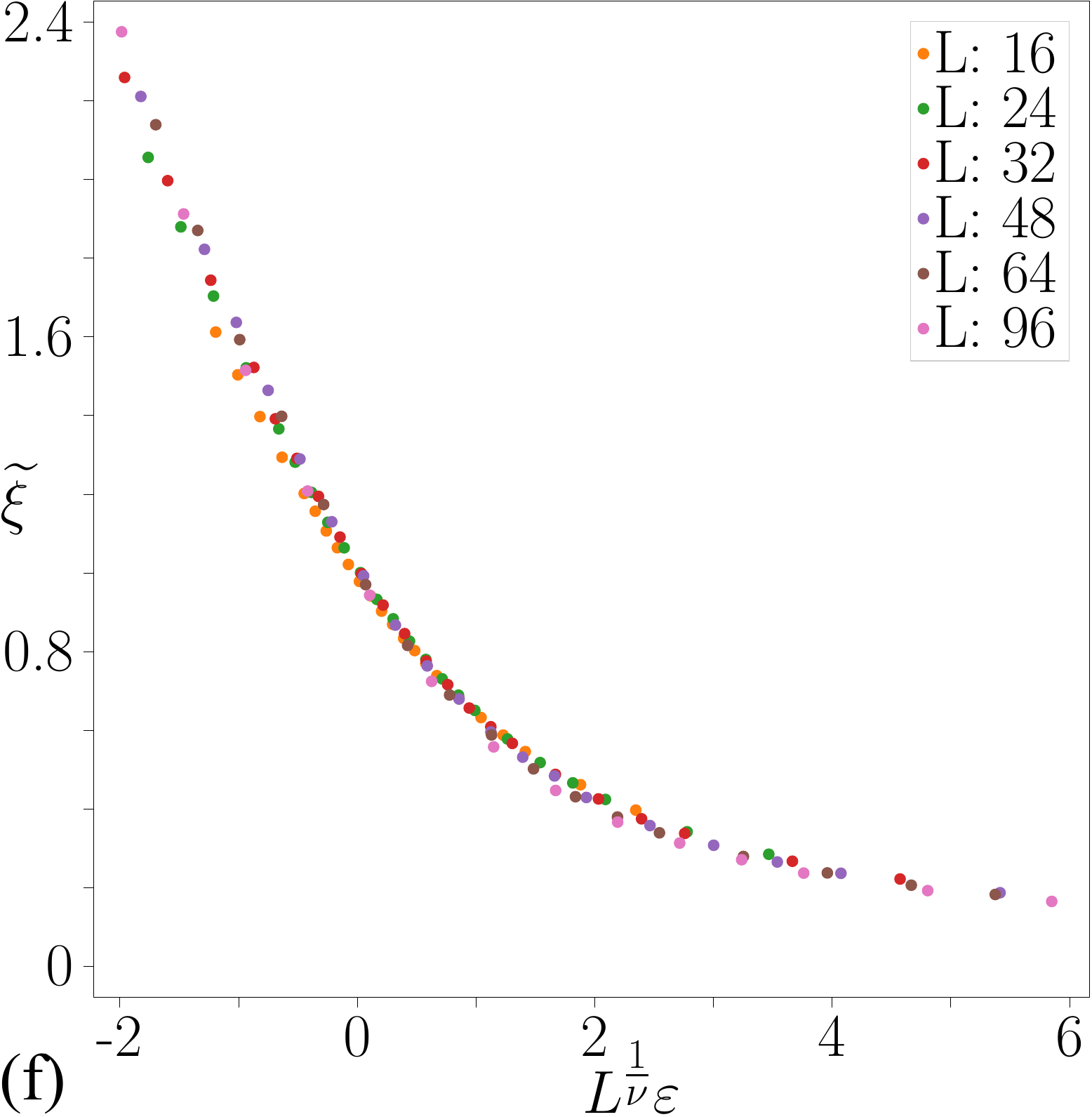}
		\caption{Critical behavior of the asynchronous sweep rule at $h=0$. In (a)-(c), we plot ensemble averages of magnetization $M$, magnetic susceptibility $\chi$ and two-point correlation function $\xi/L$, whereas in (d)-(f), we plot their respective data collapses.
  }
		\label{DatColTrAsync}
	\end{figure*}

\begin{figure*}[hpt!]
	\centering
        \includegraphics[width=0.28\textwidth]{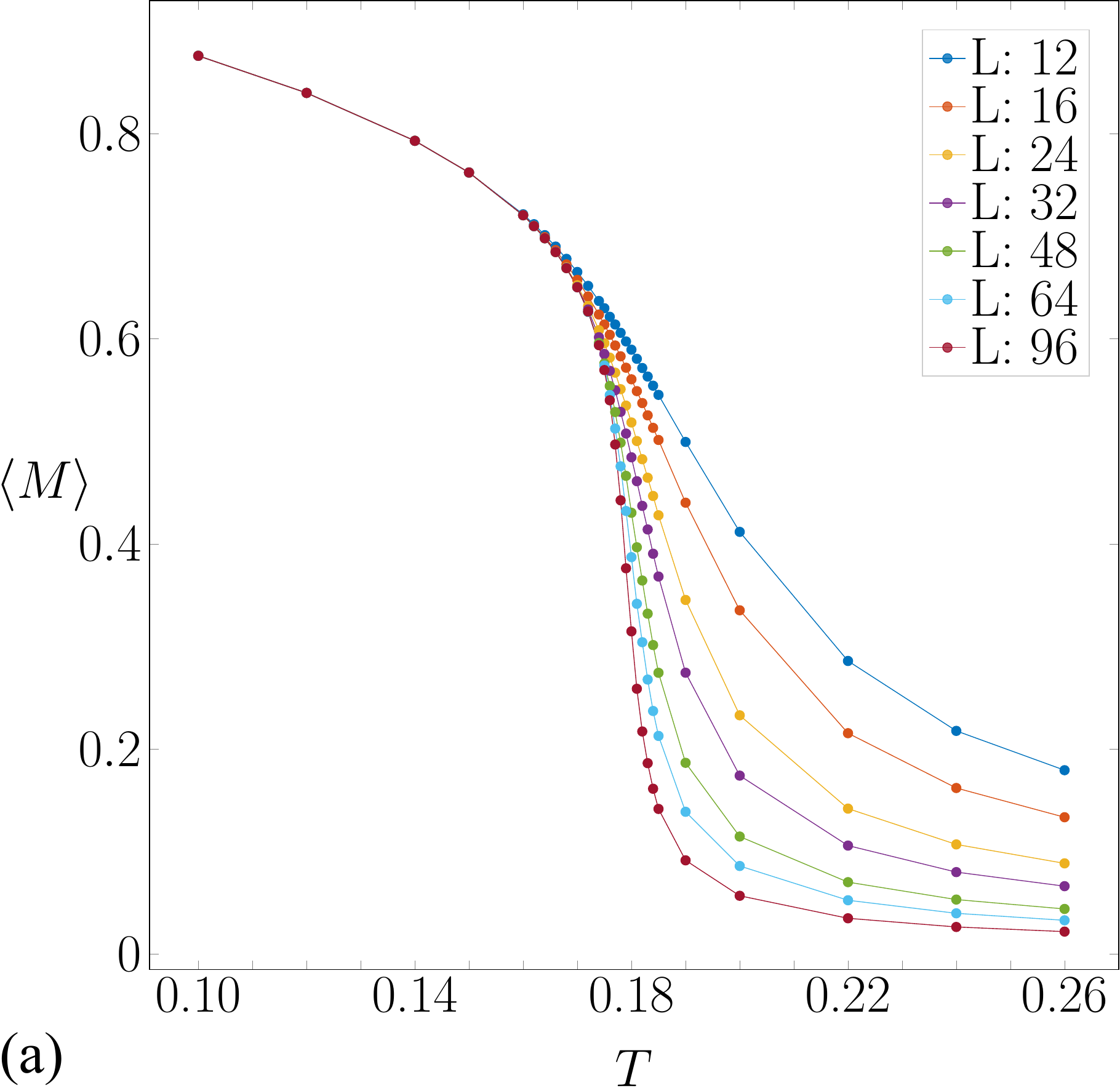}
        \includegraphics[width=0.279\textwidth]{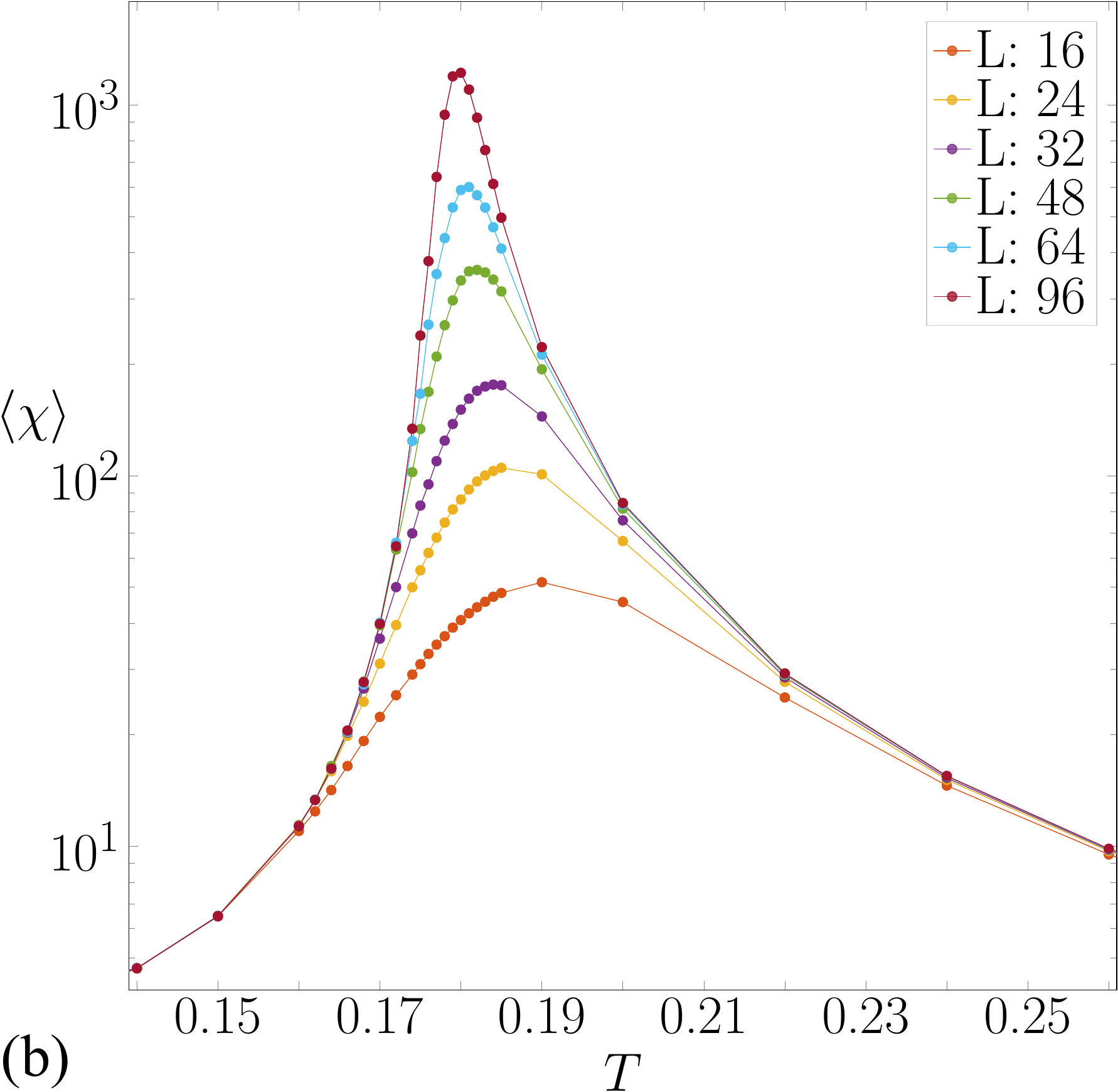}
        \includegraphics[width=0.277\textwidth]{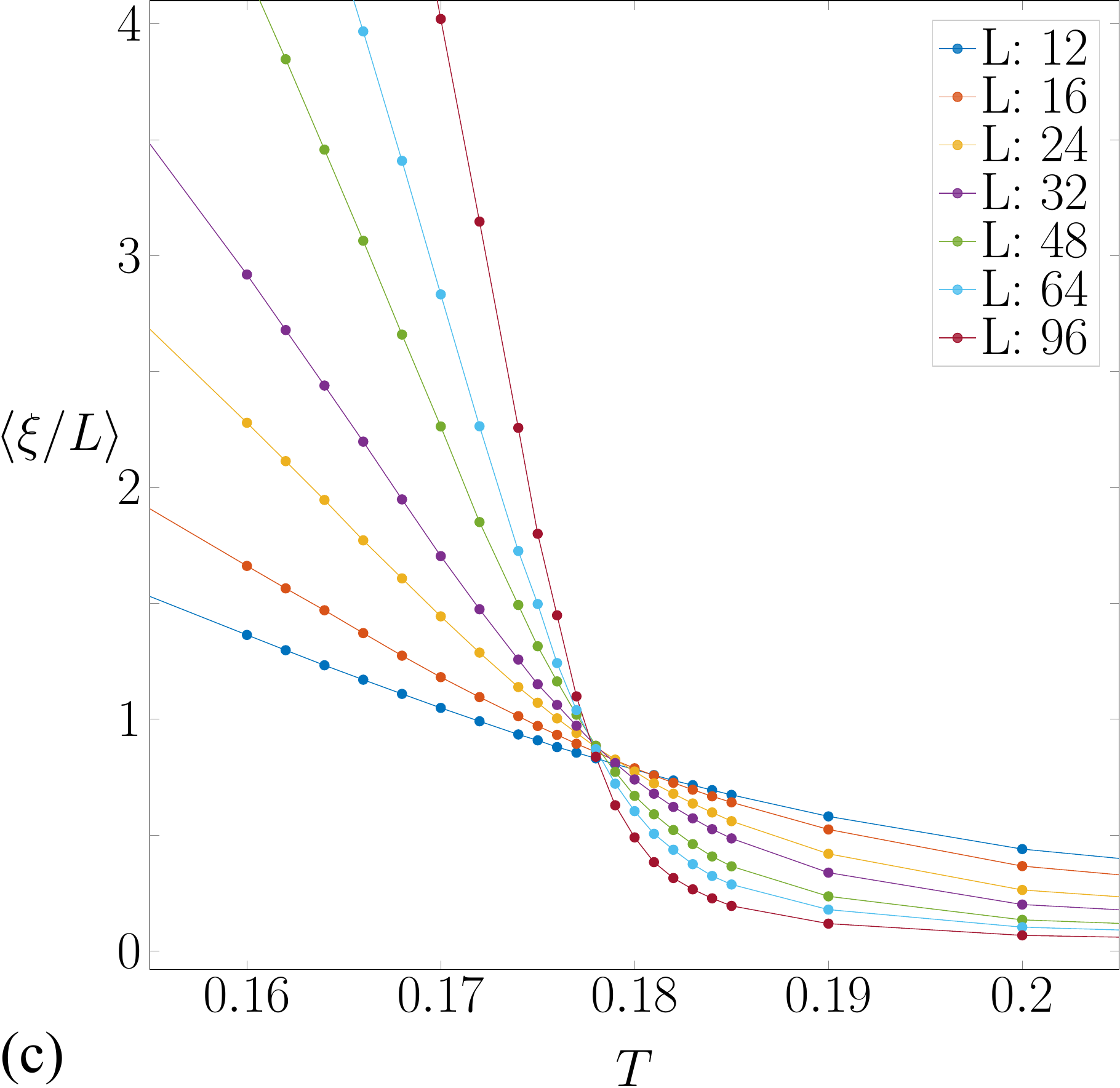}\\ [2ex]
        \includegraphics[width=0.277\textwidth]{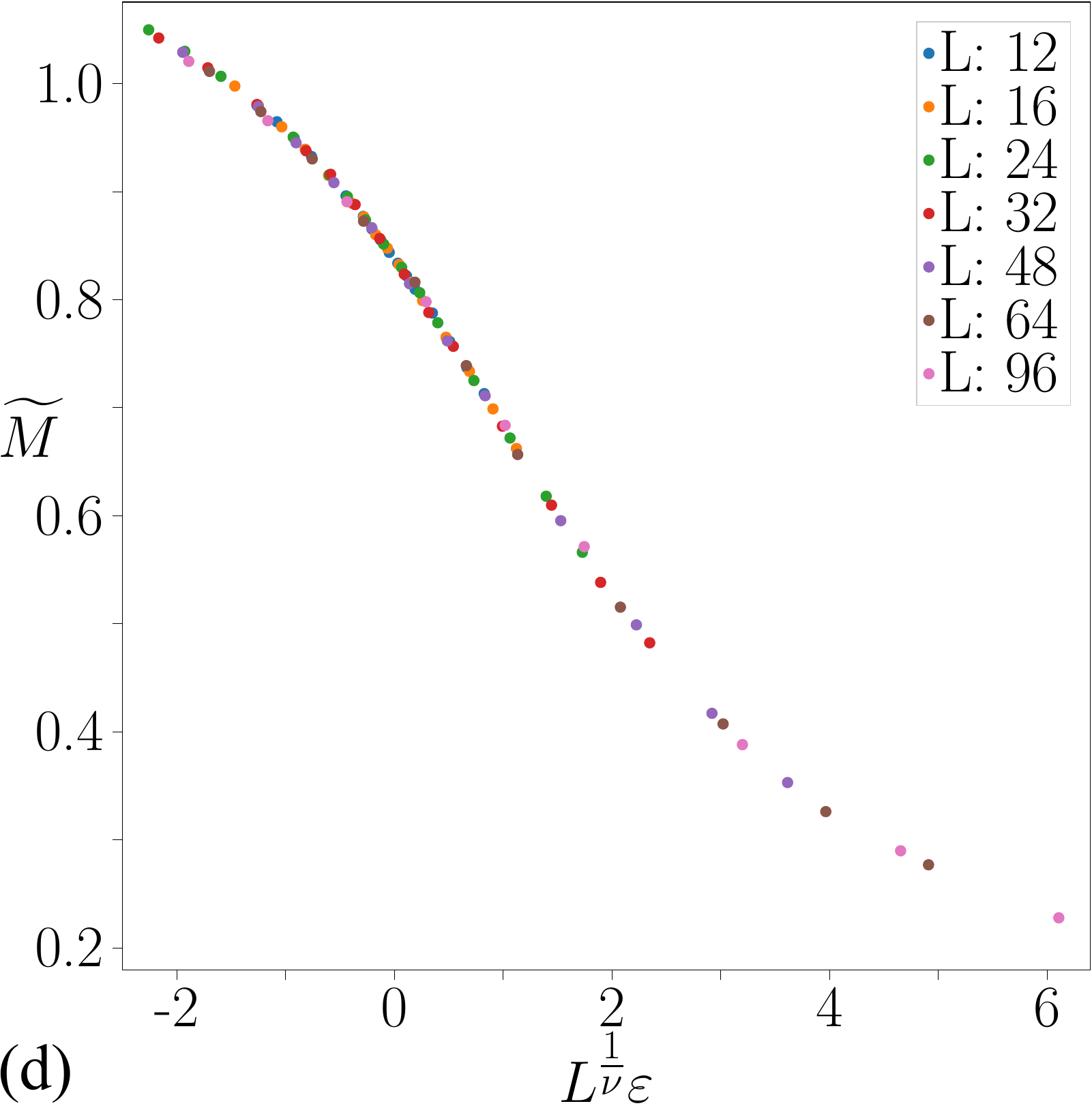}
        \includegraphics[width=0.277\textwidth]{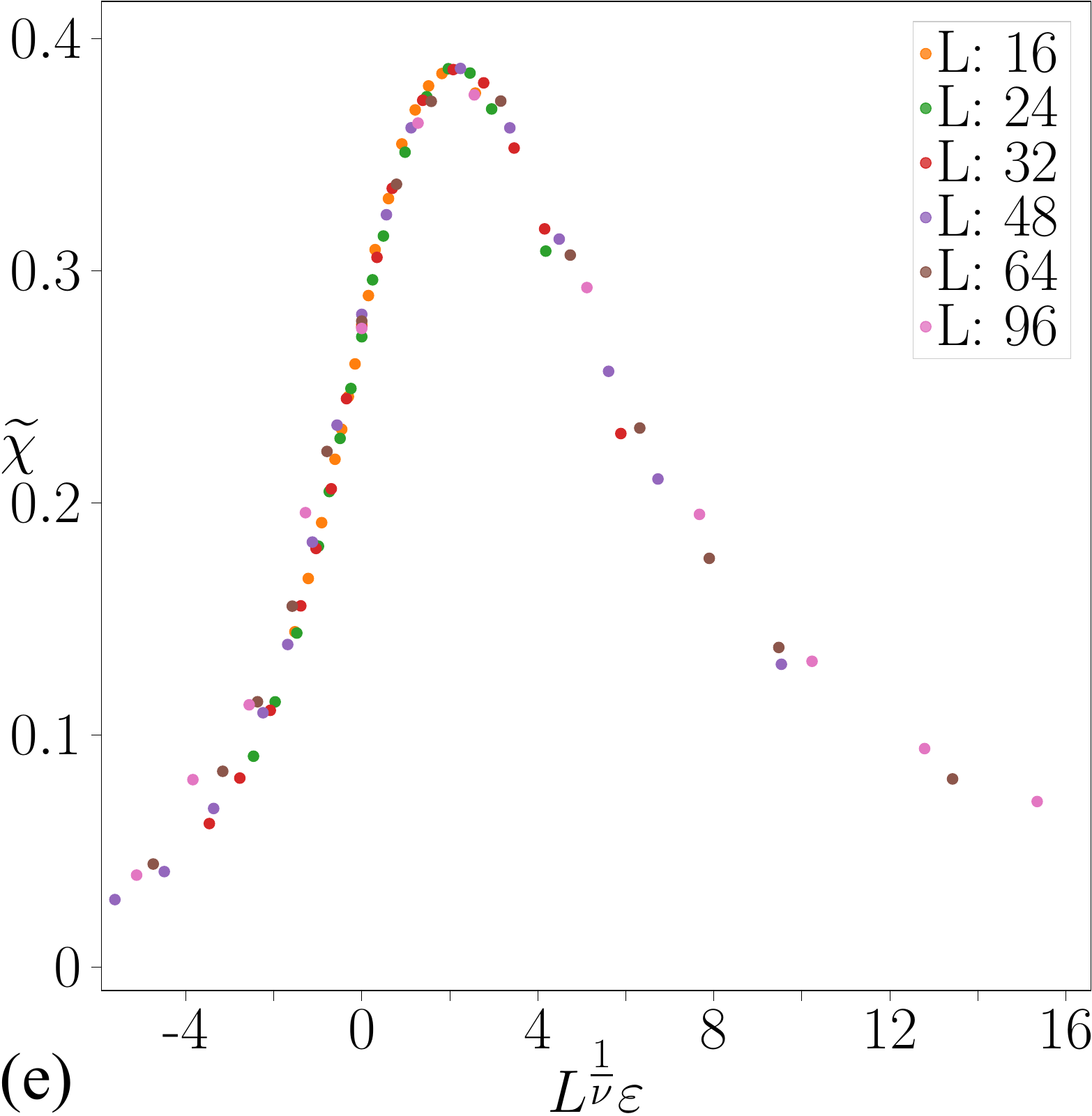}
        \includegraphics[width=0.277\textwidth]{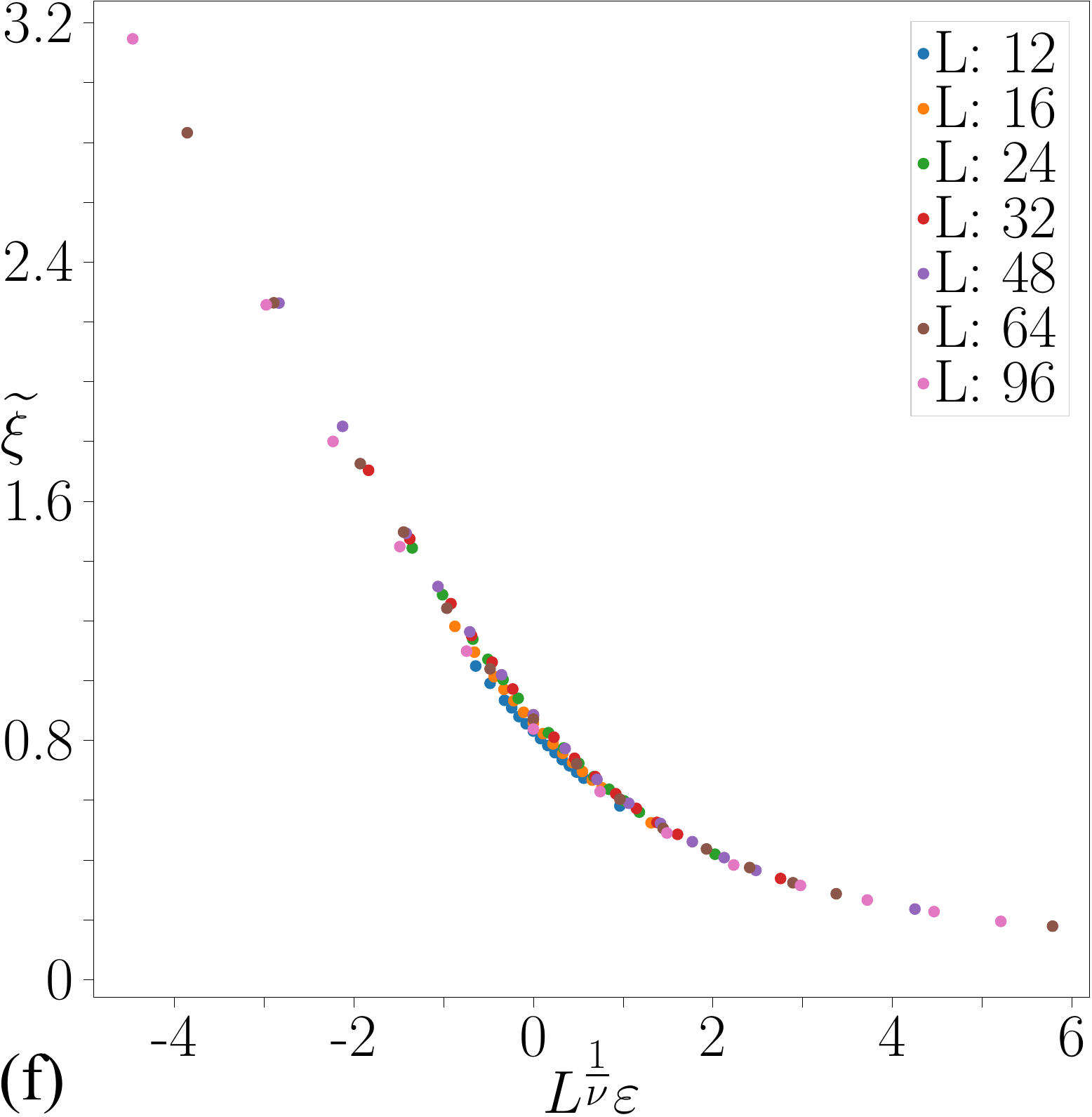}
		\caption{Critical behavior of synchronous Toom's rule at $h=0$. In (a)-(c), we plot ensemble averages of magnetization $M$, magnetic susceptibility $\chi$ and two-point correlation function $\xi/L$, whereas in (d)-(f), we plot their respective data collapses.
  }
		\label{DatColToomSync}
	\end{figure*}

\begin{figure*}[ht!]
	\centering
        \includegraphics[width=0.28\textwidth]{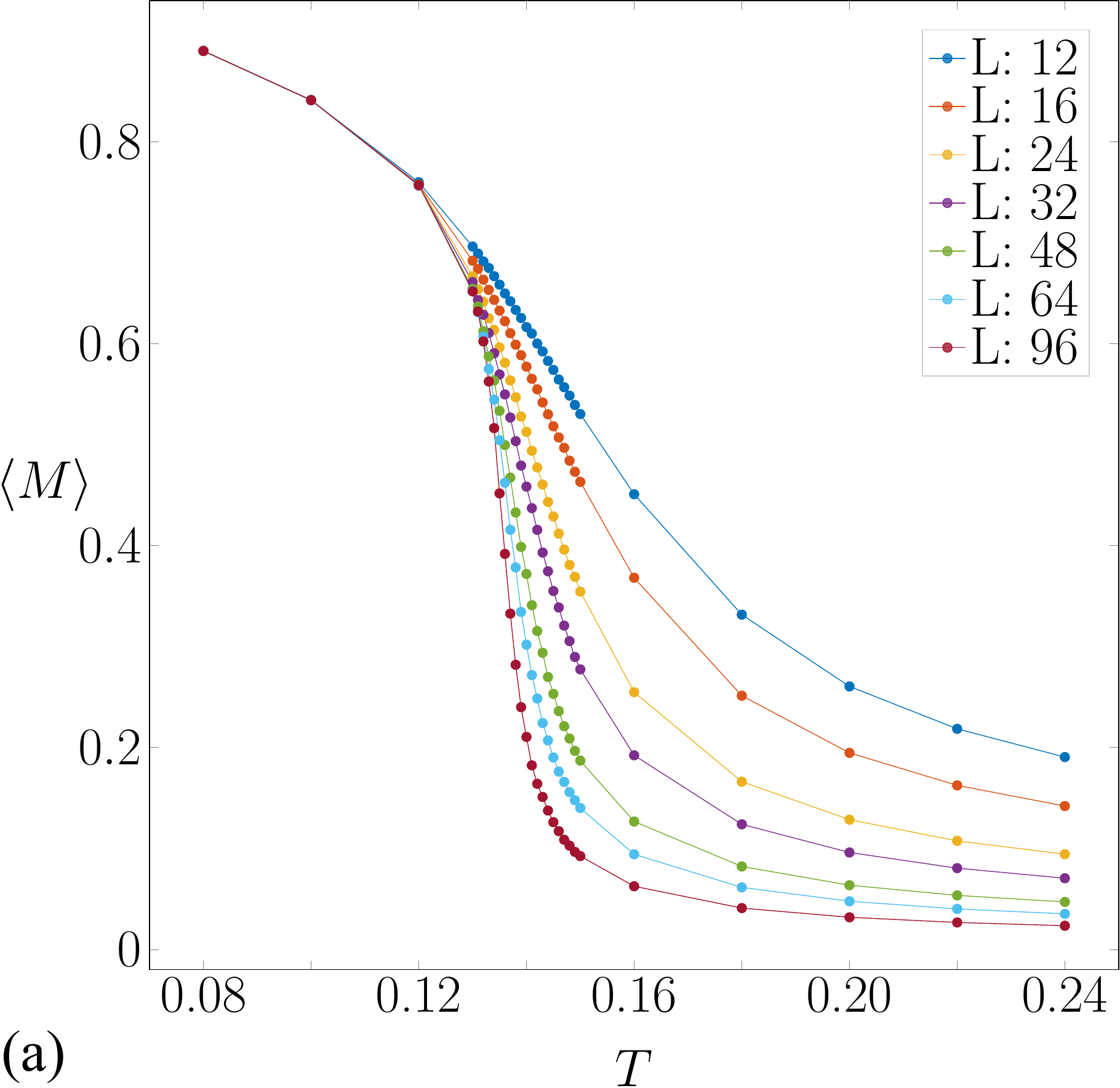}
        \includegraphics[width=0.28\textwidth]{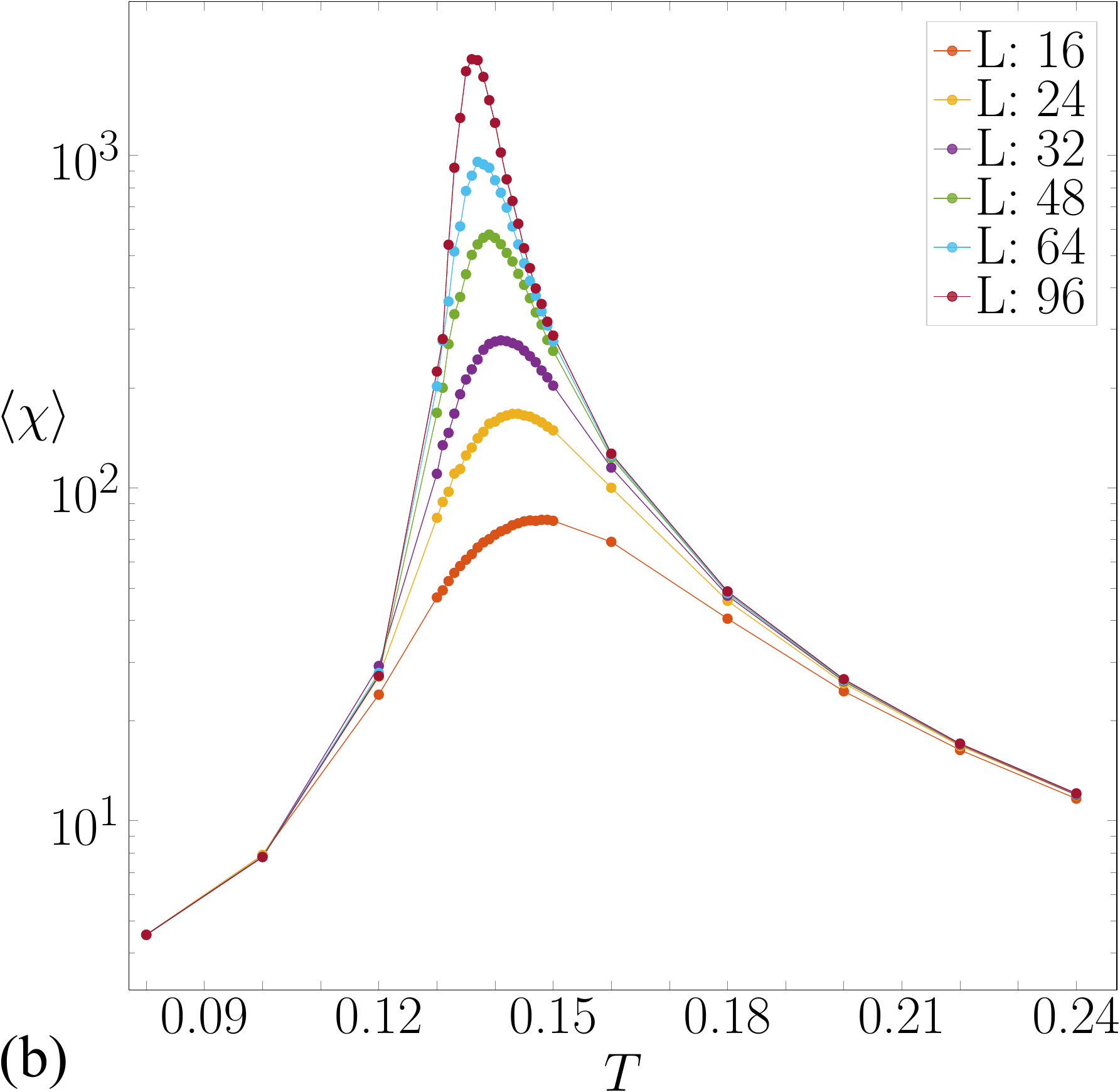}
        \includegraphics[width=0.279\textwidth]{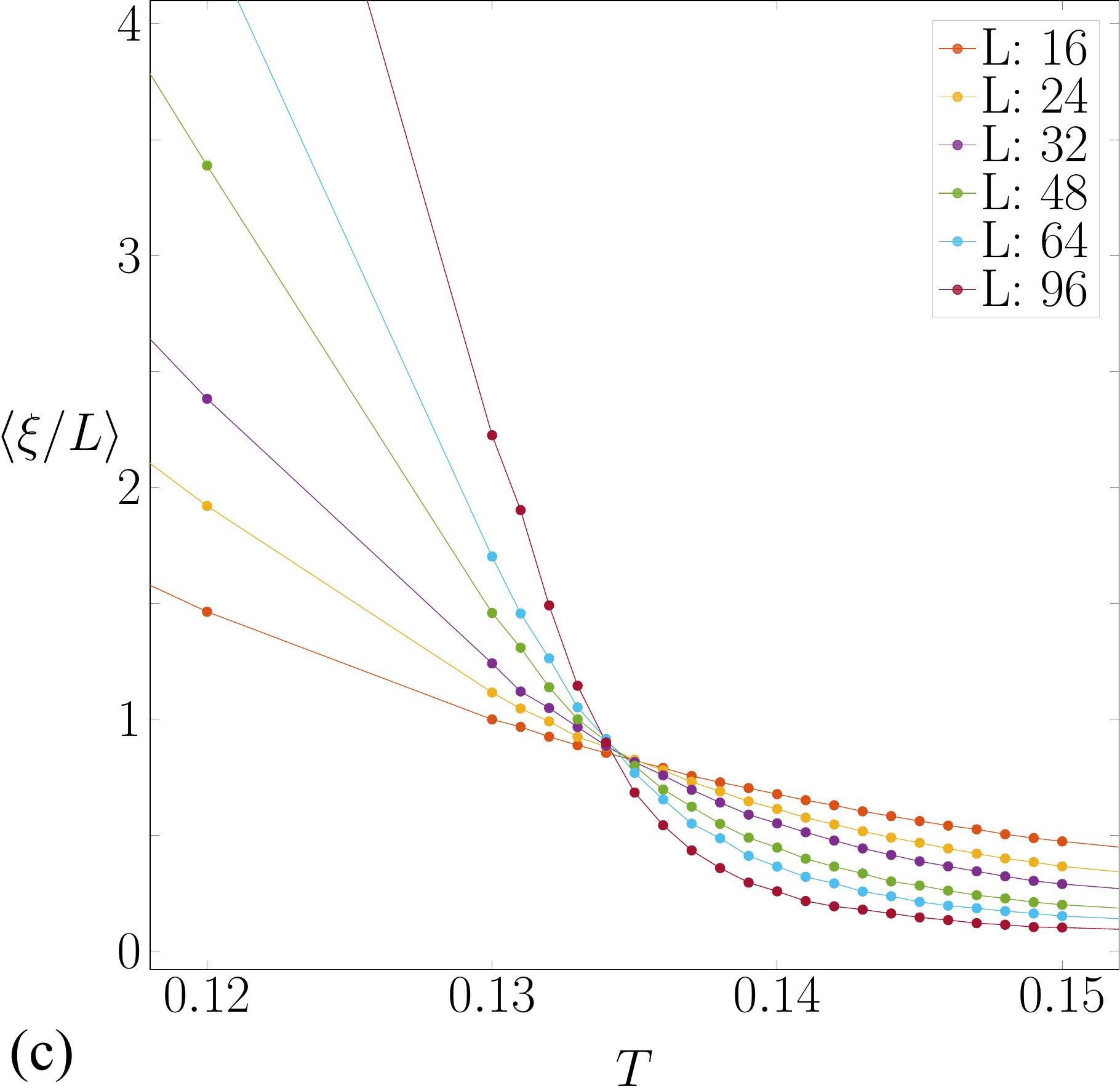}\\ [2ex]
        \includegraphics[width=0.277\textwidth]{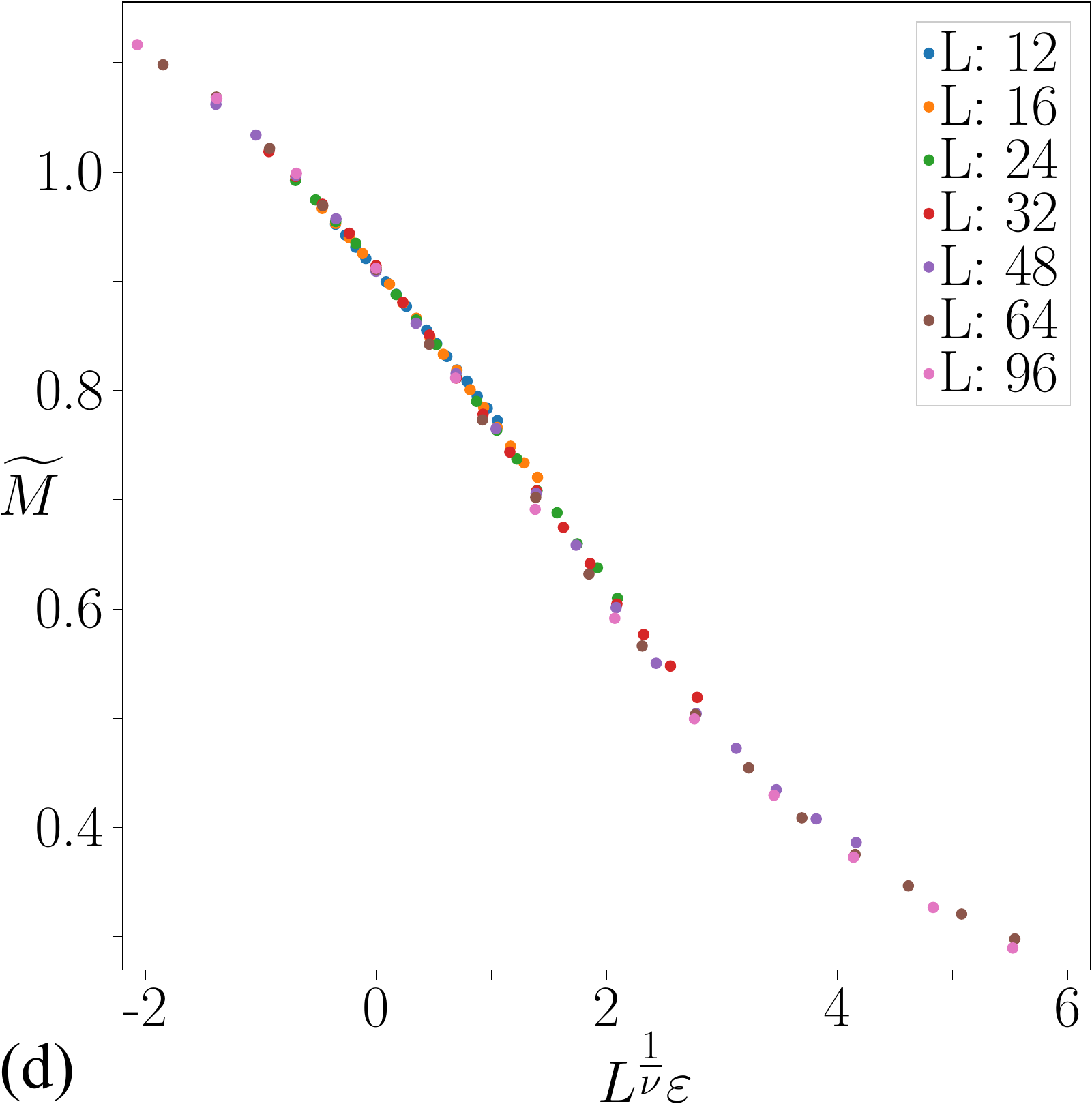}
        \includegraphics[width=0.277\textwidth]{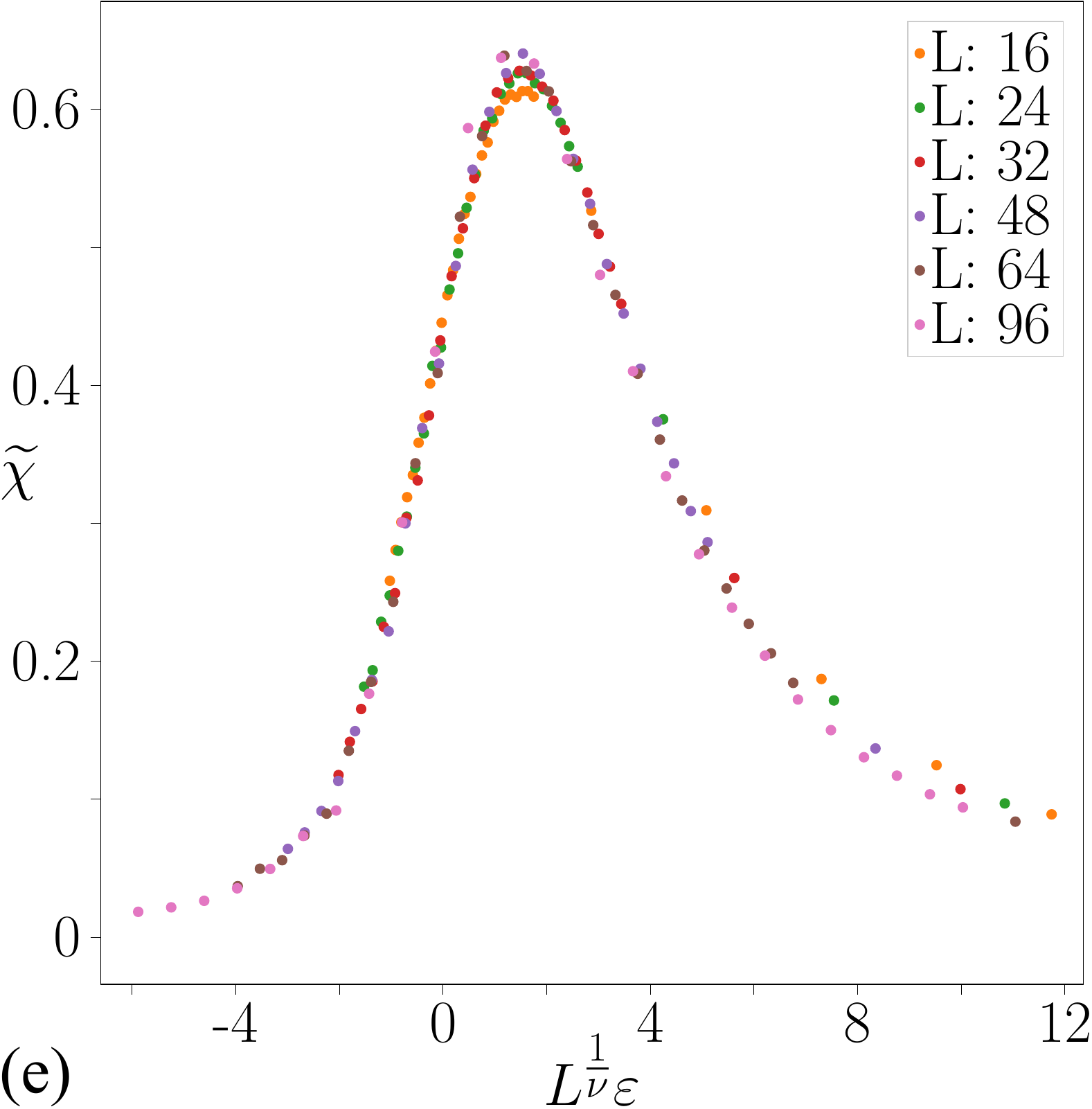}
        \includegraphics[width=0.277\textwidth]{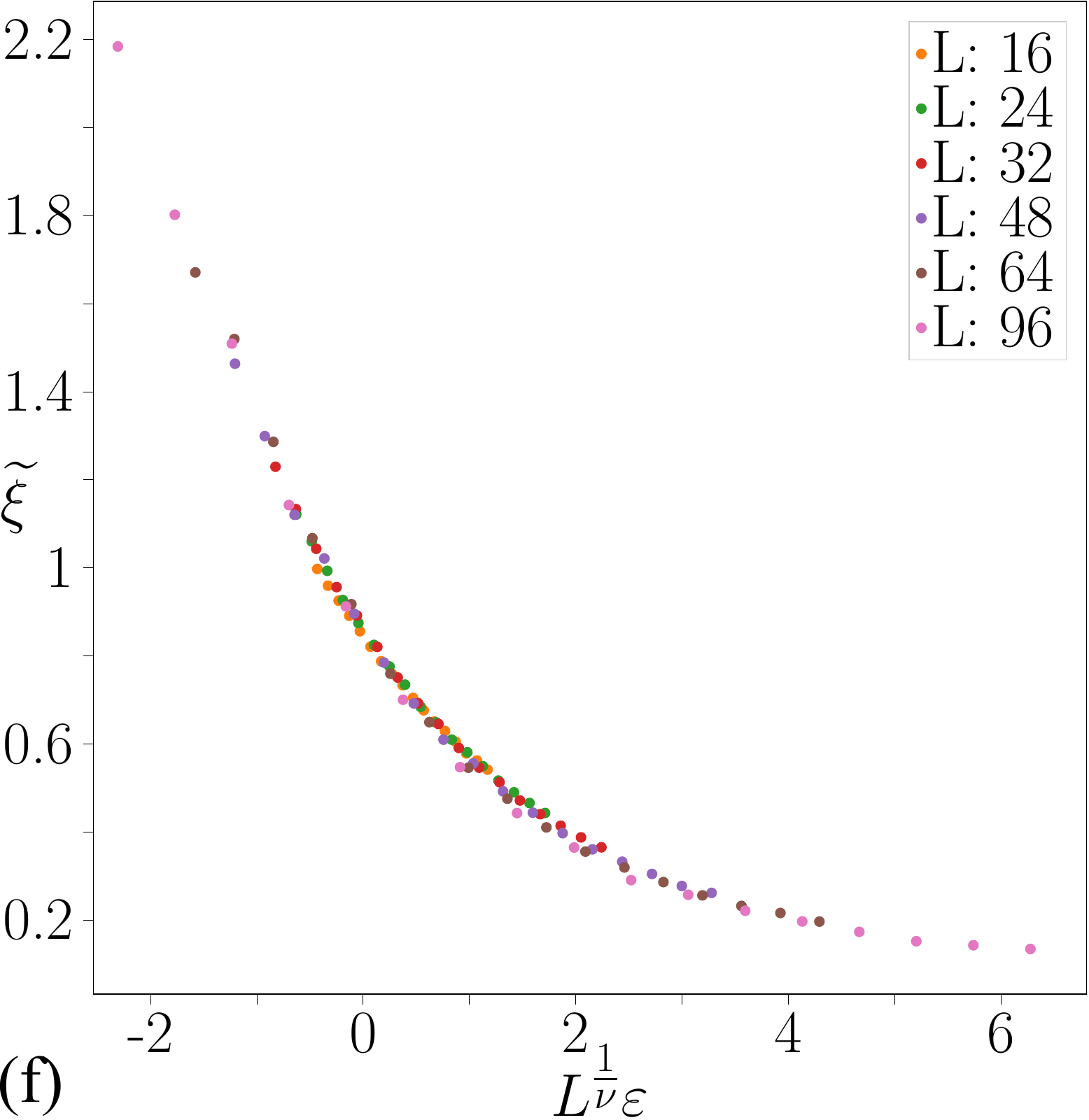}
		\caption{Critical behavior of asynchronous Toom's rule at $h=0$. In (a)-(c), we plot ensemble averages of magnetization $M$, magnetic susceptibility $\chi$ and two-point correlation function $\xi/L$, whereas in (d)-(f), we plot their respective data collapses.
  }
		\label{DatColToomAsync}
	\end{figure*}

    As discussed in Sec.~\ref{ClassMem}, the sweep rule can be implemented synchronously or asynchronously. Similarly, Toom's rule (which is equivalent to the sweep rule acting on square lattices) can also be implemented both synchronously and asynchronously. In Table~\ref{table:CritExps}, we report the critical temperature $T^0_{c}$ and critical exponents $\nu, \beta, \gamma$ obtained from the statistical-mechanical analysis of each of these variants at $h=0$ (see Figs.~\ref{DatColTrAsync},~\ref{DatColToomSync}~and~\ref{DatColToomAsync}). The critical exponents we report for Toom's rule are in agreement with those reported previously in Ref.~\cite{makowiecUniversalityClassProbabilistic2002}. In case of the synchronous variant of both sweep rule and Toom's rule, the ratios $\beta/\nu$ and $\gamma/\nu$ are equal to those of the 2D Ising model. Therefore, they belong to the weak universality class of the Ising model \cite{makowiecUniversalityClassProbabilistic2002}. The asynchronous variants of both rules have the same critical exponents as the 2D Ising model (with some discrepancies in our reported values of $\gamma$).

	\begin{table}[h!]
    \centering
    \begin{tabular}{ |c|c|c|c|c| } 
        \hline
        PCA & $\nu$ & $\beta$ & $\gamma$ & $T^{0}_{c}$\\ \hline
        synchronous sweep & $0.926(7)$ & $0.120(2)$ & $1.62(1)$ & $0.1973(1)$\\ \hline
        asynchronous sweep & $1.05(3)$  & $0.132(6)$ & $1.88(3)$ & $0.1548(1)$\\ \hline
        synchronous Toom's & $0.89(3)$  &  $0.112(6)$ &  $1.54(6)$ & $0.1779(1)$\\ \hline
        asynchronous Toom's & $1.08(2)$  & $0.130(6)$ & $1.85(6)$ & $0.1342(1)$\\ \hline
    \end{tabular}
    \caption{Critical exponents of the synchronous and asynchronous variants of the sweep rule (on triangular lattices) and Toom's rule (on square lattices).}
    \label{table:CritExps}
    \end{table}

\bibliography{citations.bib}
\end{document}